\documentclass[12pt]{article}
\usepackage{arxiv_0} % 12 pt

\graphicspath{{./Figures/}}

\usepackage[onehalfspacing]{setspace}

\usepackage[margin=1in]{geometry}

\title{
Boosting Stochastic Optimisation for High-dimensional Latent Variable Models
}
\author{Motonori Oka}
\author{Yunxiao Chen}
\author{Irini Moustaki}
\affil{Department of Statistics, \protect\\London School of Economics and Political Science}
\date{}

\begin{document}

\maketitle

\begin{abstract}
\noindent Latent variable models are widely used in social and behavioural sciences, including education, psychology, and political science. With the increasing availability of large and complex datasets, high-dimensional latent variable models have become more common. However, estimating such models via marginal maximum likelihood is computationally challenging because it requires evaluating a large number of high-dimensional integrals. Stochastic optimisation, which combines stochastic approximation and sampling techniques, has been shown to be effective. It iterates between sampling latent variables from their posterior distribution under current parameter estimates and updating the model parameters using an approximate stochastic gradient constructed from the latent variable samples.
In this paper, we investigate strategies to improve the performance of stochastic optimisation for high-dimensional latent variable models. The improvement is achieved through two strategies: a Metropolis-adjusted Langevin sampler that uses the gradient of the negative complete-data log-likelihood to sample latent variables efficiently, and a minibatch gradient technique that uses only a subset of observations when sampling latent variables and constructing stochastic gradients. Our simulation studies show that combining these strategies yields the best overall performance among competitors. An application to a personality test with 30 latent dimensions further demonstrates that the proposed algorithm scales effectively to high-dimensional settings.
\\

\end{abstract}

\section{Introduction}
Latent variable models are widely used in social and behavioural sciences, such as education, psychology, and political science \parencite{bartholomew2008analysis}. In recent years, models with many latent variables have been needed to analyse large and complex datasets. Examples include non-linear factor models for large-scale item response data \parencite[e.g.,][]{schilling_high-dimensional_2005,cai_high-dimensional_2010} and collaborative filtering \parencite[e.g.,][]{zhu2016personalized}, dynamic factor models for intensive longitudinal data \parencite[e.g.,][]{lu2015bayesian, chen2020latent}, latent space models for network data \parencite[e.g.,][]{hoff2002latent}, and multilevel models \parencite[e.g.,][]{skrondal_generalized_2004}. Statistical inference for these models is computationally challenging due to the large number of latent variables/random effects involved.

The Marginal Maximum Likelihood Estimator (MMLE) is a popular approach for estimating latent variable models \parencite[Chapter 6,][]{skrondal_generalized_2004}, alongside full Bayesian estimation \parencite[e.g.,][]{edwards2010markov} and joint maximum likelihood estimation \parencite[e.g.,][]{haberman1977maximum,chen2019joint,chen2020structured}. 
The marginal likelihood approach considers the latent variables as random and the unknown parameters as fixed. The likelihood function is obtained from the marginal distribution of the observed data, integrating out the latent variables. This integration poses computational challenges, especially in high-dimensional settings. In particular, the Expectation--Maximisation (EM) algorithm \parencite{dempster_maximum_1977,bock_marginal_1981}, often used to optimise the marginal likelihood, becomes computationally infeasible in high-dimensional settings. The EM algorithm alternates between an Expectation (E) step and a Maximisation (M) step. In the E-step, an objective function is constructed at the current parameter estimate, while the M-step updates the parameter estimate by maximising this objective function. The objective function constructed in the E-step involves numerical integrals of the latent variables under their posterior distribution determined by the current parameter estimates. The complexity of this objective function grows exponentially with the latent dimension, making both steps computationally infeasible in high-dimensional settings. 
 
Several methods have been proposed to optimise the marginal likelihood by approximating high-dimensional integrals, which can be classified into four categories here. One line of research approximates integrals analytically. A well-known method is the Laplace approximation \parencite{shun_laplace_1995,huber2004estimation} which approximates the integrals using the Laplace approximation at the current parameter values. A similar approach is the adaptive quadrature EM algorithm \parencite{schilling_high-dimensional_2005}, which uses adaptive quadrature points based on a Gaussian approximation in the E-step. 
Another line of research employs variational approximation methods to evaluate the intractable posterior distributions of latent variables in the E-step. However, all these methods may not be ideal in settings where the number of items is not sufficiently large, as they all suffer from approximation errors arising from the order of the Laplace approximation, the number of adaptive quadrature points, or the mean-field assumption underlying the variational approximation.

There are also variants of the EM algorithm that use Markov chain Monte Carlo (MCMC) methods to approximate the integrals in the E-step. Methods in this direction include the Monte Carlo EM algorithm \parencite{wei1990monte,meng1996fitting} and the Stochastic EM algorithm \parencite{diebolt1995stochastic,nielsen2000stochastic,zhang_improved_2020}. The Monte Carlo EM replaces numerical integration in the classical EM algorithm with Monte Carlo integration, generating samples of latent variables via MCMC methods. 
However, this approach introduces Monte Carlo errors to the parameter estimates in the M-step. This Monte Carlo error decays to zero only as the number of Monte Carlo samples approaches infinity. Thus, convergence to the MMLE requires the number of Monte Carlo samples to grow with the number of iterations. Accordingly, this approach can be computationally infeasible when a low Monte Carlo error tolerance is required. The Stochastic EM algorithm uses a similar idea but is computationally more efficient. It uses only a small number of Monte Carlo samples per iteration and reduces Monte Carlo error by averaging the parameter trajectory across iterations. 

Finally, we explore stochastic optimisation algorithms that combine the stochastic approximation \parencite{robbins_stochastic_1951} and sampling techniques \parencite[see;][]{gu_stochastic_1998, cai2010metropolis, cai_high-dimensional_2010, atchade_perturbed_2017, de_bortoli_efficient_2021, zhang_computation_2022}. These methods are closely related to the Stochastic EM algorithm. They iterate between a sampling step and a stochastic gradient (SG) update step. The sampling step is the same as that of the Stochastic EM algorithm. In this step, a small number of posterior samples of the latent variables are generated, typically via an MCMC sampler. Examples of an MCMC sampler in the existing stochastic optimisation algorithms include the Gibbs sampler \parencite[e.g.,][]{zhang_computation_2022} and the random-walk Metropolis-Hastings (MH) sampler \parencite[e.g.,][]{cai_high-dimensional_2010, cai2010metropolis}.
The SG update step plays a similar role to the M-step in Stochastic EM. However, instead of optimising an objective function constructed by Monte Carlo samples, the SG update step performs a one-step update on the parameters along a direction determined by the gradient of the same objective function. This update is known as stochastic approximation. The stochastic optimisation approach is computationally more efficient than Stochastic EM, as performing an SG update is substantially faster than solving an optimisation \parencite[see;][]{zhang_computation_2022}. 

This paper focuses on the estimation of high-dimensional continuous latent-variable models by proposing two strategies to improve the performance of stochastic optimisation. First, we bring attention to a Langevin-type MCMC sampler. There are various variants of a Langevin-type MCMC sampler. Examples include the unadjusted Langevin sampler \parencite{durmus2019high, roberts1996exponential}, the MH-adjusted Langevin (MALA) sampler \parencite{oliviero-durmus_geometric_2024, roberts1996exponential}, and the Hamiltonian Monte Carlo sampler \parencite{betancourt_conceptual_2017}. 
The key to a Langevin-type MCMC sampler is the construction of the Markov kernel based on the gradient of the target posterior, utilising its geometric information to improve the sampling efficiency. Compared with classical MCMC samplers, such as the random-walk MH sampler, a Langevin-type sampler can generate approximate samples more efficiently, especially in high-dimensional settings. 
In the literature on stochastic optimisation algorithms for the MMLE, \textcite{de_bortoli_efficient_2021} proposed the Stochastic Optimisation by unadjusted Langevin algorithm, which uses an unadjusted Langevin sampler in the sampling step. This study further considers the MALA sampler \parencite{oliviero-durmus_geometric_2024, roberts1996exponential}---as a boosting strategy for stochastic optimisation. The inclusion of the MH adjustment in a Langevin-type sampling substantially stabilises and improves its performance. 
Furthermore, different from a more complicated Langevin-type sampler, such as the Hamiltonian Monte Carlo sampler, the MALA sampler is easy to implement, as one only needs to tune a fixed step size of its discretised Langevin update and evaluate the gradient of the complete-data log-likelihood with respect to latent variables when generating an approximate posterior sample. 

Second, we consider using a minibatch technique to estimate high-dimensional latent variable models. The minibatch technique is commonly used in SG descent algorithms \parencite[Section 8.1.3,][]{Goodfellow-et-al-2016} and constructs the SG using a random subset of the observation units, thereby substantially reducing complexity and improving convergence speed for datasets with many observations. In addition, we explore the impact of a quasi-Newton (QN) update on the algorithms using the minibatch technique, as this is the standard approach in improving convergence speed for stochastic optimisation algorithms with fullbatch gradients that use all the observations to construct a gradient \parencite[e.g.,][]{zhang_computation_2022,cai_high-dimensional_2010}. 

We investigate and illustrate how these boosting strategies improve stochastic optimisation and evaluate their performance via simulation studies under models familiar to psychometrics, including a Multidimensional 2-Parameter Logistic (M2PL) item response theory model \parencite[Section 2.1.1.2;][]{reckase_multidimensional_2009} and a multilevel logistic regression model with random effects \parencite[Chapter 12;][]{bartholomew2008analysis}. 
Furthermore, we applied the algorithm using the boosting strategies to estimate a confirmatory M2PL model for a large-scale personality assessment dataset comprising 30,000 respondents, 300 items, and 30 latent dimensions. The estimation code was written in Julia programming language \parencite[Version 1.10.10,][]{bezanson_julia_2017}, and all the code used in this study is available on the Open Science Framework: \url{https://osf.io/3sb4t/?view_only=abd84347053a450fa88f787d168df359}.

The rest of the paper is organised as follows. In Section~\ref{sec:proposed}, we propose the boosting strategies for stochastic optimisation under a general latent variable model setting.
% In Section~\ref{sec:theory}, we establish the theoretical properties of the algorithm, showing that its iterative parameter update converges to the marginal maximum likelihood estimate as the iteration number approaches infinity. 
We conduct simulation studies in Section~\ref{sec:sim} to evaluate the performance of the algorithms with the proposed boosting strategies and compare them with alternative algorithms, followed by an application to a large-scale personality assessment dataset in Section~\ref{sec:emp}. We conclude with discussions in Section~\ref{sec:discuss}.

\section{Proposed Boosting Strategies}\label{sec:proposed}

\subsection{Problem Setup}\label{subsec:prbsetup}
We consider a latent variable model with continuous latent variables. For an observation $i$, $i = 1, ..., N$, we let $\bY_i \in \mathbb R^{J_i}$ be observed data and $\bxi_i\in \mathbb R^{K}$ be the latent variables. 
A latent variable model typically assumes that $(\bY_i, \bxi_i)$, $i =1, ..., N$, are independent, and $(\bY_i, \bxi_i)$ is assumed to follow a joint distribution with density function $f_i(\by_i,\bxi_i \vert \bbeta)$, where 
$\bbeta \in \mathcal B \subset \mathbb R^p$ denotes the unknown parameters and $\mathcal B$ is the parameter space. The goal is to estimate the unknown parameters $\bbeta$ by the Marginal Maximum Likelihood Estimator (MMLE), i.e., 
$\hat \bbeta = \argmax_{\bbeta \in \mathcal B} l(\bbeta)$, where 
\begin{align}\label{eq:loglik}
 l(\bbeta) = \sum_{i=1}^N \log\left(\int f_i(\bY_i, \bxi_i\vert \bbeta)d\bxi_i \right)   
\end{align} 
is the marginal log-likelihood function. We introduce two models as illustration examples.

\subsubsection{Multilevel Logistic Regression Model}\label{exmp:multilevel} 

We consider a two-level logistic regression model, following the same setting as in Chapter 12 of \textcite{bartholomew2008analysis}. Let the level-2 units be indexed by $i = 1, ..., N$ and the level-1 units be indexed by $j = 1, ..., J_i$. For level-1 unit $j$ within level-2 unit $i$, we observe the binary response variable $Y_{ij} \in \nami{0, 1}$. Furthermore, let $\bx_{ij} = \maru{x_{ij1}, \ldots, x_{ijK} }^\top$ be a vector of covariates for level-1 unit $j$ within level-2 unit $i$, where $K$ is the length of the vector of covariates. Here, $x_{ij1}$ is constrained to be 1 to indicate the random intercept. The level-2 random effects/latent variables are denoted as $\bxi_i = \maru{\xi_{i1}, \ldots, \xi_{iK}}^\top$. It is assumed that $Y_{i1}$, ..., $Y_{iJ_i}$ are conditionally independent given $\bxi_i$, and 
$Y_{ij}$ given $\bxi_i$ follows a Bernoulli distribution satisfying 
\begin{align}
    \bbP\kagi{Y_{ij}=1  \vert  \bmu,  \bxi_i} &= \frac{\exp\maru{\bx_{ij}^\top \bxi_i}}{1 + \exp\maru{\bx_{ij}^\top \bxi_i}}. 
\end{align}
$\xi_{i1}$ represents the random intercept term, while $\xi_{i2}$, ..., $\xi_{iK}$ represent the random slope terms. The random effects $\bxi_i$ are assumed to follow a multivariate normal distribution $\calN\maru{\bmu, \bSigma}$, where $\bmu = \maru{\mu_1, \ldots, \mu_K}^\top$ is a mean vector and $\bSigma$ is a covariance matrix that is positive-definite. To handle the positive-definite constraint of $\bSigma$, we reparameterise it by the Cholesky decomposition, 
$\bSigma = \btL \btL^\top,$
where $\btL = (l_{kk'})_{K \times K}$ is a lower triangular matrix.

To be consistent with our generic notation, we let $\bbeta = \{\mu_1, ..., \mu_K, l_{kk'}, k'=1, ..., k, k=1, ..., K\}$ be all the parameters of the model.
As the covariance matrix $\bSigma$ can be expressed as a function of the lower triangular entries of $\btL$, with slight abuse of notation, we denote it as $\bSigma(\bbeta)$.  
Then 
$f_i(\bY_{i}, \bxi_i \vert \bbeta)$ takes the form  
\begin{align}
    f_i(\bY_{i}, \bxi_i\vert \bbeta) &=  \left(\prod_{j=1}^{J_i}\frac{\exp\maru{ Y_{ij}\bx_{ij}^\top \bxi_i }}{1 + \exp\maru{\bx_{ij}^\top \bxi_i}}\right) \phi( \bxi_i \vert \bmu, \bSigma(\bbeta)),
\end{align}
where $\phi(\bxi_i \vert \bmu,\bSigma(\bbeta)  )$ is the density function for the multivariate normal distribution $\calN( \bmu, \bSigma(\bbeta))$. As no constraints are needed for the parameters in $\bbeta$, the parameter space $\mathcal B = \mathbb R^p$, where $p = K + K(K+1)/2$.  

\subsubsection{M2PL Model}
\label{exmp:m2pl}
The second model is the M2PL for confirmatory item factor analysis. In this model, each observation $i$ corresponds to a respondent, and each element of $\bY_i$ records their answer to a binary-scored item. It is assumed that different respondents receive the same set of items, and thus, $J_1 = J_2 = \cdots = J_N = J$. The latent variables $\bxi_i \in \mathbb R^K$ represent unobserved factors measured by the items. The item-factor relationship is characterised by a pre-specified $J\times K$ indicator matrix $\bQ = (q_{jk})_{J\times K}$, where $q_{jk} = 1$ if factor $k$ is directly measured by item $j$, and $q_{jk} = 0$ otherwise. 
$Y_{i1}$, ..., $Y_{iJ}$ are assumed to be conditionally independent given $\bxi_i$, and 
$Y_{ij}$ given $\bxi_i$ follows a Bernoulli distribution satisfying 
\begin{align}
    \bbP\kagi{Y_{ij}=1  \vert  d_j, \ba_j,  \bxi_i} &= \frac{\exp\maru{ d_j + \summa{k}{1}{K}a_{jk}\xi_{ik} }}{1 + \exp\maru{d_j + \summa{k}{1}{K}a_{jk}\xi_{ik}}},
\end{align}
where $a_{jk} = 0$ when $q_{jk} = 0$. The latent variables $\bxi_i$ are assumed to follow a multivariate normal distribution $\calN( \mathbf{0}, \bSigma )$, where the mean vector is constrained to be a zero vector and the diagonal entries of $\bSigma$ are constrained to be 1 for model identification. Similar to the multilevel logistic regression model in Section~\ref{exmp:multilevel}, we reparameterise $\bSigma$ by 
  $\bSigma = \btL \btL^\top$,
where $\btL = (l_{kk'})_{K \times K}$ is a lower triangular matrix satisfying $\sum_{k'=1}^k l_{kk'}^2 =1$ for all $k$.   

The model parameters are denoted as $\bbeta = \{d_1,..., d_J\} \cup \{a_{jk}: j=1, ..., J, k=1, ..., K, q_{jk} = 1\} \cup \{l_{kk'}, k'=1, ..., k, k=1, ..., K\}$. Similar to Example~\ref{exmp:multilevel}, we express $\bSigma$ as a function of $\bbeta$, denoted by $\bSigma(\bbeta)$. 
The joint density function $f_i$ takes the form
\begin{align}\label{eq:m2plpdf}
    f_i(\bY_{i}, \bxi_i \vert \bbeta)  =  \left(\prod_{j=1}^J  \frac{\exp\maru{ Y_{ij} \left(d_j + \summa{k}{1}{K} a_{jk}\xi_{ik}\right) }}{1 + \exp\maru{d_j + \summa{k}{1}{K}a_{jk}\xi_{ik}}}\right) \phi( \bxi_i \vert \mathbf{0}, \bSigma(\bbeta)).
\end{align}
The parameter space $\mathcal B = \{\bbeta \in \mathbb R^p: \sum_{k'=1}^k l_{kk'}^2 =1, k =1, ..., K\}$, where $p = J + (\sum_{j=1}^J\sum_{k=1}^K q_{jk}) + K(K+1)/2$.

\subsection{Background on Stochastic Optimisation}\label{subsec:backgroundstcopt}

A stochastic optimisation algorithm iterates between two steps: (1) sampling the latent variables from their posterior distribution based on the current model parameter estimates, and (2) updating the model parameters using an approximate SG constructed from the latent variable samples. In the subsequent sections, we introduce these steps and the strategies that can improve sampling of latent variables and model parameter updates.   

\subsubsection{Sampling the Latent Variables}
Let $t$ be the current iteration number, and let $\bbeta^{(t-1)}$ and $\bxi^{(t-1)}$ be the updated parameters and sampled latent variables from the previous iteration, respectively. 
The sampling step aims to obtain an approximate sample for each $i$ from the posterior distribution of $\bxi_i$ given $\bY_i$, evaluated at the model with parameter $\bbeta^{(t-1)}$, which is defined as
\begin{align}\label{eq:post}
   p_{\bbeta^\iter{t-1}, i} = \frac{f_i(\bY_i, \bxi_i\vert \bbeta^{(t-1)})}{\int f_i(\bY_i, \tilde \bxi_i\vert \bbeta^{(t-1)})d\tilde \bxi_i}.
\end{align} 
We also denote with $\pi_{\bbeta^\iter{t-1}} = \prod_{i=1}^{N}p_{\bbeta^\iter{t-1}, i}$ the joint posterior of latent variables given $\bY$ evaluated at the model with parameter $\bbeta^{(t-1)}$. Typically, exact sampling from the posterior distribution is infeasible.  
In that case, MCMC methods have been commonly used to generate approximate samples. An MCMC method uses $\bxi^{(t-1)}_i$ as the starting point and samples from a Markov kernel whose invariant distribution is $p_{\bbeta^\iter{t-1}, i}$. Then, $\bxi^{(t)}_i$ is obtained after one or more MCMC iterations.
The most common choice of MCMC methods in stochastic optimisation for continuous latent variable models is the random-walk MH sampler \parencite{cai_high-dimensional_2010, gu_stochastic_1998}, where the proposed value is generated by adding a Gaussian random noise, scaled by the random-walk step size, to the MCMC sample in the previous iteration. However, since this Gaussian random noise does not utilise the geometric information of the target posterior, the proposed value is not necessarily oriented towards the direction where the posterior density is higher. This feature makes the exploration of the target posterior inefficient, especially in high-dimensional parameter spaces. As a result, the random-walk MH sampler can be inefficient, leading to slow convergence in high-dimensional latent variable models.

% Second, when it comes to constructing a stochastic optimisation algorithm by the Gibbs sampler, one needs to resort to an approach in making the posterior distribution conditionally conjugate, such as the P\'olya-Gamma data augmentation \parencite{polson_bayesian_2013}. However, it requires a model-specific derivation.

\subsubsection{Updating the Model Parameters}
The model parameters are updated using an approximate SG. The approximate SG at the $t$-th iteration takes the form 
\begin{align}\label{eq:SG0}
    \bG_{\bbeta^\iter{t-1}}(\bxi^\iter{t}) = \sum_{i=1}^N \left.\left( \frac{\partial \log\nami{f_i(\bY_i, \bxi_i^\iter{t}\vert \bbeta)}}{\partial \bbeta}\right)\right\vert_{\bbeta = \bbeta^\iter{t-1}},
\end{align}
where $\bxi_i^\iter{t}$ is an MCMC sample from the previous sampling step and imputed into $\bG_{\bbeta^\iter{t-1}}(\bxi^\iter{t})$. We note that if $\bxi^{(t)}_i$ is an exact sample from $p_{\bbeta^\iter{t-1}, i}$, then $\bG_{\bbeta^\iter{t-1}}(\bxi^\iter{t})$ in \eqref{eq:SG0} is an unbiased estimator of the gradient of the marginal log-likelihood at $\bbeta^{(t-1)}$, i.e., $\mathbb E_{\pi_{\bbeta^\iter{t-1}}} \kagi{\bG_{\bbeta^\iter{t-1}}(\bxi^\iter{t})} = \nabla  l(\bbeta^{(t-1)})$, where the expectation is with respect to $\bxi^\iter{t}$.  
As $\bxi^{(t)}_i$ obtained from the sampling step is an approximate posterior sample generated by an MCMC sampler from $p_{\bbeta^\iter{t-1}, i}$, $\bG_{\bbeta^\iter{t-1}}(\bxi^\iter{t})$ is also an approximate SG evaluated at the $t$-th iteration. For simplicity, we still call $\bG_{\bbeta^\iter{t-1}}(\bxi^\iter{t})$ an SG when there is no ambiguity. 
Then, $\bbeta^{(t)}$ is obtained by an SG ascent step
\begin{align}\label{eq:par}
\bbeta^{(t)} = \bbeta^{(t-1)} + \gamma_t  {\left(\bD^{(t)}\right)}^{-1}\bG_{\bbeta^\iter{t-1}}(\bxi^\iter{t}),
\end{align}
where $\gamma_t$ is a properly chosen step size that decays to zero as $t$ increases, and $\bD^{(t)}$ is a positive definite matrix that is constructed to approximate the second-order information of the marginal log-likelihood. 
With a constrained parameter space $\mathcal B \subset \mathbb R^p$, 
such as in Section~\ref{exmp:m2pl},
a projection is needed in the parameter update step. Instead of \eqref{eq:par},  \textcite{zhang_computation_2022} suggested updating $\bbeta$ by the QN proximal update that solves a projection problem
\begin{align}\label{eq:par2}
\bbeta^{(t)} = \argmin_{\bbeta \in {\mathcal B}} \Vert \bbeta - \bbeta^{(t-1)} - \gamma_t  {(\bD^{(t)})}^{-1}\bG_{\bbeta^\iter{t-1}}(\bxi^\iter{t}) \Vert_{\bD^{(t)}}^2,  
\end{align}
where the norm $\Vert \cdot\Vert_{\bD^{(t)}}$ is defined as $\Vert \mathbf{x} \Vert_{\bD^{(t)}}^2 = \mathbf{x}^\top \bD^{(t)}\mathbf{x}$. 
In the special case where ${\bD^{(t)}}$ is diagonal and the parameter space $\mathcal B$ is a well-behaved constrained set, \eqref{eq:par2} admits a closed-form solution. Note that when $\calB = \mathbb R^{p}$, \eqref{eq:par2} reduces to the SG ascent update \eqref{eq:par}. Examples of $\bD^{(t)}$ in latent variable models include a Hessian matrix of the negative complete-data log-likelihood function \parencite{cai_high-dimensional_2010}, a diagonal approximation to a Hessian matrix of the marginal log-likelihood function \parencite{zhang_computation_2022}, and a diagonal matrix whose diagonal entries are a constant value used to rescale a step size for certain model parameters. In this study, the diagonal elements of $\bD^{(t)}$ were set to the diagonal approximation of the Hessian matrix of the marginal log-likelihood function \parencite{zhang_computation_2022} when a QN update is employed. 

\subsubsection{Determining a Final Estimate}
After completing the two aforementioned steps for a sufficient number of iterations, we obtain a final estimate of $\bbeta$. This estimate can be $\bbeta^{(t)}$ from the last iteration, or the Polyak-Ruppert trajectory average \parencite{polyak_acceleration_1992,ruppert_efficient_1988} obtained by averaging parameter updates over iterations. Under suitable regularity conditions, these estimates converge to the MMLE $\hat \bbeta$ as the number of iterations grows to infinity, even when $\bG_{\bbeta^\iter{t-1}}(\bxi^\iter{t})$ is only an approximate SG \parencite{atchade_perturbed_2017,zhang_computation_2022}.

\subsection{Boosting Strategies} 

\subsubsection{Improving Sampling Step: MALA Sampler}\label{subsubsec:sample}
We first consider improving the efficiency of the sampling step by adopting the MALA sampler, which combines a Langevin update that utilises the gradient information of the target posterior, and the MH-adjustment step. 
Consider sampling from the posterior distribution $p_{\bbeta^\iter{t-1}, i}$ in \eqref{eq:post}. The MALA sampler is based on a stochastic differential equation called the Langevin diffusion:
\begin{align}
    d\bxi_i(s) = - \nabla U_{i}(\bxi_i(s)) + \sqrt{2}d\bB_i(s),~ s \in [0, \infty),
\end{align}
where $s$ is the continuous-time index$; \bB_{i}$ is a $K$-dimensional standard Brownian motion; and $U_{i}$ is a potential function proportional to the posterior distribution of $\bxi_i$. 
In our setting, the potential function is the negative log-joint density of $(\bY_i, \bxi_i)$ given the model parameter $\bbeta^{(t-1)}$ for observation $i$, which is given by $- \log\nami{ f_i(\bY_i, \bxi_i \vert \bbeta^{(t-1)}) }$.
Under mild assumptions on the negative log-joint density, this stochastic differential equation has a strong solution for which the posterior distribution of latent variables given the model with parameter $\bbeta^{(t-1)}$ is the invariant probability measure. 
The MALA sampler is an Euler-Maruyama discrete-time approximation to the stochastic differential equation \parencite{oliviero-durmus_geometric_2024}, as sampling the solution of the stochastic differential equation is computationally challenging. 
Compared with traditional MCMC samplers used in stochastic optimisation, such as the random-walk MH sampler and the adaptive rejection Metropolis sampler, the MALA sampler converges faster under high-dimensional settings \parencite{roberts_optimal_1998}, by making use of gradient information of the posterior distribution.

We use a MALA sampler for the MCMC sampling step in stochastic optimisation. We denote the gradient of the potential function given the model parameters in the previous iteration $\bbeta^\iter{t-1}$ and evaluated at $\bxi_i^\iter{t-1}$ as
\begin{align}\label{eq:ulsgd}
    \nabla U_i^\iter{t-1} =  -\left.\frac{\partial \log\nami{f_i(\bY_i, \bxi_i \vert \bbeta^\iter{t-1})}}{\partial \bxi_i}\right\vert_{\bxi_i = \bxi_i^\iter{t-1}}. 
\end{align}
For example, under the confirmatory M2PL model, the gradient of the potential function at the $t$-th iteration is given by
\begin{align}
    \nabla U_{i}^\iter{t-1} &= \summa{j}{1}{J}\ba_j^\iter{t-1}\maru{\frac{\exp\nami{d_j^\iter{t-1} + (\ba_j^\iter{t-1})^\top \bxi_i^\iter{t-1}}}{1 + \exp\nami{d_j^\iter{t-1} + (\ba_j^\iter{t-1})^\top \bxi_i^\iter{t-1}}}  - Y_{ij}} + \nami{\bSigma(\bbeta^{(t-1)})}^{-1} \bxi_i^\iter{t-1},
\end{align}
where $a_{jk}=0$ when $q_{jk}=0$. Then, we obtain the proposed value $\bxi_i^{(*)}$ by
\begin{align} 
 \bxi_i^{(*)} = \bxi_i^{(t-1)} -h \nabla U_i^\iter{t-1} + \sqrt{2h} \bZ_i^\iter{t},   
\end{align}
where $h$ is the Euler-Maruyama discretisation step size that is fixed to a certain value, and $\bZ_i^\iter{t}$ is a white noise vector following a $K$-dimensional standard normal distribution. Then, the posterior sample at the $t$-th iteration, i.e., $\bxi_i^\iter{t}$, is obtained as
\begin{align}
    \bxi_i^\iter{t} = \bxi_i^{(t-1)} + \mathbb{I}_{\bbR^+}\nami{ \alpha_h\maru{\bxi_i^{(*)}, \bxi_i^{(t-1)}}  - W_i^\iter{t}}\maru{ \bxi_i^{(*)} -  \bxi_i^{(t-1)}},
\end{align}
where $\mathbb{I}_{\bbR^+}\nami{\cdot}$ is the indicator function taking the value of 1 if the given value is in $\bbR^+$ and $\nami{W_i^\iter{t}: t \in \bbN}$ is a sequence of independent and identically distributed uniform random variables on $[0, 1]$. The Metropolis acceptance ratio $\alpha_h\maru{\bxi_i^{(*)}, \bxi_i^{(t-1)}}$ is also defined as
\begin{align}\label{eq:mhratio}
    \alpha_h\maru{\bxi_i^{(*)}, \bxi_i^{(t-1)}} = \min\maru{1, \frac{ f_i\maru{\bY_i, \bxi_i^\iter{*}\vert \bbeta^\iter{t-1}}q_h\maru{\bxi_i^\iter{t-1} \vert \bxi_i^\iter{*}} }{ f_i\maru{\bY_i, \bxi_i^\iter{t-1}\vert \bbeta^\iter{t-1}}q_h\maru{\bxi_i^\iter{*} \vert \bxi_i^\iter{t-1}} }  }, 
\end{align}
where the transition probability $q_h\maru{\bxi_i' \vert \bxi_i}$ is given as
\begin{align}
    q_h\maru{\bxi_i' \vert \bxi_i} &= \frac{1}{\maru{4\pi h}^{K/2}}\exp\nami{-\frac{1}{4h}\norm{\bxi_i' - \maru{\bxi_i - h \left.\frac{\partial \log\nami{f_i(\bY_i, \bxi_i \vert \bbeta^\iter{t-1})}}{\partial \bxi_i}\right\vert_{\bxi_i = \bxi_i} }  }_2^2  }.
\end{align}

% \subsection{Proposed D-SOMALA Algorithm}\label{subsec:par}
\subsubsection{Improving Updating Step: Minibatch SG}\label{subsubsec:update}

We further consider improving the efficiency of the parameter update step by applying the minibatch SG. To contrast the minibatch SG with an approximate SG in \eqref{eq:SG0}, we also call the SG that involves all the observation units as the fullbatch SG. In the fullbatch SG update with \eqref{eq:SG0}, the summation is over all the observation units, which can be computationally expensive when the number of observation units is very large. As an alternative, we consider the minibatch SG. In the $t$-th iteration, we randomly draw a subset of $\nami{1, \ldots, N}$ of size $n$, denoted by $S^\iter{t}$. Then, we construct the minibatch SG as: 
\begin{align}\label{eq:SG}
    \bG_{\bbeta^\iter{t-1}}^{\text{mini}}(\bxi^\iter{t}, S^\iter{t}) = \frac{N}{n} \summa{i}{1}{N} \mathbbm{1}_{\nami{i \in S^\iter{t}}} \left.\left( \frac{\partial \log\nami{f_i(\bY_i, \bxi_i^\iter{t}\vert \bbeta)}}{\partial \bbeta}\right)\right\vert_{\bbeta = \bbeta^\iter{t-1}}.
\end{align}
 
It is easy to show that when $\bxi_i^\iter{t}$ is an exact sample from $p_{\bbeta^\iter{t-1}, i}$ for all $i$, $\bG_{\bbeta^\iter{t-1}}^{\text{mini}}(\bxi^\iter{t}, S^\iter{t})$ is an SG of the marginal log-likelihood at $\bbeta^{(t-1)}$.  
By choosing $n$ to be much smaller than $N$, one can substantially reduce the per-iteration computational complexity, as the summation in \eqref{eq:SG} reduces to $n$ terms. It also means that instead of sampling all the $\bxi_i^{(t)}$ in the sampling step, one only needs to sample all the $\bxi_i^{(t)}$ within the minibatch $S^{(t)}$. Compared with the fullbatch SG, an algorithm using the minibatch SG tends to converge faster. That is because, with the same computation budget (in terms of processed observation units) for one fullbatch SG update, the minibatch SG has already updated the parameters $N/n$ times.  

When implementing the minibatch SG update, an important decision to make is the minibatch size $n$. Theoretically, one should set $n = 1$, as it gives the fastest convergence speed because a minibatch SG with $n=1$ updates the parameters most frequently when the same number of observation units is processed. However, its implementation is neither computationally efficient nor stable for two reasons. First, CPUs and GPUs cannot exploit the full power of vectorisation, and thus processing the SG with a single observation unit is not computationally efficient \parencite[Section 12.5,][]{zhang2023dive}. Second, the variance of the minibatch SG becomes larger as the minibatch size decreases. This large variance can slow the convergence of the updated parameters around the neighbourhood of an optimum. Thus, choosing a minibatch size between 1 and $N$ strikes a trade-off between  stability and statistical convergence speed (in terms of the number of observation units processed) and computational efficiency (in terms of the average CPU/GPU processing time per sample).

\subsubsection{Combining the Proposed Boosting Strategies}\label{subsubsec:combine}
In the previous sections, we introduced two strategies to enhance stochastic optimisation: employing the MALA sampler for the sampling step and adopting the minibatch SG for parameter updates. This section summarises the algorithms derived from these strategies.

First, the stochastic optimisation algorithm that replaces the random-walk MH sampler with the MALA sampler is termed the \emph{Stochastic Optimisation by MALA} (SOMALA). Its QN variant is also called the QN-SOMALA. Similarly, those using the random-walk MH sampler are referred to as the \emph{Stochastic Optimisation by random-walk MH} (SOMH) and QN-SOMH. Furthermore, we refer to the SOMALA with the minibatch SG as the D-SOMALA and call its QN variant the QN-D-SOMALA. Their counterparts of the random-walk MH sampler are also named analogously. The pseudocode of the algorithms using the MALA sampler is in Supplementary Materials A.3.

Next, we summarise the algorithms that can be derived from the proposed boosting strategies, together with their counterparts using the random-walk MH sampler in Table~\ref{tab:algorithms}. It should be noted that the components and structure of the SOMH are the same as those of the MH Robbins-Monro algorithm \parencite{cai_high-dimensional_2010}, as this algorithm also adopts the random-walk MH sampler for the posterior sampling of latent variables in the stochastic imputation step. Accordingly, the QN-SOMH is listed as the baseline algorithm in Table~\ref{tab:algorithms}, as it is the same as the MH Robbins-Monro algorithm with a QN update and is the standard stochastic optimisation algorithm within the psychometric community. 

Here, the QN-SOMALA and QN-SOMH can be viewed as special cases of the QN-D-SOMALA and QN-D-SOMH with batch size $n=N$. The D-SOMALA and D-SOMH update model parameters using only the first-order information of the objective function. The algorithms that use a QN update employ the diagonal approximation to the Hessian of the marginal log-likelihood function \parencite{zhang_computation_2022} in the model parameter update. The implementation details of stochastic optimisation algorithms are provided in the Supplementary Materials A.4. 

Lastly, the theoretical guarantee of the convergence to the MMLE under the algorithms using the MALA sampler can also be established in a similar manner as in Theorem 5 of \textcite{de_bortoli_efficient_2021} by verifying that some related conditions on the stability of a Markov process driven by the MALA sampler are satisfied uniformly in the model parameter $\bbeta$. The examples of such conditions include the ergodicity condition in $V$-norm uniformly in model parameter $\bbeta$ and the condition to control the distance between the invariant distribution of the Markov kernel and the target distribution uniformly in the model parameter $\bbeta$ \parencite{de_bortoli_efficient_2021}. Easily verifiable conditions for the geometric ergodicity of the MALA sampler have also been proposed by \textcite{oliviero-durmus_geometric_2024}.

\begin{table}[htbp]
  \centering
  \caption{Algorithms that can be derived by the proposed boosting strategies}
  \begin{adjustbox}{max width=\textwidth}
    \begin{tabular}{rccccc}
    \toprule
          & \multicolumn{2}{c}{MCMC Samplers} & \multicolumn{2}{c}{Type of Stochastic Gradients} & \multirow{2}[2]{*}{QN Update} \\
          \cmidrule{2-5} & random-walk MH & MALA  & Fullbatch & Minibatch &  \\
    \midrule
    QN-SOMH & $\checkmark$ &       & $\checkmark$ &       & $\checkmark$ \\
    \hline\hline \addlinespace[2pt] 
    D-SOMALA &       & $\checkmark$ &       & $\checkmark$ &  \\
    D-SOMH & $\checkmark$ &       &       & $\checkmark$ &  \\
    QN-D-SOMALA &       & $\checkmark$ &       & $\checkmark$ & $\checkmark$ \\
    QN-D-SOMH & $\checkmark$ &       &       & $\checkmark$ & $\checkmark$ \\
    QN-SOMALA &       & $\checkmark$ & $\checkmark$ &       & $\checkmark$ \\
    \bottomrule
    \end{tabular}%
        \end{adjustbox}
  \label{tab:algorithms}%
          \begin{tablenotes}
 \item \footnotesize{\textit{Note.} \\
 SOMALA: Stochastic Optimisation by MALA\\
 SOMH: Stochastic Optimisation by random-walk MH\\
 D-SOMALA: Doubly Stochastic Optimisation by MALA\\
 D-SOMH: Doubly Stochastic Optimisation by random-walk MH\\
 QN: quasi-Newton
 }
 \end{tablenotes}
\end{table}%

\section{Simulation Study}\label{sec:sim}

We investigate the impact of the proposed boosting strategies on the convergence and accuracy of stochastic optimisation algorithms using simulated datasets from the two latent variable models introduced in Section~\ref{subsec:prbsetup}. Specifically, the algorithms in comparison were developed by considering the following choices of constructing stochastic optimisation algorithms: (1) whether an MCMC sampler is the MALA or random-walk MH samplers; (2) whether the SG of model parameters is based on a minibatch or fullbatch of observations; and (3) whether a model parameter update is performed using the SG with second-order information (i.e., QN update) or only the SG. As shown in Table \ref{tab:algorithms} in the previous section, six algorithms are derived and considered in the simulation study. 
We note that the fullbatch SG algorithms without a QN update are not considered in the simulation study. 

The purpose of this comparison is threefold:
\begin{enumerate}
    \item To show that the algorithms with the MALA sampler converge faster and provide more accurate parameter estimates than those with the random-walk MH sampler, especially when the latent dimension $K$ is high.
    \item To show that the minibatch SG algorithms converge faster and provide comparable or more accurate parameter estimates than the fullbatch SG algorithms.
    \item To show whether faster convergence can be achieved by a QN update even in the minibatch SG algorithms. 
\end{enumerate}
 
\subsection{Simulation Settings}
The following simulation settings are considered for the multilevel logistic regression and M2PL models. 

\subsubsection{Multilevel logistic regression model.} 
In the multilevel logistic regression model, we explore two different scenarios: lower-dimensional and higher-dimensional settings. In the lower-dimensional setting, there are $10,000$ level-2 units ($N=10,000$), with ten level-1 units observed within each level-2 unit ($J_i=10$). We set the number of covariates to $K=5$, including the covariate with value 1 to denote the random intercept, resulting in five latent variables. In the higher-dimensional setting, there are $10,000$ level-2 units ($N=10,000$), with twenty level-1 units observed within each level-2 unit ($J_i=20$). The number of covariates was set to $K=10$, resulting in ten latent variables. 

In both settings, the values of covariates, except for the one to denote a random intercept, were generated from a multivariate normal distribution with mean vector $\mathbf{0}_{K-1}$ and a correlation matrix whose off-diagonal elements were set to 0.25. For the fixed effects of the model, the true intercept was set to 0.3, and the true slopes were generated from $\text{Uniform}(0.1,1.1)$. The true values of these fixed effects are given in Supplementary Materials A.1.1. The diagonal and off-diagonal elements of the covariance matrix $\bSigma$ of the random effects were set to 0.1 and 0.05, respectively. Finally, 100 datasets were generated for each setting to compare the stochastic optimisation algorithms. 

\subsubsection{M2PL model.} Under the confirmatory M2PL model, we consider both lower- and higher-dimensional settings. In the lower-dimensional setting, we set the number of respondents and items to $N=10,000$ and $J=50$, respectively, and the number of latent variables to $K=5$. In the higher-dimensional setting, we set the number of respondents and items to $N=10,000$ and 
$J=200$ and the number of latent variables to $K=10$. 
In both settings, the diagonal and off-diagonal elements of the correlation matrix $\bSigma$ of latent variables were set to 1.0 and 0.5, respectively. 
The indicator matrix $\bQ$ in the lower-dimensional setting was specified to include three identity matrices and three sets of all the possible patterns of a row vector of $\bQ$ measuring two latent factors and concatenate them with randomly chosen five possible row-vector patterns measuring three latent factors. 
That in the higher-dimensional setting was specified to include three identity matrices and three sets of all the possible patterns of a row vector of $\bQ$ measuring two latent factors and concatenate them with randomly chosen 35 possible row-vector patterns measuring three latent factors. The visual illustration of the indicator matrices $\bQ$ in the lower- and higher-dimensional settings is provided in Supplementary Materials A.1.4.

The values of the intercept parameters $d_j$ were generated from $\text{Uniform}(-1, 1)$, while the values of the non-zero factor loading parameters $a_{jk}$ were generated from $\text{Uniform}(0.5, 1.5)$. Due to space constraints, the specific values of these item parameters are not given in the Supplementary Materials; however, the values are available on the Open Science Framework: \url{https://osf.io/3sb4t/?view_only=abd84347053a450fa88f787d168df359}. The latent variables were generated from the multivariate normal distribution $\calN(\mathbf{0}_K, \bSigma)$. Lastly, 100 datasets were generated for each setting.

We consider three different minibatch sizes, i.e., $n = 250$, $500$, and $1,000$, for the D-SOMALA and D-SOMH and their QN variants. These minibatch sizes were chosen because each corresponds to $ 2.5\%$, $ 5.0\%$, or $10.0\%$ of the original sample size (i.e., $N=10,000$). 
Additionally, we fix the implementation details to ensure a relatively fair comparison between the algorithms as follows: 

\begin{enumerate}
    \item All the algorithms start with the same initial values for model parameters, $\bbeta^{(0)}$ and latent variables, $\bxi^{(0)}$. Specifically, initial values are obtained by randomly sampling the model parameters and latent variables from distributions that differ from the ones used to generate the data; see Supplementary Materials A.1.2 for details.
    \item The step size for a model parameter update is set to $\gamma_t = t^{-0.51}$ for all algorithms following the results from \textcite{zhang_computation_2022}.
    The step size $h$ of the MALA sampler is set to a fixed value that does not change over iterations, and the value of $h$ is tuned for all the algorithms using the MALA sampler from four candidate values. Those using the random-walk MH sampler do not involve $h$ but require specifying their random-walk step sizes. Those were also tuned from four candidate values. All details of the tuning procedure are provided in Supplementary Materials A.1.3.
    
    Moreover, as the minibatch SG updates the parameters $N/n$ times using the same computational budget (in terms of processed observation units) for one fullbatch SG update, we let $\gamma_t$ decay after $N/n$ model parameter updates in the minibatch SG algorithms. With this decaying schedule, the step sizes decay at a similar rate for all algorithms in terms of processed observation units. This means that after processing $N$ observation units in model parameter updates, the step sizes for all the algorithms will decay.

    Regarding the choice of $\bD^\iter{t}$, the algorithms with a QN update adopt the same updating procedure of the diagonal matrix introduced in \textcite{zhang_computation_2022}, which approximates the Hessian matrix of the marginal log-likelihood function. 
    In addition, the step size related to the covariance matrix of the multilevel logistic regression model was rescaled by 0.05 for the minibatch SG algorithms to prevent instability during their optimisation.
    
    All the computations were performed on a desktop computer equipped with an Intel(R) Xeon(R) Gold 6246R CPU @ 3.40GHz.
\end{enumerate}

\subsection{Evaluation Criteria}
To evaluate the convergence speed of the algorithms, we use a similar evaluation scheme to the one in \textcite[Section 12.5;][]{zhang2023dive}. 
Specifically, we save the values of updated model parameters every second during the iterations until convergence is achieved. Then, the iterate-averaged values of the updated model parameters were computed after the epoch from which the trajectory of the updated model parameters started to stabilise. The epoch from which iterate averaging was applied was determined by examining the trajectory of the updated model parameters on a simulated dataset, with iterate averaging starting at 1,000 epochs for the multilevel logistic regression model and at 500 epochs for the confirmatory M2PL model.

Subsequently, we computed the error using the saved updated parameter values and their iterate-averaged values and created an error trajectory as a function of the elapsed time. More precisely, we evaluated the mean absolute error (MAE) using the model parameters updated at a specific time point, denoted by $\tilde{t}$. If $\tilde{t}$ is at a time point after iterate averaging has started, the MAE was computed based on the iterate-averaged value of the model parameters updated at $\tilde{t}$. Our specific time points range from $\tilde{t} = 1, 2, \ldots$ to the time point when the convergence was achieved. As different algorithms exhibit different convergence behaviours, the time point at which convergence was achieved was determined by the algorithm that showed the fastest convergence.

In the multilevel logistic regression model, we computed the absolute errors (AE) of the mean vector of random effects $\bmu$ and the covariance matrix $\bSigma = (\sigma_{kk'})_{K \times K}$ reparameterised by $\btL$ at the $\tilde{t}$-th time point in the $r$-th simulated dataset given by
\begin{align}
   \text{AE}_{\bmu}^{(\tilde{t}), (r)} &= \frac{1}{K}\sum_{k=1}^K\abs{ \mu_k^{(\tilde{t}), (r)} - \mu_k^{*}}\\
    \text{AE}_{\bSigma}^{(\tilde{t}), (r)} &= \frac{1}{K^2}\sum_{k=1}^K\sum_{k'=1}^K\abs{ \sigma_{kk'}^{(\tilde{t}), (r)} - \sigma_{kk'}^*},
\end{align}
where the parameter with superscript ‘‘$*$'' denotes its true value and the parameter with superscript $(\tilde{t}), (r)$ denotes its value at the $\tilde{t}$-th time point in the $r$-th simulated dataset. 
Then, for example, the MAE for $\bmu$ at the $\tilde{t}$-th time point was computed as
% \begin{align}
%    \text{MAE}_{\bmu}^{(\tilde{t})} &= \frac{1}{100} \sum_{r=1}^{100} \text{AE}_{\bmu}^{(\tilde{t}), (r)}\\
%     \text{MAE}_{\bSigma}^{(\tilde{t})} &= \frac{1}{100} \sum_{r=1}^{100} \text{AE}_{\bSigma}^{(\tilde{t}), (r)},
% \end{align}
\begin{align}
   \text{MAE}_{\bmu}^{(\tilde{t})} &= \frac{1}{100} \sum_{r=1}^{100} \text{AE}_{\bmu}^{(\tilde{t}), (r)},
\end{align}
where $100$ is the number of generated datasets in the simulation study. 

In the confirmatory M2PL model, the AEs of the intercept parameters $d_1, \ldots, d_J$, factor loading parameters $\ba_1, \ldots, \ba_J$, and the correlation matrix $\bSigma = (\sigma_{kk'})_{K \times K}$ reparameterised by $\btL$ at the $\tilde{t}$-th time point in the $r$-th simulated dataset were computed as
\begin{align}
    \text{AE}_{d_1, \ldots, d_J}^{(\tilde{t}), (r)} &= \frac{1}{J}\sum_{j=1}^J\abs{ d_j^{(\tilde{t}), (r)} - d_j^{*}},\\
    \text{AE}_{\ba_1, \ldots, \ba_J}^{(\tilde{t}), (r)} &=  \frac{1}{\sum_{j=1}^J\sum_{k=1}^K q_{jk}}\sum_{j=1}^J\sum_{k=1}^K q_{jk}\abs{ a_{jk}^{(\tilde{t}), (r)} - a_{jk}^*}, \\
    \text{AE}_{\bSigma}^{(\tilde{t}), (r)} &= \frac{1}{K(K-1)}\sum_{k=1}^K\sum_{k'=1, k\neq k'}^K\abs{ \sigma_{kk'}^{(\tilde{t}), (r)} - \sigma_{kk'}^*}.
\end{align}
Subsequently, for example, the MAE for $d_1,\ldots,d_J$ at the $\tilde{t}$-th time point was computed as 
% Subsequently, the MAE for the confirmatory M2PL model at the $\tilde{t}$-th time point was computed as
% \begin{align}
%     \text{MAE}_{d_1, \ldots, d_J}^{(\tilde{t})} &= \frac{1}{100} \sum_{r=1}^{100} \text{AE}_{d_1, \ldots, d_J}^{(\tilde{t}), (r)} \\
%     \text{MAE}_{\ba_1, \ldots, \ba_J}^{(\tilde{t})} &= \frac{1}{100} \sum_{r=1}^{100} \text{AE}_{\ba_1, \ldots, \ba_J}^{(\tilde{t}), (r)} \\
%     \text{MAE}_{\bSigma}^{(\tilde{t})} &= \frac{1}{100} \sum_{r=1}^{100} \text{AE}_{\bSigma}^{(\tilde{t}), (r)} .
% \end{align}
\begin{align}
    \text{MAE}_{d_1, \ldots, d_J}^{(\tilde{t})} &= \frac{1}{100} \sum_{r=1}^{100} \text{AE}_{d_1, \ldots, d_J}^{(\tilde{t}), (r)}.
\end{align} 

For both models, we reported the trajectories of the MAEs averaged over the relevant model parameters. For example, the averaged MAE for the confirmatory M2PL model at time point $\tilde{t}$ was computed as $\frac{1}{3}\maru{\text{MAE}_{d_1, \ldots, d_J}^{(\tilde{t})} + \text{MAE}_{\ba_1, \ldots, \ba_J}^{(\tilde{t})} + \text{MAE}_{\bSigma}^{(\tilde{t})}}$. The trajectories of the MAEs for each individual model parameter, which were not averaged over the relevant parameters, are provided in Supplementary Materials A.2.

For better interpretability of the results, Figures~\ref{fig:mlogreg} and~\ref{fig:m2pl} present the trajectories of the averaged MAEs to compare the algorithms that rely on different MCMC samplers while keeping the other specifications fixed (i.e., minibatch size and the use of a quasi-Newton update). For example, these figures show the comparison between D-SOMALA and D-SOMH when the minibatch size is set to $n=250$. Due to space limitations, the figures for the other comparisons are provided in Supplementary Materials A.2. 

\subsection{Results}
We present the results for the multilevel logistic regression model in Tables~\ref{tab:mlogreg_K5} and~\ref{tab:mlogreg_K10}, which report the trajectories of the averaged MAEs across the relevant model parameters over 2,000 seconds for $K=5$ and $10$. Tables~\ref{tab:m2pl_K5} and~\ref{tab:m2pl_K10} provide the corresponding results for the confirmatory M2PL model, showing the trajectories of the averaged MAEs over 500 seconds for $K=5$ and 1,000 seconds for $K=10$. Visual illustrations of the averaged MAE trajectories to compare D-SOMALA and D-SOMH with a minibatch size of $n=250$ are shown in Figures~\ref{fig:mlogreg} and~\ref{fig:m2pl}.

\subsubsection{General Findings}\label{subsubsec:gfindings}
\begin{figure}[htbp]
  \centering
    \includegraphics[width=\linewidth]{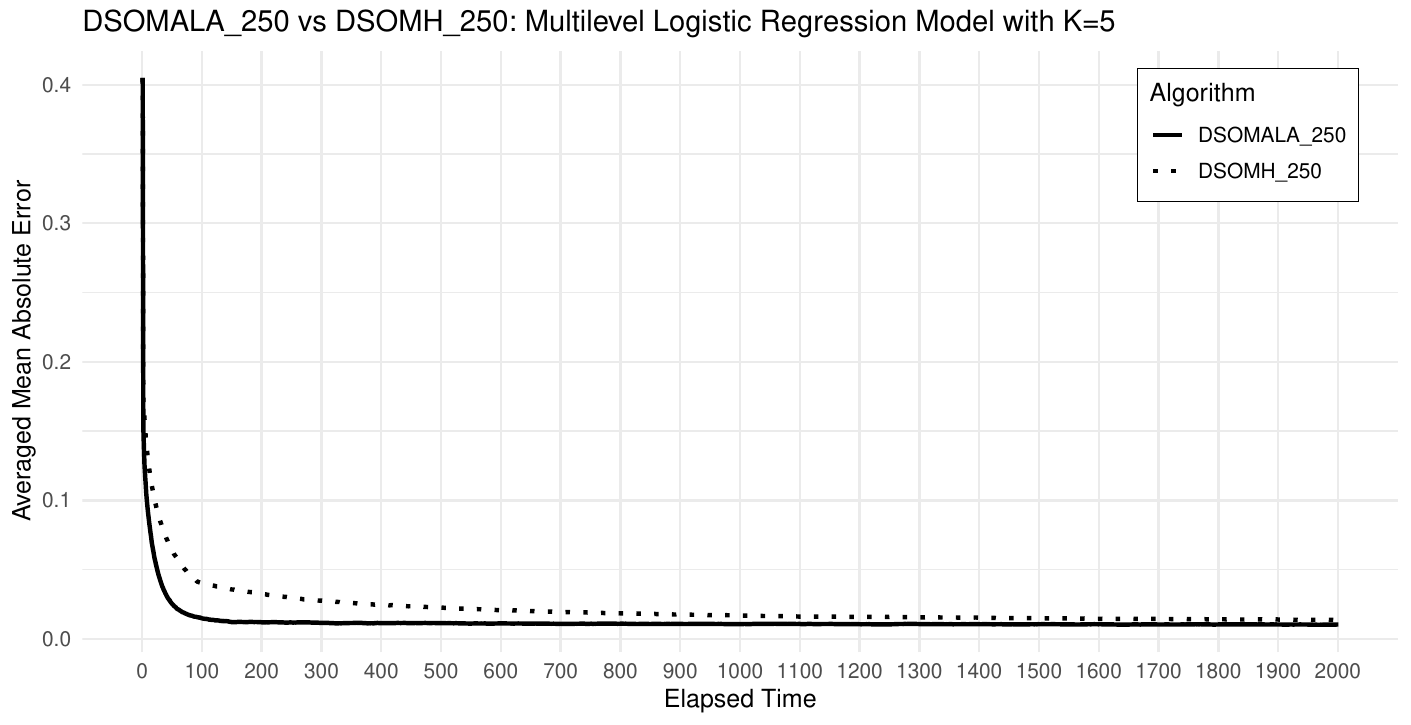}
    \includegraphics[width=\linewidth]{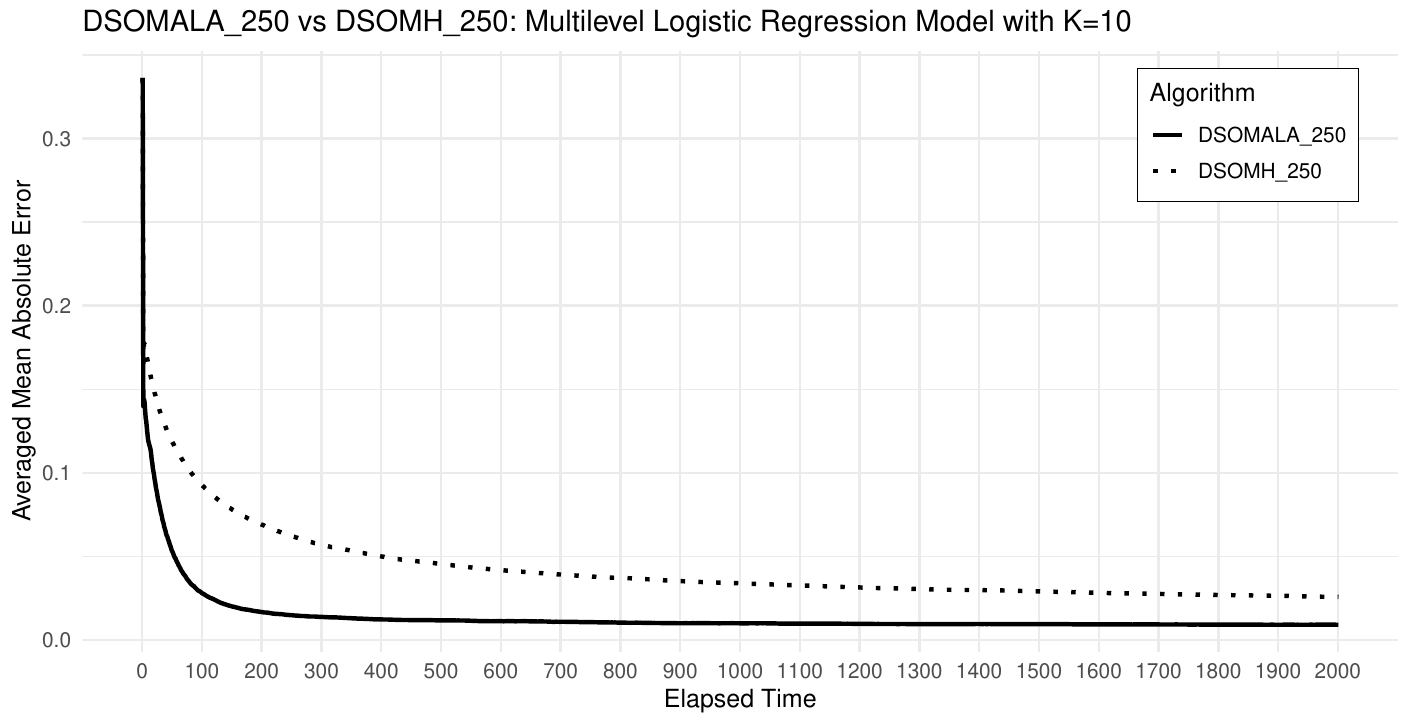}
    \caption{The trajectories of averaged MAEs for the multilevel logistic regression model. The legends indicate the type of algorithm. The number after the underbar ‘‘\_'' represents the minibatch size.}
    \label{fig:mlogreg}
\end{figure}

\begin{figure}[htbp]
  \centering
    \includegraphics[width=\linewidth]{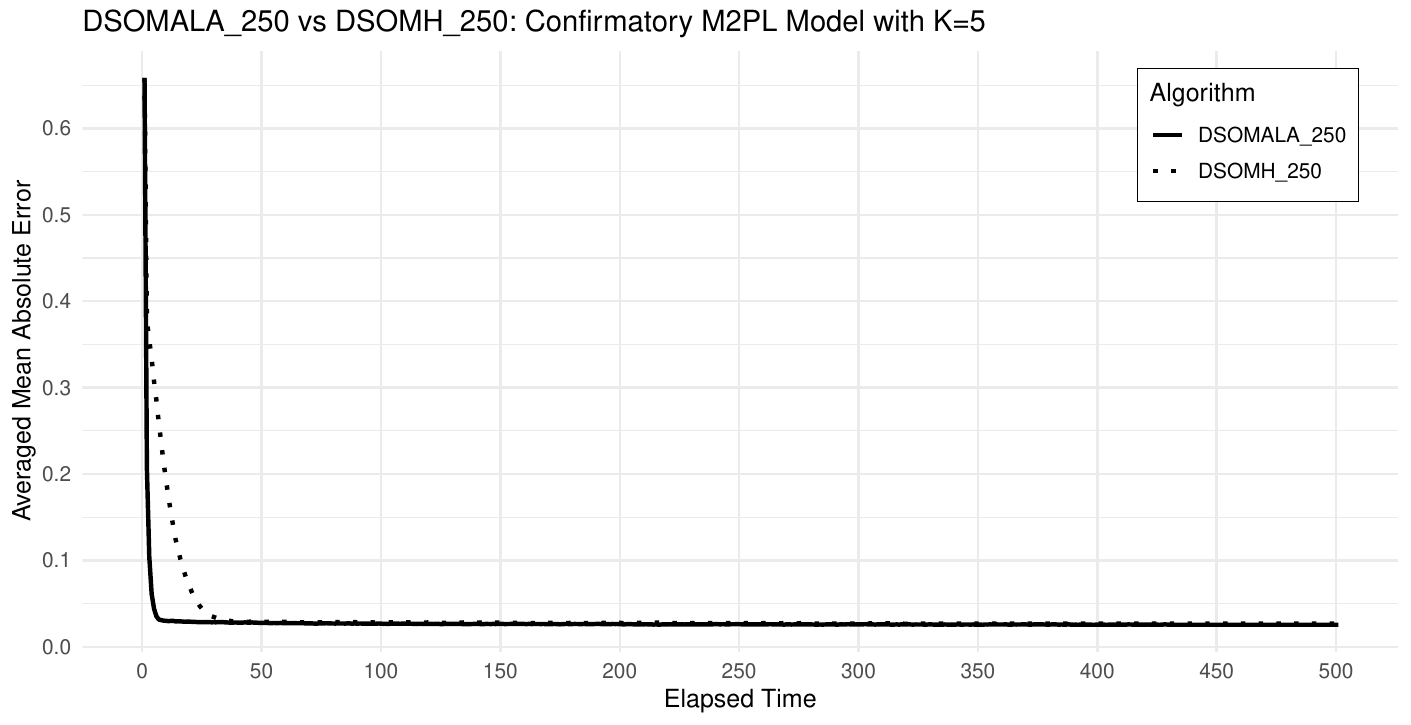}
    \includegraphics[width=\linewidth]{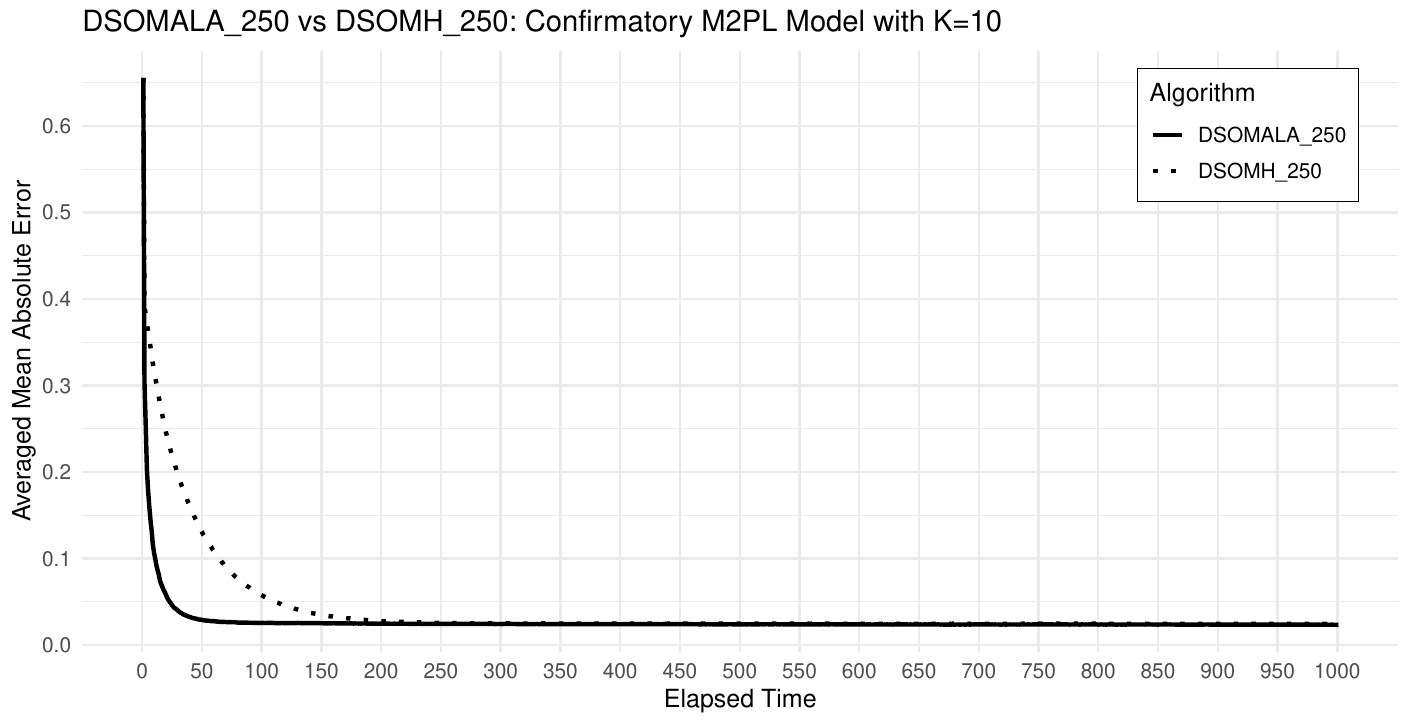}
    \caption{The trajectories of averaged MAEs for the confirmatory M2PL model. The legends indicate the type of algorithm. The number after the underbar ‘‘\_'' represents the minibatch size.}
    \label{fig:m2pl}
\end{figure}

Here, we report the general findings from both simulations. 
To begin with, we examine the figures that show the trajectories of the averaged MAEs for comparing D-SOMALA and D-SOMH with a minibatch size of $n=250$. As shown in Figures~\ref{fig:mlogreg} and~\ref{fig:m2pl}, the trajectories obtained under D-SOMALA ($n=250$) converge the fastest and achieve the highest estimation accuracy across all model settings. This advantage of the MALA sampler is more pronounced in the multilevel logistic regression model than in the confirmatory M2PL model, and more evident in the higher-dimensional settings than in the lower-dimensional ones. The similar patterns were found in the other comparisons provided in Supplementary Materials A.2.

Subsequently, we examine Tables \ref{tab:mlogreg_K5} to \ref{tab:m2pl_K10}, which show the trajectories of the averaged MAEs for all the algorithms being compared and the two latent variable models. First, we focus on the choice of MCMC samplers by comparing algorithms that use the MALA sampler (D-SOMALA) with those that use the random-walk MH sampler (D-SOMH).
The algorithms using the MALA sampler generally converged faster and more accurately than those using the random-walk MH sampler in both the lower- and higher-dimensional settings for both models. The efficiency gain stems from the use of gradient information, which yields more informed proposal values and makes posterior sampling more efficient. There is only one exception in the comparison between the QN-SOMALA and QN-SOMH in the lower-dimensional setting of the multilevel logistic regression model. This is because the computational overhead of computing the gradient outweighs the advantage of producing the more informed proposed value, as the MALA sampler requires evaluating the gradient of latent variables twice per iteration to construct its transition probability.

Second, we examine the impact of using the minibatch SG instead of the fullbatch SG (i.e., QN-SOMALA and QN-SOMH). This impact was different between the algorithms using the MALA and random-walk MH samplers. Specifically, the minibatch SG algorithms using the MALA sampler generally converged faster and more accurately than their fullbatch SG counterpart. In contrast, the minibatch SG algorithms using the random-walk MH sampler did not consistently present faster convergence and more accurate estimates than their fullbatch SG counterpart. For example, in higher-dimensional settings, minibatch SG algorithms using the random-walk MH sampler tended to converge faster and more accurately than their fullbatch SG counterparts, especially with smaller minibatches such as $n=250$ and $500$. However, in the lower-dimensional setting, these algorithms did not generally converge faster in the early stage of elapsed time and did not provide more accurate estimates at the end of elapsed time than their fullbatch SG counterpart.  

Lastly, we investigate the impact of incorporating a QN update (QN-D-SOMALA, QN-SOMALA, QN-D-SOMH, QN-SOMH) into the minibatch SG algorithms. A QN update generally did not improve convergence in the minibatch SG algorithms. This is because the variance of the second-order derivatives of the complete-data log-likelihood function becomes large in the algorithms with smaller minibatch sizes, leading to instability in the minibatch SG parameter update. The exception was found only in the comparison between the D-SOMH and its QN variants under the lower-dimensional setting of the confirmatory M2PL model, where the QN-D-SOMH converged substantially faster than the D-SOMH in the early stage of elapsed time, although the accuracy of the D-SOMH became better than the QN-D-SOMH at the end of elapsed time.

\begin{table}[htbp]
  \centering
  \caption{The trajectories of the averaged MAEs across the model parameters for the multilevel logistic regression model with $K=5, J=10, N=10,000$}
   \resizebox{\linewidth}{!}{%
    \begin{tabular}{cccccccccccc}
    \toprule
    \multicolumn{2}{c}{\multirow{2}[3]{*}{Methods}} & \multicolumn{10}{c}{Elapsed Time (seconds)} \\
\cmidrule{3-12}    \multicolumn{2}{c}{} & 0     & 50    & 100   & 150   & 200   & 250   & 500   & 1000  & 1500  & 2000 \\
    \midrule
    \multicolumn{1}{l}{D-SOMALA} & \multicolumn{1}{l}{$n=250$} & .4050 & .0249 & .0148 & .0122 & .0120 & .0118 & .0114 & .0107 & .0106 & .0104 \\
    \multicolumn{1}{l}{D-SOMALA} & \multicolumn{1}{l}{$n=500$} & .4050 & .0345 & .0187 & .0140 & .0129 & .0124 & .0114 & .0108 & .0105 & .0105 \\
    \multicolumn{1}{l}{D-SOMALA} & \multicolumn{1}{l}{$n=1,000$} & .4050 & .0533 & .0317 & .0226 & .0184 & .0175 & .0149 & .0126 & .0117 & .0111 \\
    \hdashline \addlinespace[2pt] 
    \multicolumn{1}{l}{D-SOMH} & \multicolumn{1}{l}{$n=250$} & .4050 & .0633 & .0404 & .0357 & .0325 & .0296 & .0226 & .0169 & .0149 & .0137 \\
    \multicolumn{1}{l}{D-SOMH} & \multicolumn{1}{l}{$n=500$} & .4050 & .0706 & .0448 & .0404 & .0372 & .0350 & .0275 & .0210 & .0184 & .0165 \\
    \multicolumn{1}{l}{D-SOMH} & \multicolumn{1}{l}{$n=1,000$} & .4050 & .0828 & .0567 & .0492 & .0463 & .0440 & .0366 & .0289 & .0249 & .0223 \\
    \hdashline \addlinespace[2pt] 
    \multicolumn{1}{l}{QN-D-SOMALA} & \multicolumn{1}{l}{$n=250$} & .4050 & .0348 & .0187 & .0149 & .0134 & .0129 & .0122 & .0116 & .0111 & .0111 \\
    \multicolumn{1}{l}{QN-D-SOMALA} & \multicolumn{1}{l}{$n=500$} & .4050 & .0377 & .0208 & .0156 & .0134 & .0128 & .0119 & .0113 & .0109 & .0108 \\
    \multicolumn{1}{l}{QN-D-SOMALA} & \multicolumn{1}{l}{$n=1,000$} & .4050 & .0596 & .0361 & .0258 & .0202 & .0172 & .0146 & .0128 & .0119 & .0111 \\
    \hdashline \addlinespace[2pt] 
    \multicolumn{1}{l}{QN-D-SOMH} & \multicolumn{1}{l}{$n=250$} & .4050 & .0700 & .0454 & .0379 & .0351 & .0325 & .0259 & .0207 & .0184 & .0167 \\
    \multicolumn{1}{l}{QN-D-SOMH} & \multicolumn{1}{l}{$n=500$} & .4050 & .0768 & .0504 & .0411 & .0382 & .0366 & .0298 & .0240 & .0207 & .0190 \\
    \multicolumn{1}{l}{QN-D-SOMH} & \multicolumn{1}{l}{$n=1,000$} & .4050 & .0939 & .0670 & .0524 & .0457 & .0443 & .0382 & .0309 & .0273 & .0249 \\
    \hdashline \addlinespace[2pt] 
    \multicolumn{2}{c}{QN-SOMALA} & .4050 & .0964 & .0696 & .0530 & .0477 & .0434 & .0300 & .0190 & .0151 & .0132 \\
    \multicolumn{2}{c}{QN-SOMH} & .4050 & .0733 & .0479 & .0401 & .0349 & .0315 & .0216 & .0158 & .0136 & .0125 \\
    \bottomrule
    \end{tabular}%
    }
  \label{tab:mlogreg_K5}%
\end{table}%

\begin{table}[htbp]
  \centering
  \caption{The trajectories of the averaged MAEs across the model parameters for the multilevel logistic regression model with $K=10, J=20, N=10,000$}
    \resizebox{\linewidth}{!}{%
        \begin{tabular}{cccccccccccc}
        \toprule
    \multicolumn{2}{c}{\multirow{2}[3]{*}{Methods}} & \multicolumn{10}{c}{Elapsed Time (seconds)} \\
\cmidrule{3-12}    \multicolumn{2}{c}{} & 0     & 50    & 100   & 150   & 200   & 250   & 500   & 1000  & 1500  & 2000 \\
    \midrule
    \multicolumn{1}{l}{D-SOMALA} & \multicolumn{1}{l}{$n=250$} & .3363 & .0525 & .0278 & .0200 & .0166 & .0148 & .0117 & .0099 & .0094 & .0091 \\
    \multicolumn{1}{l}{D-SOMALA} & \multicolumn{1}{l}{$n=500$} & .3363 & .0689 & .0388 & .0261 & .0206 & .0175 & .0121 & .0097 & .0090 & .0088 \\
    \multicolumn{1}{l}{D-SOMALA} & \multicolumn{1}{l}{$n=1,000$} & .3363 & .0865 & .0602 & .0447 & .0361 & .0307 & .0188 & .0127 & .0106 & .0096 \\
    \hdashline \addlinespace[2pt] 
    \multicolumn{1}{l}{D-SOMH} & \multicolumn{1}{l}{$n=250$} & .3363 & .1180 & .0923 & .0779 & .0689 & .0621 & .0454 & .0339 & .0291 & .0257 \\
    \multicolumn{1}{l}{D-SOMH} & \multicolumn{1}{l}{$n=500$} & .3363 & .1224 & .0960 & .0837 & .0750 & .0686 & .0506 & .0379 & .0319 & .0282 \\
    \multicolumn{1}{l}{D-SOMH} & \multicolumn{1}{l}{$n=1,000$} & .3363 & .1309 & .1073 & .0964 & .0878 & .0813 & .0624 & .0464 & .0392 & .0347 \\
    \hdashline \addlinespace[2pt] 
    \multicolumn{1}{l}{QN-D-SOMALA} & \multicolumn{1}{l}{$n=250$} & .3363 & .0533 & .0283 & .0203 & .0171 & .0148 & .0120 & .0104 & .0101 & .0097 \\
    \multicolumn{1}{l}{QN-D-SOMALA} & \multicolumn{1}{l}{$n=500$} & .3363 & .0659 & .0360 & .0246 & .0192 & .0165 & .0119 & .0098 & .0093 & .0091 \\
    \multicolumn{1}{l}{QN-D-SOMALA} & \multicolumn{1}{l}{$n=1,000$} & .3363 & .0795 & .0503 & .0355 & .0274 & .0231 & .0144 & .0106 & .0096 & .0091 \\
    \hdashline \addlinespace[2pt] 
    \multicolumn{1}{l}{QN-D-SOMH} & \multicolumn{1}{l}{$n=250$} & .3363 & .1268 & .0995 & .0834 & .0727 & .0658 & .0470 & .0345 & .0291 & .0259 \\
    \multicolumn{1}{l}{QN-D-SOMH} & \multicolumn{1}{l}{$n=500$} & .3363 & .1296 & .1006 & .0853 & .0754 & .0687 & .0497 & .0361 & .0299 & .0263 \\
    \multicolumn{1}{l}{QN-D-SOMH} & \multicolumn{1}{l}{$n=1,000$} & .3363 & .1349 & .1079 & .0932 & .0839 & .0766 & .0573 & .0423 & .0356 & .0312 \\
    \hdashline \addlinespace[2pt] 
    \multicolumn{2}{c}{QN-SOMALA} & .3363 & .1192 & .0967 & .0804 & .0703 & .0626 & .0432 & .0266 & .0191 & .0149 \\
    \multicolumn{2}{c}{QN-SOMH} & .3363 & .1323 & .1044 & .0912 & .0818 & .0747 & .0538 & .0379 & .0313 & .0276 \\
    \bottomrule
    \end{tabular}%
    }
  \label{tab:mlogreg_K10}%
\end{table}%

\begin{table}[htbp]
  \centering
  \caption{The trajectories of the averaged MAEs across the model parameters for the confirmatory M2PL model with $K=5, J=50, N=10,000$}
  \resizebox{\linewidth}{!}{%
     \begin{tabular}{cccccccccccc}
    \toprule
    \multicolumn{2}{c}{\multirow{2}[4]{*}{Methods}} & \multicolumn{10}{c}{Elapsed Time (seconds)} \\
\cmidrule{3-12}    \multicolumn{2}{c}{} & 0     & 20    & 40    & 60    & 80    & 100   & 200   & 300   & 400   & 500 \\
    \midrule
    \multicolumn{1}{l}{D-SOMALA} & \multicolumn{1}{l}{$n=250$} & .6586 & .0289 & .0279 & .0274 & .0270 & .0267 & .0261 & .0261 & .0256 & .0256 \\
    \multicolumn{1}{l}{D-SOMALA} & \multicolumn{1}{l}{$n=500$} & .6586 & .0269 & .0265 & .0260 & .0258 & .0259 & .0256 & .0254 & .0251 & .0252 \\
    \multicolumn{1}{l}{D-SOMALA} & \multicolumn{1}{l}{$n=1,000$} & .6586 & .0258 & .0255 & .0251 & .0251 & .0250 & .0250 & .0247 & .0247 & .0248 \\
    \hdashline \addlinespace[2pt] 
    \multicolumn{1}{l}{D-SOMH} & \multicolumn{1}{l}{$n=250$} & .6586 & .0622 & .0294 & .0288 & .0286 & .0281 & .0276 & .0270 & .0269 & .0269 \\
    \multicolumn{1}{l}{D-SOMH} & \multicolumn{1}{l}{$n=500$} & .6586 & .0698 & .0285 & .0272 & .0270 & .0271 & .0266 & .0260 & .0265 & .0265 \\
    \multicolumn{1}{l}{D-SOMH} & \multicolumn{1}{l}{$n=1,000$} & .6586 & .0933 & .0335 & .0269 & .0263 & .0261 & .0260 & .0258 & .0257 & .0256 \\
    \hdashline \addlinespace[2pt] 
    \multicolumn{1}{l}{QN-D-SOMALA} & \multicolumn{1}{l}{$n=250$} & .6586 & .0382 & .0355 & .0342 & .0337 & .0331 & .0321 & .0314 & .0316 & .0312 \\
    \multicolumn{1}{l}{QN-D-SOMALA} & \multicolumn{1}{l}{$n=500$} & .6586 & .0320 & .0304 & .0297 & .0291 & .0293 & .0290 & .0290 & .0285 & .0283 \\
    \multicolumn{1}{l}{QN-D-SOMALA} & \multicolumn{1}{l}{$n=1,000$} & .6586 & .0305 & .0295 & .0282 & .0280 & .0278 & .0271 & .0274 & .0270 & .0271 \\
    \hdashline \addlinespace[2pt] 
    \multicolumn{1}{l}{QN-D-SOMH} & \multicolumn{1}{l}{$n=250$} & .6586 & .0365 & .0342 & .0331 & .0328 & .0325 & .0325 & .0322 & .0316 & .0313 \\
    \multicolumn{1}{l}{QN-D-SOMH} & \multicolumn{1}{l}{$n=500$} & .6586 & .0354 & .0311 & .0308 & .0310 & .0305 & .0303 & .0305 & .0301 & .0294 \\
    \multicolumn{1}{l}{QN-D-SOMH} & \multicolumn{1}{l}{$n=1,000$} & .6586 & .0366 & .0294 & .0290 & .0288 & .0293 & .0289 & .0287 & .0286 & .0284 \\
    \hdashline \addlinespace[2pt] 
    \multicolumn{2}{c}{QN-SOMALA} & .6586 & .0284 & .0261 & .0252 & .0248 & .0247 & .0244 & .0245 & .0244 & .0244 \\
    \multicolumn{2}{c}{QN-SOMH} & .6586 & .0446 & .0265 & .0255 & .0251 & .0251 & .0252 & .0252 & .0251 & .0250 \\
    \bottomrule
    \end{tabular}%
    }
  \label{tab:m2pl_K5}%
\end{table}%

\begin{table}[htbp]
  \centering
  \caption{The trajectories of the averaged MAEs across the model parameters for the confirmatory M2PL model with $K=10, J=200, N=10,000$}
      \resizebox{\linewidth}{!}{%     
    \begin{tabular}{cccccccccccc}
    \toprule
    \multicolumn{2}{c}{\multirow{2}[4]{*}{Methods}} & \multicolumn{10}{c}{Elapsed Time (seconds)} \\
\cmidrule{3-12}    \multicolumn{2}{c}{} & 0     & 40    & 80    & 120   & 160   & 200   & 400   & 600   & 800   & 1000 \\
    \midrule
    \multicolumn{1}{l}{D-SOMALA} & \multicolumn{1}{l}{$n=250$} & .6556 & .0319 & .0260 & .0255 & .0253 & .0248 & .0241 & .0240 & .0238 & .0236 \\
    \multicolumn{1}{l}{D-SOMALA} & \multicolumn{1}{l}{$n=500$} & .6556 & .0334 & .0250 & .0238 & .0235 & .0233 & .0231 & .0229 & .0228 & .0228 \\
    \multicolumn{1}{l}{D-SOMALA} & \multicolumn{1}{l}{$n=1,000$} & .6556 & .0554 & .0340 & .0279 & .0254 & .0243 & .0229 & .0227 & .0226 & .0224 \\
    \hdashline \addlinespace[2pt] 
    \multicolumn{1}{l}{D-SOMH} & \multicolumn{1}{l}{$n=250$} & .6556 & .1571 & .0756 & .0449 & .0328 & .0278 & .0249 & .0248 & .0247 & .0247 \\
    \multicolumn{1}{l}{D-SOMH} & \multicolumn{1}{l}{$n=500$} & .6556 & .1633 & .0836 & .0512 & .0365 & .0294 & .0238 & .0237 & .0236 & .0236 \\
    \multicolumn{1}{l}{D-SOMH} & \multicolumn{1}{l}{$n=1,000$} & .6556 & .1767 & .1000 & .0652 & .0471 & .0372 & .0239 & .0228 & .0228 & .0228 \\
    \hdashline \addlinespace[2pt] 
    \multicolumn{1}{l}{QN-D-SOMALA} & \multicolumn{1}{l}{$n=250$} & .6556 & .0379 & .0338 & .0327 & .0321 & .0313 & .0296 & .0292 & .0283 & .0285 \\
    \multicolumn{1}{l}{QN-D-SOMALA} & \multicolumn{1}{l}{$n=500$} & .6556 & .0328 & .0292 & .0281 & .0280 & .0275 & .0269 & .0266 & .0263 & .0262 \\
    \multicolumn{1}{l}{QN-D-SOMALA} & \multicolumn{1}{l}{$n=1,000$} & .6556 & .0309 & .0265 & .0263 & .0259 & .0258 & .0250 & .0250 & .0249 & .0247 \\
    \hdashline \addlinespace[2pt] 
    \multicolumn{1}{l}{QN-D-SOMH} & \multicolumn{1}{l}{$n=250$} & .6556 & .1958 & .1067 & .0683 & .0508 & .0415 & .0295 & .0284 & .0284 & .0281 \\
    \multicolumn{1}{l}{QN-D-SOMH} & \multicolumn{1}{l}{$n=500$} & .6556 & .1934 & .1052 & .0682 & .0506 & .0411 & .0283 & .0269 & .0266 & .0268 \\
    \multicolumn{1}{l}{QN-D-SOMH} & \multicolumn{1}{l}{$n=1,000$} & .6556 & .1988 & .1133 & .0747 & .0554 & .0448 & .0286 & .0258 & .0254 & .0252 \\
    \hdashline \addlinespace[2pt] 
    \multicolumn{2}{c}{QN-SOMALA} & .6556 & .0681 & .0466 & .0381 & .0338 & .0314 & .0271 & .0260 & .0253 & .0250 \\
    \multicolumn{2}{c}{QN-SOMH} & .6556 & .2400 & .1605 & .1291 & .1099 & .0971 & .0596 & .0421 & .0342 & .0300 \\
    \bottomrule
    \end{tabular}%
    }
  \label{tab:m2pl_K10}%
\end{table}%

\subsubsection{Comparative Findings}
Here, we summarise the findings of the multilevel logistic regression and confirmatory M2PL models. 

First, the advantage of the MALA sampler was more salient in the multilevel logistic regression model than in the confirmatory M2PL model. Overall, the algorithms using the MALA sampler exhibited substantially faster convergence than those using the random-walk MH sampler in both models. Although their accuracy at the end of the elapsed time was substantially better than that of the random-walk MH sampler in the multilevel logistic regression model, this superiority was less pronounced in the confirmatory M2PL model. The outperformance of the MALA sampler over he random walk MH sampler in the multilevel model may be due to the posterior distribution of the latent variables/random effects being flat. In the simulation study, the number of level-1 units in the multilevel logistic regression model is either $J=10$ or $20$.
On the other hand, the number of items for the confirmatory M2PL model is either $J=50$ or $200$. Thus, the posterior distribution of the latent variables in the multilevel logistic regression model is expected to be flatter than that of the confirmatory M2PL model, since the size of $J$ is associated with the degree of uncertainty in latent variable estimation. Because of the flatness,  the random-walk MH sampler tends to accept proposed values even when they are far from regions of higher posterior density. This, in turn, results in slower convergence for algorithms using the random-walk MH sampler compared with those using the MALA sampler in the multilevel logistic regression model. 

Second, faster convergence and higher accuracy were generally achieved with minibatch SG in both the multilevel logistic regression model and the higher-dimensional confirmatory M2PL model. However, its advantage was less clear in the lower-dimensional setting of the confirmatory M2PL model. For example, in this setting, although the convergence behaviour is similar between the minibatch SG and fullbatch SG algorithms, the accuracy of the minibatch SG algorithms at the end of elapsed time was slightly worse than that of the fullbatch SG algorithms. 
We also observed that, under the lower-dimensional confirmatory M2PL model, the D-SOMH showed substantially slower early-stage convergence than the QN-SOMH. In contrast, the convergence behaviours of the D-SOMALA and QN-SOMALA remained comparable. 
These findings indicate that minibatch SG facilitates faster convergence in high-dimensional settings; however, its impact varies with model complexity and the choice of MCMC sampler. The results also suggest that when minibatch SG is coupled with the MALA sampler, its advantage is more stable than when it is combined with the random-walk MH sampler.

Lastly, there was no notable difference in the impact of using a QN update between the two models. As discussed in Section~\ref{subsubsec:gfindings}, the QN update generally did not improve convergence in the minibatch SG algorithms, except in the comparison between D-SOMH and its QN variants under the lower-dimensional setting of the confirmatory M2PL model.

\section{Application to International Personality Item Pool NEO Personality Inventory}\label{sec:emp}
In this section, we employ the D-SOMALA to fit the confirmatory M2PL model to personality assessment data obtained from the International Personality Item Pool -- Neuroticism, Extraversion, and Openness (IPIP-NEO) personality inventory \parencite{johnson_measuring_2014}. Our focus will be on examining the relationship between the measurement items and their corresponding personality traits, as well as the correlation structure of these latent traits.

\subsection{Data Description}
The dataset used in this study was obtained from the repository of \textcite{johnson_measuring_2014} on the Open Science Framework: \url{https://osf.io/wxvth/}. It includes 307,313 responses to the IPIP-NEO-300 inventory, which measures the Big Five personality traits: openness to experience (O), conscientiousness (C), extraversion (E), agreeableness (A), and neuroticism (N). According to \textcite{johnson_measuring_2014}, each personality trait comprises six facets, yielding 30 personality traits in the inventory. For example, openness to experience includes the following six facets: Imagination (O1), Artistic interests (O2), Emotionality (O3), Adventurousness (O4), Intellect (O5), and Liberalism (O6). Each facet is measured by 10 items, yielding a total of 300 items. For this empirical illustration, a randomly selected subset of 30,000 cases was used, with each case completing all 300 items. 
The items were originally measured on a five-point Likert scale. In the analysis, the items were dichotomised by computing the median of each item and coding an item response greater than or equal to the corresponding item's median as 1, and 0 otherwise.

\subsection{Estimation Settings}
We applied the confirmatory M2PL model to this dataset and adopted the D-SOMALA for parameter estimation. The specifications for the parameter estimation are as follows. Similar to the simulation study, we proceeded with a tuning procedure to determine the step size in the MALA sampler, with candidate values as follows: $h = \nami{0.01, 0.05, 0.1, 0.2}$. 
The tuning procedure was implemented in parallel with four cores, where we ran 500 epochs with $n=1,000$, $\gamma_t = t^{-0.51}$, and $h\in \nami{0.01, 0.05, 0.1, 0.2}$ and monitored the trajectory of the value of the negative complete-data log-likelihood function. We then averaged the negative complete-data log-likelihood over the updated model parameters. We sampled latent variables from the last fifty epochs, and chose the value of $h$ that yielded the minimum among the four candidate values. The value of $h=0.01$ was chosen from the tuning procedure. 
Regarding the starting values in the tuning procedure, the initial values of the latent variables were set to the standardised observed sum scores computed from items measuring the same factor. The initial values of intercept parameters were set to zero, while those of non-zero factor loading parameters were set to 1. The initial value of the correlation matrix was based on the correlations computed from the initial values of latent variables.

Subsequently, we used the updated model parameters and the latent variables sampled at the chosen value of $h=0.01$ in the last iteration of the tuning procedure as the starting values for further analysis. In the analysis, we ran the D-SOMALA with $n=1,000$ until the convergence criterion introduced in Supplementary Materials A.4 was satisfied. Specifically, the algorithm was terminated when the value of $\text{DIFF}_{\text{\scriptsize MAX}}(\cdot)$ remained below 0.05 ten times consecutively within the window size of $\tilde{w}=50$.
The step sizes of the D-SOMALA were specified as $\gamma_t = t^{-0.51}$ and $h = 0.01$. We then report the average of the Polyak-Ruppert trajectories over the last 500 epochs as the final estimate.

Lastly, the entries of the diagonal elements of $\bD^ \iter{t}$ corresponding to the correlation matrix were specified so that the corresponding step sizes were rescaled by 0.1 for improved stability of the optimisation procedure. The other entries were set to 1, meaning that no rescaling was applied to the other model parameters. All the computations were performed on a desktop computer equipped with an Intel(R) Xeon(R) Gold 6246R CPU @ 3.40GHz.

\subsection{Results}

\begin{figure}[htbp]
\color{black}
 \begin{center}
 \includegraphics[width=\linewidth]{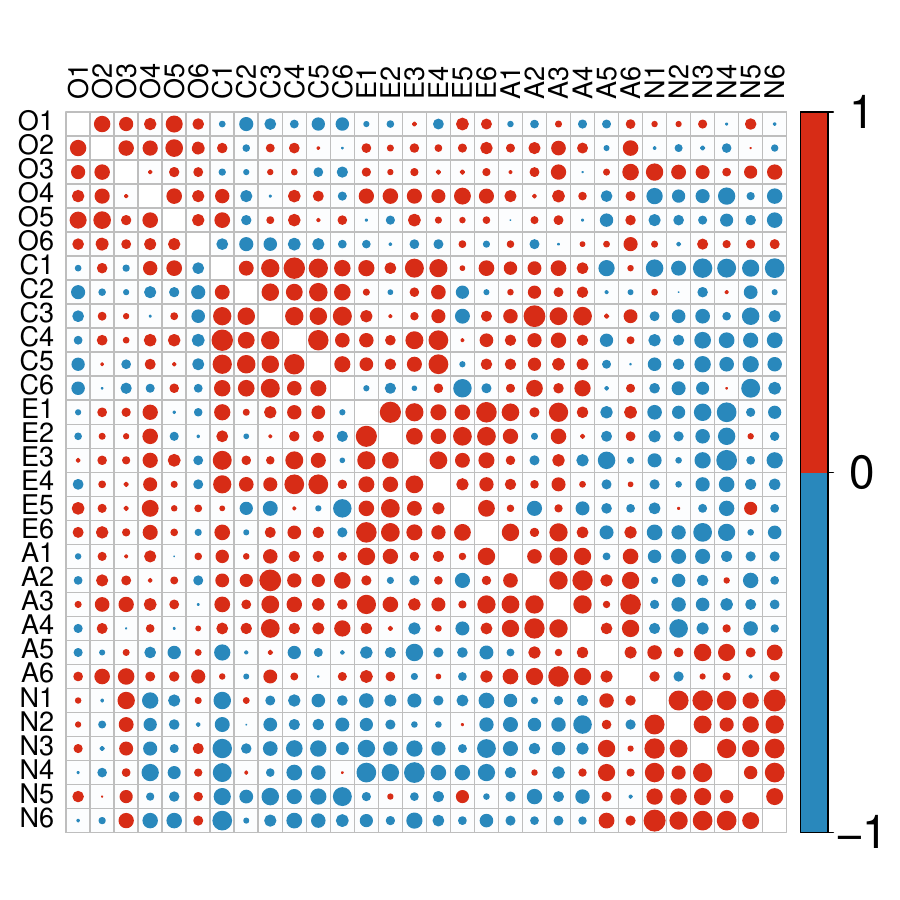}
 \caption{Estimated latent factor correlation matrix, IPIP-NEO dataset. Red and blue circles indicate  positive and negative correlation coefficients, respectively, with the size of each circle corresponding to the magnitude of the coefficient.}
\label{fig:cormat}
 \end{center}
\end{figure}

The estimated factor loadings and latent factor correlation matrix are given in Table \ref{tab:IPIP300loadings} and Figure \ref{fig:cormat}, respectively. In Table \ref{tab:IPIP300loadings}, only
the non-zero factor loadings corresponding to the indicator matrix $\bQ$ are shown. The indicator matrix $\bQ$ was set to have a simple structure in which each item measures only one factor, and each factor is measured by ten items. The values in the ten columns under ‘‘Factor Loading Parameters'' represent the factor 
loadings of each item related to the relevant latent trait.  

Our analysis revealed patterns similar to those reported in a previous study that examined the same personality inventory \parencite{zhang_improved_2020}. It should be noted that the dataset in our study differs from that used in the previous study, which had a sample size of 7,325 \parencite{johnson_ascertaining_2005}. The dataset used in this study is taken from \textcite{johnson_measuring_2014}. All the estimated factor loadings in Table  \ref{tab:IPIP300loadings} are positive, which is consistent with the results reported in \textcite{zhang_improved_2020}.

Additionally, Figure \ref{fig:cormat} shows large positive correlation coefficients in the diagonal blocks formed by facets belonging to the same personality trait. This finding aligns with the design of the personality inventory and is consistent with \textcite{zhang_improved_2020}. Furthermore, the correlations among the facets of different personality traits are consistent with the findings of \textcite{zhang_improved_2020}. For example, most of the facets of N1-N6 negatively correlate with those of E1-E6 and C1-C6.

We have also found the factor loadings related to the sixth and seventh items (‘‘Think highly of myself'' and ‘‘Have a high opinion of myself''), which measure ‘‘A5 (modesty)'' to be very large (6.10 and 6.92). Similarly, the factor loadings in the study by \textcite{zhang_improved_2020} were also large (4.19 and 4.41). When computing the polychoric correlation between the sixth and seventh items, we found an extremely large correlation coefficient (0.947), indicating the high sample correlation between these two items. 

The D-SOMALA took 24 minutes and 25 seconds to tune the MALA sampler step size (i.e., $h$) and 26 minutes and 33 seconds to estimate the model parameters. Despite the large sample size, the number of items, and the latent dimensions, the algorithm 
exhibited high scalability. Additionally, two irregularities in the dataset contributed to slower parameter estimation. First, there were two unusually large factor loadings. Second, the correlation matrix contained large values. These features usually lead to slow parameter learning, requiring many iterations for the algorithm to converge. Therefore, the short processing time in such a large-scale, complex setting demonstrates the high scalability of D-SOMALA.

\begin{table}[htbp]
  \centering
  \caption{Estimated factor loading parameters of the confirmatory M2PL model, IPIP-NEO dataset}
    \resizebox{\linewidth}{!}{%
    \begin{tabular}{rcccccccccc}
    \toprule
    Latent Traits & \multicolumn{10}{c}{Factor Loading Parameters} \\
       \midrule
    O1    & 1.67  & 1.77  & 2.70  & 2.31  & 1.42  & 1.03  & 2.36  & 1.60  & 1.50  & 1.65  \\
    O2    & 2.86  & 0.77  & 1.50  & 0.95  & 1.32  & 3.26  & 1.62  & 2.23  & 0.56  & 0.93  \\
    O3    & 2.10  & 1.18  & 0.58  & 0.31  & 0.47  & 2.23  & 2.44  & 1.20  & 1.63  & 1.57  \\
    O4    & 1.22  & 1.00  & 0.86  & 1.43  & 1.43  & 2.48  & 2.50  & 1.09  & 0.70  & 1.15  \\
    O5    & 1.11  & 1.81  & 1.52  & 1.20  & 1.16  & 1.66  & 1.97  & 2.15  & 2.27  & 1.85  \\
    O6    & 1.55  & 0.76  & 1.43  & 0.78  & 1.95  & 0.67  & 1.29  & 1.46  & 1.73  & 1.01  \\
          &       &       &       &       &       &       &       &       &       &  \\
    C1    & 1.66  & 1.85  & 1.90  & 1.47  & 1.80  & 2.18  & 1.32  & 1.37  & 1.62  & 1.07  \\
    C2    & 1.91  & 1.86  & 0.86  & 1.69  & 1.45  & 1.56  & 1.72  & 1.55  & 1.19  & 1.44  \\
    C3    & 1.54  & 1.45  & 0.97  & 1.67  & 1.25  & 1.78  & 1.57  & 1.42  & 1.54  & 1.30  \\
    C4    & 1.31  & 2.35  & 1.61  & 1.39  & 1.78  & 1.27  & 0.83  & 1.47  & 2.00  & 2.02  \\
    C5    & 1.68  & 1.47  & 2.23  & 1.98  & 1.54  & 2.59  & 2.19  & 2.23  & 2.89  & 1.43  \\
    C6    & 0.80  & 0.91  & 0.71  & 2.84  & 2.01  & 1.31  & 2.36  & 1.49  & 2.94  & 1.14  \\
          &       &       &       &       &       &       &       &       &       &  \\
    E1    & 2.19  & 1.68  & 3.09  & 2.61  & 1.62  & 1.46  & 1.98  & 2.27  & 1.13  & 1.69  \\
    E2    & 2.32  & 2.00  & 1.56  & 1.43  & 1.22  & 2.05  & 1.96  & 2.48  & 2.80  & 1.68  \\
    E3    & 2.55  & 2.17  & 1.21  & 1.06  & 2.27  & 2.06  & 1.63  & 1.44  & 1.15  & 1.29  \\
    E4    & 1.84  & 1.99  & 1.60  & 1.14  & 0.60  & 0.87  & 0.75  & 0.72  & 0.64  & 0.78  \\
    E5    & 2.20  & 2.05  & 1.94  & 1.54  & 1.62  & 1.94  & 1.11  & 1.53  & 1.06  & 0.93  \\
    E6    & 1.86  & 2.59  & 0.96  & 1.28  & 2.28  & 2.14  & 1.35  & 1.23  & 1.11  & 0.96  \\
          &       &       &       &       &       &       &       &       &       &  \\
    A1    & 2.29  & 2.24  & 2.63  & 1.71  & 1.89  & 1.01  & 3.30  & 1.62  & 1.20  & 1.69  \\
    A2    & 0.92  & 1.11  & 0.76  & 1.51  & 1.00  & 1.63  & 1.06  & 1.15  & 2.40  & 1.46  \\
    A3    & 1.65  & 1.21  & 1.81  & 2.00  & 1.11  & 1.34  & 1.47  & 1.29  & 1.53  & 1.74  \\
    A4    & 0.64  & 0.60  & 0.80  & 1.20  & 1.10  & 1.07  & 1.72  & 2.05  & 1.51  & 1.07  \\
    A5    & 0.71  & 0.81  & 0.93  & 0.63  & 1.48  & 6.10  & 6.92  & 0.81  & 0.62  & 0.83  \\
    A6    & 1.42  & 1.75  & 0.91  & 1.09  & 1.52  & 1.40  & 0.75  & 1.13  & 0.97  & 0.92  \\
          &       &       &       &       &       &       &       &       &       &  \\
    N1    & 1.65  & 1.53  & 1.74  & 2.51  & 1.59  & 1.58  & 1.51  & 1.31  & 1.31  & 1.21  \\
    N2    & 2.73  & 2.31  & 2.46  & 1.59  & 2.35  & 2.82  & 2.37  & 2.32  & 1.52  & 1.18  \\
    N3    & 2.34  & 2.80  & 2.84  & 2.73  & 1.61  & 1.91  & 1.40  & 1.83  & 2.04  & 2.05  \\
    N4    & 1.47  & 1.17  & 1.96  & 1.62  & 1.41  & 0.95  & 1.32  & 1.16  & 1.50  & 1.42  \\
    N5    & 0.97  & 1.20  & 1.35  & 1.25  & 0.59  & 1.41  & 1.45  & 1.49  & 0.81  & 0.60  \\
    N6    & 2.20  & 1.64  & 2.02  & 1.21  & 1.72  & 2.10  & 1.30  & 1.67  & 1.66  & 2.11  \\
    \bottomrule
    \end{tabular}%
    }
  \label{tab:IPIP300loadings}%
      \begin{tablenotes}
 \item \footnotesize{\textit{Note.} The indicator matrix of the IPIP-NEO dataset has a simple structure. Accordingly, the values of each row represent the estimated factor loading parameters of ten items that load on the latent trait in that row.}
 \end{tablenotes}
\end{table}%

\section{Discussions} \label{sec:discuss}
The paper proposes two boosting strategies for stochastic optimisation for marginal maximum likelihood estimation. The first boosting strategy aims to improve the efficiency of the latent variable sampling step by employing the MALA sampler, which capitalises on the gradient of latent variables to produce more informed proposed values. The second boosting strategy aims to improve the efficiency of the parameter update step by applying the minibatch technique to the SG. 

The simulation study revealed the following general patterns. First, across both models studied here and the dimensional settings, the algorithms using the MALA sampler consistently outperformed those using the random-walk MH sampler in terms of convergence speed and accuracy. The advantage of the MALA sampler became more pronounced in higher-dimensional settings, where using the gradient information led to more efficient exploration of the parameter space in high-dimensional latent variables. 
Second, the minibatch SG algorithms generally accelerated convergence, especially when the minibatch size was small (e.g., $n=250$ and $500$). This faster convergence was more pronounced in the higher-dimensional setting than in the lower-dimensional setting. 
Lastly, a QN update rarely improved convergence speed and accuracy in the minibatch SG algorithms. This is because the minibatch second-order derivatives are noisy with small minibatches, leading to instability in parameter updates, and because evaluating them incurs additional computational costs.
These findings show that D-SOMALA — a combination of the MALA sampler and the minibatch SG — is the best-performing construction of a stochastic optimisation algorithm among competitors. 
Moreover, our real data analysis demonstrated that the D-SOMALA scales well to the large-scale dataset with 30,000 respondents, 300 items, and 30 latent dimensions, completing the tuning and estimation procedures and successfully converging to a sensible solution within 50 minutes. 

Both the simulation studies and the real-data example demonstrate that the proposed algorithm with the boosting strategies is highly effective for estimating high-dimensional latent variable models. Furthermore, because it can be applied to any model with continuous latent variables whose parameter space has no constraints, as long as gradients with respect to the latent variables are available, it applies to a wide range of models that accommodate various data types and structures. Potential applications include latent regression models \parencite{von2010stochastic} and bi-factor models \parencite{gibbons_full-information_1992,gibbons_full-information_2007,cai_generalized_2011}.

Regarding the specification of hyperparameters, D-SOMALA requires a decaying step size for the model parameter update ($\gamma_t$), a fixed step size for the MALA sampler ($h$), and a minibatch size. As suggested by \textcite{zhang_computation_2022}, the decaying step size for the model parameter update can be set to $\gamma_t = t^{-0.51}$. The fixed step size for the MALA sampler can be tuned by monitoring the trajectory of the complete-data log-likelihood values over a small number of epochs. This procedure, described in Supplementary Materials A.1, is computationally inexpensive and, as shown in our real-data analysis, can be carried out efficiently. With respect to the minibatch size, although algorithms with smaller minibatches tend to converge faster in the early stages, they do not necessarily yield better convergence in the later stages. For instance, minibatch sizes of $n=250$ and $500$ yielded the smallest MAEs at the end of the timeframe in the multilevel logistic regression model, whereas $n=1,000$ yielded the smallest MAEs in the confirmatory M2PL model. Since the optimal minibatch size can depend on model complexity, it may also be tuned in the same manner as the step sizes.

While the proposed algorithm is promising for addressing the estimation problem in high-dimensional latent variable models, there is still room for improvement and exploration in several directions. One limitation is that the current algorithm cannot be directly applied to models with discrete latent variables because of the MALA sampler's nature. This is important because many psychometric models involve discrete latent variables \parencite[e.g.,][]{von_davier_partially_2004, henson_defining_2009}. Therefore, it would be valuable to extend the current algorithms to a more general setting that allows for both continuous and discrete latent variables.
There have been recent developments in Langevin-like methods for MCMC analysis of models involving discrete variables \parencite{zanella2019informed,zhang2022langevin} that use the geometric information of the target distribution to propose MCMC updates. These methods may be incorporated into the current framework to accommodate both continuous and discrete latent variables. 

It is also important to consider extending the proposed algorithm to settings in which the continuous latent variables are subject to equality and inequality constraints, which are commonly encountered in psychometric models. One example is the mixed membership model \parencite{erosheva2004mixed,airoldi2015handbook}, where latent variables lie on a probability simplex. The MALA sampler does not handle sampling
in constrained spaces. However, recent advancements in MCMC sampling have led to the development of mirror-Langevin methods \parencite{2018_mirrorlangevin} that  enable efficient sampling from constrained spaces, including the probability simplex. These methods combine the MALA sampler with mirror descent methods for constrained optimisation \parencite{nemirovskij1983problem} to achieve sampling in constrained spaces. 
It is possible to integrate mirror-Langevin methods into the current algorithm and to develop a theoretical guarantee for the resulting stochastic optimisation algorithm. We are working on incorporating these extensions into the D-SOMALA estimation framework. 

Finally, we note that the estimation framework can be readily extended to a stochastic proximal-gradient update step for regularised marginal maximum-likelihood estimation, where the objective function includes a LASSO-type non-smooth penalty term.

In summary, this study investigates the boosting strategies for stochastic optimisation algorithms and proposes the best-performing construction of a stochastic optimisation algorithm---the D-SOMALA---combining the MALA sampler and the minibatch SG. The D-SOMALA is easy to implement, potentially applicable to a wide range of latent variable models, and scales effectively as the dimension of the latent space increases. It paves the way for estimating high-dimensional latent variable models, which are becoming more prevalent in social and behavioural research.

\printbibliography

@article{atchade_perturbed_2017,
	title = {On perturbed proximal gradient algorithms},
	volume = {18},
	url = {https://jmlr.org/papers/volume18/15-038/15-038.pdf},
	journal = {Journal of Machine Learning Research},
	author = {Atchade, Yves F and Fort, Gersende and Moulines, Eric},
	year = {2017},
	pages = {1--33},
}

@article{zhang_improved_2020,
	title = {An improved stochastic {EM} algorithm for large-scale full-information item factor analysis},
	volume = {73},
	issn = {0007-1102, 2044-8317},
	shorttitle = {An improved stochastic {\textbackslash}textlessspan style="font-variant},
	url = {https://onlinelibrary.wiley.com/doi/10.1111/bmsp.12153},
	doi = {10.1111/bmsp.12153},
	number = {1},
	urldate = {2021-10-24},
	journal = {British Journal of Mathematical and Statistical Psychology},
	author = {Zhang, Siliang and Chen, Yunxiao and Liu, Yang},
	month = feb,
	year = {2020},
	pages = {44--71},
}

@article{zhang_computation_2022,
	title = {Computation for Latent Variable Model Estimation: A Unified Stochastic Proximal Framework},
	volume = {87},
	issn = {0033-3123, 1860-0980},
	url = {https://link.springer.com/10.1007/s11336-022-09863-9},
	doi = {10.1007/s11336-022-09863-9},
	shorttitle = {Computation for Latent Variable Model Estimation},
	pages = {1473--1502},
	number = {4},
	journaltitle = {Psychometrika},
	shortjournal = {Psychometrika},
	author = {Zhang, Siliang and Chen, Yunxiao},
	urldate = {2024-06-12},
	date = {2022-12},
	langid = {english},
}

@article{betancourt_conceptual_2017,
	title = {A {Conceptual} {Introduction} to {Hamiltonian} {Monte} {Carlo}},
	url = {https://arxiv.org/abs/1701.02434},
	doi = {10.48550/ARXIV.1701.02434},
	urldate = {2023-02-21},
	journal = {arXiv},
	author = {Betancourt, Michael},
	year = {2017},
}

@article{shun_laplace_1995,
	title = {Laplace {Approximation} of {High} {Dimensional} {Integrals}},
	volume = {57},
	issn = {00359246},
	url = {https://onlinelibrary.wiley.com/doi/10.1111/j.2517-6161.1995.tb02060.x},
	doi = {10.1111/j.2517-6161.1995.tb02060.x},
	number = {4},
	urldate = {2023-02-22},
	journal = {Journal of the Royal Statistical Society: Series B (Methodological)},
	author = {Shun, Zhenming and McCullagh, Peter},
	month = nov,
	year = {1995},
	pages = {749--760},
}

@article{schilling_high-dimensional_2005,
	title = {High-dimensional maximum marginal likelihood item factor analysis by adaptive quadrature},
	rights = {http://www.springer.com/tdm},
	issn = {0033-3123, 1860-0980},
	url = {http://link.springer.com/10.1007/s11336-003-1141-x},
	doi = {10.1007/s11336-003-1141-x},
	journaltitle = {Psychometrika},
	shortjournal = {Psychometrika},
	author = {Schilling, Stephen and Bock, R. Darrell.},
	urldate = {2024-06-12},
	date = {2005-10-05},
	langid = {english},
    volume = {70},
    number = {3}, 
    pages = {533--555}
}

@article{gu_stochastic_1998,
	title = {A stochastic approximation algorithm with {Markov} chain {Monte}-{Carlo} method for incomplete data estimation problems},
	volume = {95},
	issn = {0027-8424, 1091-6490},
	url = {https://pnas.org/doi/full/10.1073/pnas.95.13.7270},
	doi = {10.1073/pnas.95.13.7270},
	number = {13},
	urldate = {2023-02-22},
	journal = {Proceedings of the National Academy of Sciences},
	author = {Gu, Ming Gao and Kong, Fan Hui},
	month = jun,
	year = {1998},
	pages = {7270--7274},
}

@book{reckase_multidimensional_2009,
	address = {New York, NY},
	title = {Multidimensional {Item} {Response} {Theory}},
	url = {http://link.springer.com/10.1007/978-0-387-89976-3},
	urldate = {2023-02-24},
	publisher = {Springer New York},
	author = {Reckase, M.D.},
	year = {2009},
	doi = {10.1007/978-0-387-89976-3},
}

@article{de_bortoli_efficient_2021,
	title = {Efficient stochastic optimisation by unadjusted {Langevin} {Monte} {Carlo}: {Application} to maximum marginal likelihood and empirical {Bayesian} estimation},
	volume = {31},
	issn = {0960-3174, 1573-1375},
	shorttitle = {Efficient stochastic optimisation by unadjusted {Langevin} {Monte} {Carlo}},
	url = {https://link.springer.com/10.1007/s11222-020-09986-y},
	doi = {10.1007/s11222-020-09986-y},
	language = {en},
	number = {3},
	urldate = {2023-08-21},
	journal = {Statistics and Computing},
	author = {De Bortoli, Valentin and Durmus, Alain and Pereyra, Marcelo and Vidal, Ana F.},
	month = may,
	year = {2021},
	pages = {1--18},
}

@article{cai_high-dimensional_2010,
	title = {High-dimensional {exploratory} {item} {factor} {analysis} by a {Metropolis}–{Hastings} {Robbins}–{Monro} {Algorithm}},
	volume = {75},
	issn = {0033-3123, 1860-0980},
	url = {https://link.springer.com/10.1007/s11336-009-9136-x},
	doi = {10.1007/s11336-009-9136-x},
	language = {en},
	number = {1},
	urldate = {2023-09-20},
	journal = {Psychometrika},
	author = {Cai, Li},
	month = mar,
	year = {2010},
	pages = {33--57},
}

@article{roberts_optimal_1998,
	title = {Optimal {scaling} of {discrete} {approximations} to {Langevin} {diffusions}},
	volume = {60},
	issn = {1369-7412, 1467-9868},
	url = {https://academic.oup.com/jrsssb/article/60/1/255/7083121},
	doi = {10.1111/1467-9868.00123},
	language = {en},
	number = {1},
	urldate = {2023-09-20},
	journal = {Journal of the Royal Statistical Society Series B: Statistical Methodology},
	author = {Roberts, Gareth O. and Rosenthal, Jeffrey S.},
	month = jan,
	year = {1998},
	pages = {255--268},
}

@article{gronau_tutorial_2017,
	title = {A tutorial on bridge sampling},
	volume = {81},
	issn = {00222496},
	url = {https://linkinghub.elsevier.com/retrieve/pii/S0022249617300640},
	doi = {10.1016/j.jmp.2017.09.005},
	pages = {80--97},
	journaltitle = {Journal of Mathematical Psychology},
	shortjournal = {Journal of Mathematical Psychology},
	author = {Gronau, Quentin F. and Sarafoglou, Alexandra and Matzke, Dora and Ly, Alexander and Boehm, Udo and Marsman, Maarten and Leslie, David S. and Forster, Jonathan J. and Wagenmakers, Eric-Jan and Steingroever, Helen},
	urldate = {2023-11-13},
	date = {2017-12},
	langid = {english},
}

@book{bishop_pattern_2006,
	location = {New York},
	title = {Pattern recognition and machine learning},
	isbn = {978-0-387-31073-2},
	series = {Information science and statistics},
	pagetotal = {738},
	publisher = {Springer},
	author = {Bishop, Christopher M.},
	date = {2006},
}

@article{durmus2019high,
	title = {High-dimensional {Bayesian} inference via the unadjusted {Langevin} algorithm},
	volume = {25},
	issn = {1350-7265},
	url = {https://projecteuclid.org/journals/bernoulli/volume-25/issue-4A/High-dimensional-Bayesian-inference-via-the-unadjusted-Langevin-algorithm/10.3150/18-BEJ1073.full},
	doi = {10.3150/18-BEJ1073},
	number = {4A},
pages={2854--2882},
	journaltitle = {Bernoulli},
	shortjournal = {Bernoulli},
	author = {Durmus, Alain and Moulines, Éric},
	urldate = {2024-06-12},
	date = {2019-11-01},
}

@article{cai2010metropolis,
	title = {{Metropolis}-{Hastings} {Robbins}-{Monro} Algorithm for Confirmatory Item Factor Analysis},
	volume = {35},
	rights = {http://journals.sagepub.com/page/policies/text-and-data-mining-license},
	issn = {1076-9986, 1935-1054},
	url = {http://journals.sagepub.com/doi/10.3102/1076998609353115},
	doi = {10.3102/1076998609353115},
	pages = {307--335},
	number = {3},
	journaltitle = {Journal of Educational and Behavioral Statistics},
	shortjournal = {Journal of Educational and Behavioral Statistics},
	author = {Cai, Li},
	urldate = {2024-06-12},
	date = {2010-06},
	langid = {english},
}

@article{chen2020latent,
	title = {A Latent {Gaussian} process model for analysing intensive longitudinal data},
	volume = {73},
	issn = {0007-1102, 2044-8317},
	url = {https://bpspsychub.onlinelibrary.wiley.com/doi/10.1111/bmsp.12180},
	doi = {10.1111/bmsp.12180},
	pages = {237--260},
	number = {2},
	journaltitle = {British Journal of Mathematical and Statistical Psychology},
	shortjournal = {Brit J Math \& Statis},
	author = {Chen, Yunxiao and Zhang, Siliang},
	urldate = {2024-06-12},
	date = {2020-05},
	langid = {english},
}

@article{zhu2016personalized,
	title = {Personalized Prediction and Sparsity Pursuit in Latent Factor Models},
	volume = {111},
	issn = {0162-1459, 1537-274X},
	url = {https://www.tandfonline.com/doi/full/10.1080/01621459.2014.999158},
	doi = {10.1080/01621459.2014.999158},
	pages = {241--252},
	number = {513},
	journaltitle = {Journal of the American Statistical Association},
	shortjournal = {Journal of the American Statistical Association},
	author = {Zhu, Yunzhang and Shen, Xiaotong and Ye, Changqing},
	urldate = {2024-06-12},
	date = {2016-01-02},
	langid = {english},
}

@article{hoff2002latent,
	title = {Latent Space Approaches to Social Network Analysis},
	volume = {97},
	issn = {0162-1459, 1537-274X},
	url = {http://www.tandfonline.com/doi/abs/10.1198/016214502388618906},
	doi = {10.1198/016214502388618906},
	pages = {1090--1098},
	number = {460},
	journaltitle = {Journal of the American Statistical Association},
	shortjournal = {Journal of the American Statistical Association},
	author = {Hoff, Peter D and Raftery, Adrian E and Handcock, Mark S},
	urldate = {2024-06-12},
	date = {2002-12},
	langid = {english},
}

@article{haberman1977maximum,
	title = {Maximum Likelihood Estimates in Exponential Response Models},
	volume = {5},
	issn = {0090-5364},
	url = {https://projecteuclid.org/journals/annals-of-statistics/volume-5/issue-5/Maximum-Likelihood-Estimates-in-Exponential-Response-Models/10.1214/aos/1176343941.full},
	doi = {10.1214/aos/1176343941},
	number = {5},
	journaltitle = {The Annals of Statistics},
	shortjournal = {Ann. Statist.},
	author = {Haberman, Shelby J.},
	urldate = {2024-06-12},
	date = {1977-09-01},
    pages = {815--841}
}

@article{chen2019joint,
	title = {Joint Maximum Likelihood Estimation for High-Dimensional Exploratory Item Factor Analysis},
	volume = {84},
	issn = {0033-3123, 1860-0980},
	url = {http://link.springer.com/10.1007/s11336-018-9646-5},
	doi = {10.1007/s11336-018-9646-5},
	pages = {124--146},
	number = {1},
	journaltitle = {Psychometrika},
	shortjournal = {Psychometrika},
	author = {Chen, Yunxiao and Li, Xiaoou and Zhang, Siliang},
	urldate = {2024-06-12},
	date = {2019-03},
	langid = {english},
}

@article{chen2020structured,
	title = {Structured Latent Factor Analysis for Large-scale Data: Identifiability, Estimability, and Their Implications},
	volume = {115},
	issn = {0162-1459, 1537-274X},
	url = {https://www.tandfonline.com/doi/full/10.1080/01621459.2019.1635485},
	doi = {10.1080/01621459.2019.1635485},
	shorttitle = {Structured Latent Factor Analysis for Large-scale Data},
	pages = {1756--1770},
	number = {532},
	journaltitle = {Journal of the American Statistical Association},
	shortjournal = {Journal of the American Statistical Association},
	author = {Chen, Yunxiao and Li, Xiaoou and Zhang, Siliang},
	urldate = {2024-06-12},
	date = {2020-10-01},
	langid = {english},
}

@article{lu2015bayesian,
	title = {Bayesian analysis of ambulatory blood pressure dynamics with application to irregularly spaced sparse data},
	volume = {9},
	issn = {1932-6157},
	url = {https://projecteuclid.org/journals/annals-of-applied-statistics/volume-9/issue-3/Bayesian-analysis-of-ambulatory-blood-pressure-dynamics-with-application-to/10.1214/15-AOAS846.full},
	doi = {10.1214/15-AOAS846},
	number = {3},
    pages={1601--1620},
	journaltitle = {The Annals of Applied Statistics},
	shortjournal = {Ann. Appl. Stat.},
	author = {Lu, Zhao-Hua and Chow, Sy-Miin and Sherwood, Andrew and Zhu, Hongtu},
	urldate = {2024-06-12},
	date = {2015-09-01},
}

@book{bartholomew2008analysis,
  title={Analysis of multivariate social science data},
  author={Bartholomew, David J and Steele, Fiona and Galbraith, J and Moustaki, Irini},
  year={2008},
  publisher={CRC Press}
}

@article{huber2004estimation,
	title = {Estimation of Generalized Linear Latent Variable Models},
	volume = {66},
	rights = {https://academic.oup.com/journals/pages/open\_access/funder\_policies/chorus/standard\_publication\_model},
	issn = {1369-7412, 1467-9868},
	url = {https://academic.oup.com/jrsssb/article/66/4/893/7109366},
	doi = {10.1111/j.1467-9868.2004.05627.x},
	pages = {893--908},
	number = {4},
	journaltitle = {Journal of the Royal Statistical Society Series B: Statistical Methodology},
	author = {Huber, Philippe and Ronchetti, Elvezio and Victoria-Feser, Maria-Pia},
	urldate = {2024-06-12},
	date = {2004-11-01},
	langid = {english},
}

@article{wei1990monte,
	title = {A {Monte Carlo} Implementation of the {EM} Algorithm and the Poor Man's Data Augmentation Algorithms},
	volume = {85},
	issn = {0162-1459, 1537-274X},
	url = {http://www.tandfonline.com/doi/abs/10.1080/01621459.1990.10474930},
	doi = {10.1080/01621459.1990.10474930},
	pages = {699--704},
	number = {411},
	journaltitle = {Journal of the American Statistical Association},
	shortjournal = {Journal of the American Statistical Association},
	author = {Wei, Greg C. G. and Tanner, Martin A.},
	urldate = {2024-06-12},
	date = {1990-09},
	langid = {english},
}

@article{meng1996fitting,
	title = {Fitting Full-Information Item Factor Models and an Empirical Investigation of Bridge Sampling},
	volume = {91},
	issn = {0162-1459, 1537-274X},
	url = {http://www.tandfonline.com/doi/abs/10.1080/01621459.1996.10476995},
	doi = {10.1080/01621459.1996.10476995},
	pages = {1254--1267},
	number = {435},
	journaltitle = {Journal of the American Statistical Association},
	shortjournal = {Journal of the American Statistical Association},
	author = {Meng, Xiao-Li and Schilling, Stephen},
	urldate = {2024-06-12},
	date = {1996-09},
	langid = {english},
}

@incollection{diebolt1995stochastic,
  title={Stochastic {EM}: method and application},
  author={Diebolt, Jean and Ip, Eddie HS},
  booktitle={Markov chain {Monte Carlo} in practice},
  year={1995},
  publisher={CRC Press}
}

@article{nielsen2000stochastic,
	title = {The Stochastic {EM} Algorithm: Estimation and Asymptotic Results},
	volume = {6},
	issn = {13507265},
	url = {https://www.jstor.org/stable/3318671?origin=crossref},
	doi = {10.2307/3318671},
	shorttitle = {The Stochastic {EM} Algorithm},
	pages = {457--489},
	number = {3},
	journaltitle = {Bernoulli},
	shortjournal = {Bernoulli},
	author = {Nielsen, Søren Feodor},
	urldate = {2024-06-12},
	date = {2000-06},
}

@article{roberts1996exponential,
	title = {Exponential Convergence of {Langevin} Distributions and Their Discrete Approximations},
	volume = {2},
	issn = {13507265},
	url = {https://www.jstor.org/stable/3318418?origin=crossref},
	doi = {10.2307/3318418},
	pages = {341--363},
	number = {4},
	journaltitle = {Bernoulli},
	shortjournal = {Bernoulli},
	author = {Roberts, Gareth O. and Tweedie, Richard L.},
	urldate = {2024-06-12},
	date = {1996-12},
}

@article{dempster_maximum_1977,
	title = {Maximum Likelihood from Incomplete Data Via the \textit{{EM}} Algorithm},
	volume = {39},
	issn = {0035-9246, 2517-6161},
	url = {https://rss.onlinelibrary.wiley.com/doi/10.1111/j.2517-6161.1977.tb01600.x},
	doi = {10.1111/j.2517-6161.1977.tb01600.x},
	pages = {1--22},
	number = {1},
	journaltitle = {Journal of the Royal Statistical Society: Series B (Methodological)},
	shortjournal = {Journal of the Royal Statistical Society: Series B (Methodological)},
	author = {Dempster, A. P. and Laird, N. M. and Rubin, D. B.},
	urldate = {2024-01-14},
	date = {1977-09},
	langid = {english},
}

@article{bock_marginal_1981,
	title = {Marginal maximum likelihood estimation of item parameters: Application of an {EM} algorithm},
	volume = {46},
	issn = {0033-3123, 1860-0980},
	url = {http://link.springer.com/10.1007/BF02293801},
	doi = {10.1007/BF02293801},
	shorttitle = {Marginal maximum likelihood estimation of item parameters},
	pages = {443--459},
	number = {4},
	journaltitle = {Psychometrika},
	shortjournal = {Psychometrika},
	author = {Bock, R. Darrell and Aitkin, Murray},
	urldate = {2024-01-14},
	date = {1981-12},
	langid = {english},
}

@article{robbins_stochastic_1951,
	title = {A Stochastic Approximation Method},
	volume = {22},
	issn = {0003-4851},
	url = {http://projecteuclid.org/euclid.aoms/1177729586},
	doi = {10.1214/aoms/1177729586},
	pages = {400--407},
	number = {3},
	journaltitle = {The Annals of Mathematical Statistics},
	shortjournal = {Ann. Math. Statist.},
	author = {Robbins, Herbert and Monro, Sutton},
	urldate = {2024-01-14},
	date = {1951-09},
	langid = {english},
}

@book{Goodfellow-et-al-2016,
    title={Deep Learning},
    author={Ian Goodfellow and Yoshua Bengio and Aaron Courville},
    publisher={MIT Press},
    year={2016}
}

@article{johnson_measuring_2014,
	title = {Measuring thirty facets of the Five Factor Model with a 120-item public domain inventory: Development of the {IPIP}-{NEO}-120},
	volume = {51},
	issn = {00926566},
	url = {https://linkinghub.elsevier.com/retrieve/pii/S0092656614000506},
	doi = {10.1016/j.jrp.2014.05.003},
	shorttitle = {Measuring thirty facets of the Five Factor Model with a 120-item public domain inventory},
	pages = {78--89},
	journaltitle = {Journal of Research in Personality},
	shortjournal = {Journal of Research in Personality},
	author = {Johnson, John A.},
	urldate = {2024-02-16},
	date = {2014-08},
	langid = {english},
}

@book{skrondal_generalized_2004,
	location = {Boca Raton},
	title = {Generalized latent variable modeling: multilevel, longitudinal, and structural equation models},
	isbn = {978-1-58488-000-4},
	series = {Chapman \& Hall/{CRC} interdisciplinary statistics series},
	shorttitle = {Generalized latent variable modeling},
	pagetotal = {508},
	publisher = {Chapman \& Hall/{CRC}},
	author = {Skrondal, Anders and Rabe-Hesketh, S.},
	date = {2004},
}

@article{johnson_ascertaining_2005,
	title = {Ascertaining the validity of individual protocols from Web-based personality inventories},
	volume = {39},
	rights = {https://www.elsevier.com/tdm/userlicense/1.0/},
	issn = {00926566},
	url = {https://linkinghub.elsevier.com/retrieve/pii/S0092656604000856},
	doi = {10.1016/j.jrp.2004.09.009},
	pages = {103--129},
	number = {1},
	journaltitle = {Journal of Research in Personality},
	shortjournal = {Journal of Research in Personality},
	author = {Johnson, John A.},
	urldate = {2024-04-18},
	date = {2005-02},
	langid = {english},
}

@article{von_davier_partially_2004,
	title = {Partially Observed Mixtures of {IRT} Models: An Extension of the Generalized Partial-Credit Model},
	volume = {28},
	rights = {http://journals.sagepub.com/page/policies/text-and-data-mining-license},
	issn = {0146-6216, 1552-3497},
	url = {http://journals.sagepub.com/doi/10.1177/0146621604268734},
	doi = {10.1177/0146621604268734},
	shorttitle = {Partially Observed Mixtures of {IRT} Models},
	pages = {389--406},
	number = {6},
	journaltitle = {Applied Psychological Measurement},
	shortjournal = {Applied Psychological Measurement},
	author = {von Davier, Matthias and Yamamoto, Kentaro},
	urldate = {2024-04-27},
	date = {2004-11},
	langid = {english},
}

@article{polyak_acceleration_1992,
	title = {Acceleration of Stochastic Approximation by Averaging},
	volume = {30},
	issn = {0363-0129, 1095-7138},
	url = {http://epubs.siam.org/doi/10.1137/0330046},
	doi = {10.1137/0330046},
	pages = {838--855},
	number = {4},
	journaltitle = {{SIAM} Journal on Control and Optimization},
	shortjournal = {{SIAM} J. Control Optim.},
	author = {Polyak, B. T. and Juditsky, A. B.},
	urldate = {2020-09-26},
	date = {1992-07},
	langid = {english}
}

@article{ruppert_efficient_1988,
	title = {Efficient estimations from a slowly convergent {R}obbins-{M}onro process},
	url = {https://hdl.handle.net/1813/8664},
	journaltitle = {Technical Report, Cornell University Operations Research and Industrial Engineering},
	author = {Ruppert, D},
	date = {1988}
}

@book{zhang2023dive,
	title = {Dive into deep learning},
	isbn = {978-1-00-938943-3},
	pagetotal = {548},
	publisher = {Cambridge University Press},
	author = {Zhang, Aston and Lipton, Zachary and Li, Mu and Smola, Alexander J.},
	date = {2024},
	doi = {10.1017/9781009389426},
}

@article{zanella2019informed,
	title = {Informed Proposals for Local {MCMC} in Discrete Spaces},
	volume = {115},
	issn = {0162-1459, 1537-274X},
	url = {https://www.tandfonline.com/doi/full/10.1080/01621459.2019.1585255},
	doi = {10.1080/01621459.2019.1585255},
	pages = {852--865},
	number = {530},
	journaltitle = {Journal of the American Statistical Association},
	shortjournal = {Journal of the American Statistical Association},
	author = {Zanella, Giacomo},
	urldate = {2024-06-12},
	date = {2020-04-02},
	langid = {english},
}

@inproceedings{zhang2022langevin,
  title = 	 {A {L}angevin-like Sampler for Discrete Distributions},
  author =       {Zhang, Ruqi and Liu, Xingchao and Liu, Qiang},
  booktitle = 	 {Proceedings of the 39th International Conference on Machine Learning},
  pages = 	 {26375--26396},
  year = 	 {2022},
  editor = 	 {Chaudhuri, Kamalika and Jegelka, Stefanie and Song, Le and Szepesvari, Csaba and Niu, Gang and Sabato, Sivan},
  volume = 	 {162},
  series = 	 {Proceedings of Machine Learning Research},
  publisher =    {PMLR},
  url = 	 {https://proceedings.mlr.press/v162/zhang22t.html}
}

@article{erosheva2004mixed,
	title = {Mixed-membership models of scientific publications},
	volume = {101},
	issn = {0027-8424, 1091-6490},
	url = {https://pnas.org/doi/full/10.1073/pnas.0307760101},
	doi = {10.1073/pnas.0307760101},
	pages = {5220--5227},
	number = {suppl\_1},
	journaltitle = {Proceedings of the National Academy of Sciences},
	shortjournal = {Proc. Natl. Acad. Sci. U.S.A.},
	author = {Erosheva, Elena and Fienberg, Stephen and Lafferty, John},
	urldate = {2024-06-01},
	date = {2004-04-06},
	langid = {english},
}

@book{airoldi2015handbook,
  title={Handbook of mixed membership models and their applications},
  author={Airoldi, Edoardo M and Blei, David M and Erosheva, Elena A and Fienberg, Stephen E},
  year={2015},
  publisher={CRC Press}
}

@book{nemirovskij1983problem,
  title={Problem complexity and method efficiency in optimization},
  author={Nemirovskij, Arkadij Semenovi{\v{c}} and Yudin, David Borisovich},
  year={1983},
  publisher={Wiley-Interscience}
}

@article{henson_defining_2009,
	title = {Defining a Family of Cognitive Diagnosis Models Using Log-Linear Models with Latent Variables},
	volume = {74},
	issn = {0033-3123, 1860-0980},
	url = {http://link.springer.com/10.1007/s11336-008-9089-5},
	doi = {10.1007/s11336-008-9089-5},
	pages = {191--210},
	number = {2},
	journaltitle = {Psychometrika},
	shortjournal = {Psychometrika},
	author = {Henson, Robert A. and Templin, Jonathan L. and Willse, John T.},
	urldate = {2020-09-26},
	date = {2009-06},
	langid = {english},
}

@inproceedings{2018_mirrorlangevin,
 author = {Hsieh, Ya-Ping and Kavis, Ali and Rolland, Paul and Cevher, Volkan},
 booktitle = {Advances in Neural Information Processing Systems},
 editor = {S. Bengio and H. Wallach and H. Larochelle and K. Grauman and N. Cesa-Bianchi and R. Garnett},
 pages = {},
 publisher = {Curran Associates, Inc.},
 title = {Mirrored {Langevin} Dynamics},
 url = {https://proceedings.neurips.cc/paper_files/paper/2018/file/6490791e7abf6b29a381288cc23a8223-Paper.pdf},
 volume = {31},
 year = {2018}
}

@article{bezanson_julia_2017,
	title = {Julia: A Fresh Approach to Numerical Computing},
	volume = {59},
	issn = {0036-1445, 1095-7200},
	url = {https://epubs.siam.org/doi/10.1137/141000671},
	doi = {10.1137/141000671},
	shorttitle = {Julia},
	pages = {65--98},
	number = {1},
	journaltitle = {{SIAM} Review},
	shortjournal = {{SIAM} Rev.},
	author = {Bezanson, Jeff and Edelman, Alan and Karpinski, Stefan and Shah, Viral B.},
	urldate = {2020-09-26},
	date = {2017-01},
	langid = {english}
}

@article{edwards2010markov,
  title={A {Markov chain Monte Carlo} approach to confirmatory item factor analysis},
  author={Edwards, Michael C},
  journal={Psychometrika},
  volume={75},
  number={3},
  pages={474--497},
  year={2010},
  publisher={Springer}
}

@article{newton_approximate_1994,
	title = {Approximate {Bayesian} {Inference} with the {Weighted} {Likelihood} {Bootstrap}},
	volume = {56},
	copyright = {https://academic.oup.com/journals/pages/open\_access/funder\_policies/chorus/standard\_publication\_model},
	url = {https://academic.oup.com/jrsssb/article/56/1/3/7035883},
	doi = {10.1111/j.2517-6161.1994.tb01956.x},
	language = {en},
	number = {1},
	urldate = {2024-11-20},
	journal = {Journal of the Royal Statistical Society Series B: Statistical Methodology},
	author = {Newton, Michael A. and Raftery, Adrian E.},
	month = jan,
	year = {1994},
	pages = {3--26},
}

@book{busemeyer_model_2015,
	title = {Model {Comparison} and the {Principle} of {Parsimony}},
	volume = {1},
	url = {https://academic.oup.com/edited-volume/41261/chapter/350847102},
	language = {en},
	urldate = {2024-11-22},
	publisher = {Oxford University Press},
	author = {Vandekerckhove, Joachim and Matzke, Dora and Wagenmakers, Eric-Jan},
	editor = {Busemeyer, Jerome R. and Wang, Zheng and Townsend, James T. and Eidels, Ami},
	month = dec,
	year = {2015},
	doi = {10.1093/oxfordhb/9780199957996.013.14},
}

@article{neal_annealed_2001,
	title = {Annealed importance sampling},
	volume = {11},
	issn = {09603174},
	url = {http://link.springer.com/10.1023/A:1008923215028},
	doi = {10.1023/A:1008923215028},
	number = {2},
	urldate = {2024-11-22},
	journal = {Statistics and Computing},
	author = {Neal, Radford M.},
	year = {2001},
	pages = {125--139},
}

@article{von2010stochastic,
  title={Stochastic approximation methods for latent regression item response models},
  author={von Davier, Matthias and Sinharay, Sandip},
  journal={Journal of Educational and Behavioral Statistics},
  volume={35},
  number={2},
  pages={174--193},
  year={2010},
  publisher={SAGE Publications Sage CA: Los Angeles, CA}
}

@article{gibbons_full-information_1992,
	title = {Full-information item bi-factor analysis},
	volume = {57},
	copyright = {http://www.springer.com/tdm},
	issn = {0033-3123, 1860-0980},
	url = {http://link.springer.com/10.1007/BF02295430},
	doi = {10.1007/BF02295430},
	language = {en},
	number = {3},
	urldate = {2024-12-13},
	journal = {Psychometrika},
	author = {Gibbons, Robert D. and Hedeker, Donald R.},
	month = sep,
	year = {1992},
	pages = {423--436},
}

@article{gibbons_full-information_2007,
	title = {Full-{Information} {Item} {Bifactor} {Analysis} of {Graded} {Response} {Data}},
	volume = {31},
	copyright = {https://journals.sagepub.com/page/policies/text-and-data-mining-license},
	issn = {0146-6216, 1552-3497},
	url = {https://journals.sagepub.com/doi/10.1177/0146621606289485},
	doi = {10.1177/0146621606289485},
	language = {en},
	number = {1},
	urldate = {2024-12-13},
	journal = {Applied Psychological Measurement},
	author = {Gibbons, Robert D. and Bock, R. Darrell and Hedeker, Donald and Weiss, David J. and Segawa, Eisuke and Bhaumik, Dulal K. and Kupfer, David J. and Frank, Ellen and Grochocinski, Victoria J. and Stover, Angela},
	month = jan,
	year = {2007},
	pages = {4--19},
}

@article{cai_generalized_2011,
	title = {Generalized full-information item bifactor analysis.},
	volume = {16},
	issn = {1939-1463, 1082-989X},
	url = {https://doi.apa.org/doi/10.1037/a0023350},
	doi = {10.1037/a0023350},
	language = {en},
	number = {3},
	urldate = {2024-12-13},
	journal = {Psychological Methods},
	author = {Cai, Li and Yang, Ji Seung and Hansen, Mark},
	month = sep,
	year = {2011},
	pages = {221--248},
}

@article{oliviero-durmus_geometric_2024,
	title = {On geometric convergence for the {Metropolis-adjusted Langevin} algorithm under simple conditions},
	volume = {111},
	rights = {https://academic.oup.com/pages/standard-publication-reuse-rights},
	issn = {0006-3444, 1464-3510},
	url = {https://academic.oup.com/biomet/article/111/1/273/7288171},
	doi = {10.1093/biomet/asad060},
	pages = {273--289},
	number = {1},
	journaltitle = {Biometrika},
	author = {Oliviero-Durmus, Alain and Moulines, Éric},
	urldate = {2025-03-10},
	date = {2024-02-12},
	langid = {english},
}
% \input{0_revised_Supplementary_Materials}

% \documentclass[12pt]{article}
% \usepackage{arxiv_0} % 12 pt

% \graphicspath{{./Figures/}}
% % \usepackage{arxiv_1}
% % \linespread{1.2}

% \usepackage{pdflscape}

% \begin{document}

\section*{Supplementary Materials}

\addcontentsline{toc}{subsection}{A.1 Additional Details on the Settings in the Simulation Study}
\subsection*{A.1 Additional Details on the Settings in the Simulation Study}\label{supm:A1}

\addcontentsline{toc}{subsubsection}{A.1.1 True Values in the Simulation Study}
\subsubsection*{A.1.1 True Values in the Simulation Study}\label{supm:A11}

\paragraph{Multilevel Logistic Regression Model.}
In the lower-dimensional setting with $K=5$, the true mean vector of random effects was generated from $\text{Uniform}\maru{0.1, 1.1}$ and set to $$\mu^* = \maru{0.300, 1.060, 0.950, 0.129, 0.826}.$$ In the higher dimensional setting with $K=10$, the true mean vector was generated from $\text{Uniform}\maru{0.1, 1.1}$ and set to $$\mu^* = \maru{0.300, 1.060, 0.950, 0.129, 0.826, 0.857, 0.193, 0.809, 0.844, 0.301}.$$ These values are rounded off to three decimal places. In both settings, the true values for the diagonal and off-diagonal elements of the covariance matrix $\bSigma$ were set to 0.1 and 0.05, respectively.

\paragraph{M2PL Model.}
The number of intercept and factor loading parameters is too large to present the specific values in the Supplementary Materials; we have provided them on the Open Science Framework: \url{https://osf.io/3sb4t/?view_only=abd84347053a450fa88f787d168df359}. In both the lower- and higher-dimensional settings, the true values for the diagonal and off-diagonal elements of correlation matrix $\bSigma$ were set to 1.0 and 0.5, respectively.

\addcontentsline{toc}{subsubsection}{A.1.2 Initial Values}
\subsubsection*{A.1.2 Initial Values}\label{supm:A12}
\paragraph{Multilevel Logistic Regression Model.}

The initial values of latent variables were generated from the standard multivariate normal distribution $\calN\maru{\mathbf{0}_{K}, 0.5\btI_K }$, the sign of these generated values were set to the same as that of the corresponding true values. The initial values of the mean vector $\bmu$ were generated from $\text{Uniform}\maru{0, 1.5}$.  
Regarding the initial value of the covariance matrix $\bSigma$, the identity matrix was assigned.

\paragraph{M2PL Model.}
Similarly, the initial values of latent variables were generated from the standard multivariate normal distribution $\calN\maru{\mathbf{0}_{K}, 5\btI_K }$, the sign of these generated values were set to the same as that of the corresponding true values. The initial values of intercept $d_j$ were generated from the standard normal distribution, and those of factor loading $\ba_j$ were generated from $\text{Uniform}\maru{0, 2}$.
In order to align with the multilevel logistic regression model, we also set the initial value of the correlation matrix $\bSigma$ to be the identity matrix.

\addcontentsline{toc}{subsubsection}{A.1.3 Candidate Values of the Constant Factors for the Step Sizes for the MALA Sampling and the Random-walk Step Size for the Random-walk MH Sampling}
\subsubsection*{A.1.3 Candidate Values of the Constant Factors for the Step Sizes for the MALA Sampling and the Random-walk Step Size for the Random-walk MH Sampling}\label{supm:A13}

Due to the large number of simulation conditions, we only tuned the step sizes of the MALA sampler in the QN-SOMALA and D-SOMALA with $n=250$ and the random-walk step size of the random-walk MH sampler the QN-SOMH and D-SOMH with $n=250$. The other algorithms with the minibatch SG and MALA sampler used the step size that was tuned under the D-SOMALA with $n=250$. Those with the minibatch SG and random-walk MH sampler also used the step size that was tuned under the D-SOMH with $n=250$. For the step size in a model parameter update, we assigned $\mu_1 = 1.0$ for all the algorithms in consideration.

Regarding the step size of the MALA step, we consider the following five candidate values: $\mu_2 = \nami{0.01, 0.05, 0.1, 0.2}$. With respect to the random-walk MH sampling, the random-walk step size corresponds to the standard deviation of Gaussian noise $\epsilon$ with normal distribution $\calN(0, \sigma^2)$, and the SOMH and D-SOMH algorithm generates a proposed latent variable by $\tilde{\xi}_{ik} = \xi_{ik}^\iter{t-1} + \epsilon$, where $\tilde{\xi}_{ik}$ is the $k$-th dimension of the proposed value of latent variables, and $\xi_{ik}^\iter{t-1}$ is the $k$-th dimension of the value of latent variables sampled in the previous iteration. Accordingly, we consider the random-walk step size candidate values: $\sigma^2 = \nami{0.1, 0.2, 0.3, 0.4}$.

Next, we ran all the algorithms for one thousand epochs in total given different candidate values under each simulation condition for one simulated dataset. 
% Then, we computed the absolute error of the last updated value for each of model parameters and averaged the absolute error across them. Subsequently, the candidate value that provided the smallest value of the averaged absolute error was used as the hyperparameter value in the simulation study. 
Then, for each candidate value of the hyperparameter, we computed the values of the negative complete-data log-likelihood function using the updated model parameters and sample latent variables from the last fifty epochs. 
Subsequently, we averaged these values and chose the candidate value that provided the smallest averaged value of the negative complete-data log-likelihood function as the hyperparameter value in the simulation study.

% Table \ref{tab:stepsize} shows the step sizes chosen from the candidate values. 

% \begin{table}[htbp]
%   \centering
%   \caption{Step sizes in the simulation study}
%     \begin{tabular}{cccc}
%     \toprule
%       &  &  &  \\
%     \bottomrule
%     \end{tabular}%
%     \begin{tablenotes}
%  \item \footnotesize{\textit{Note.} In the column of $h$, $\sigma^2$, the value presented in the rows of ‘‘SOMH'' and ‘D-SOMH'' is the random-walk step size $\sigma^2$. }
%  \end{tablenotes}
%   \label{tab:stepsize}%
% \end{table}%

\clearpage
\addcontentsline{toc}{subsubsection}{A.1.4 Q-matrices Used in the Simulation Study}
\subsubsection*{A.1.4 Q-matrices Used in the Simulation Study}\label{supm:A14}
\begin{figure}[htbp]
\color{black}
 \begin{center}
 %\begin{tabular}{c}
 \centering
 \includegraphics[width=\linewidth]{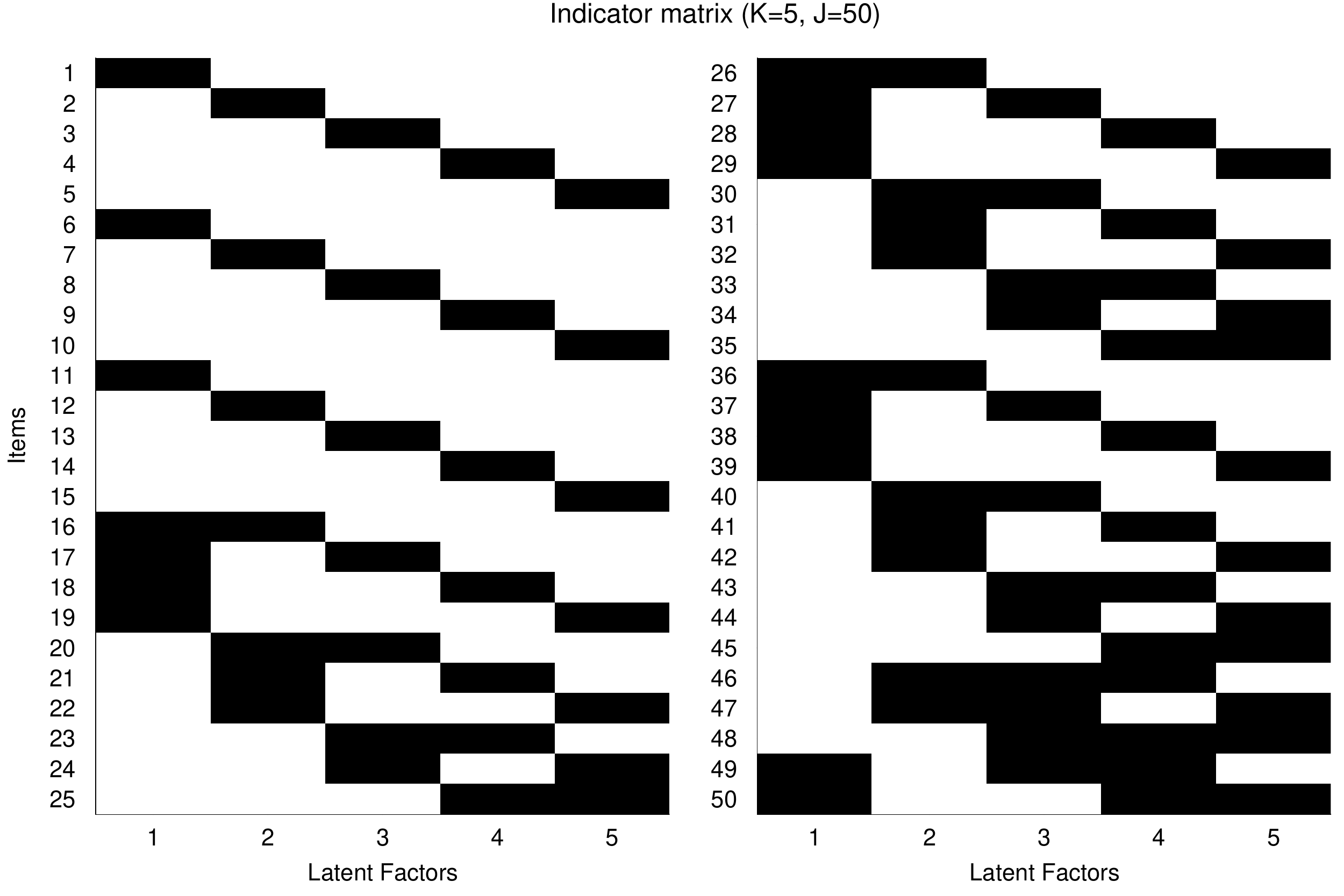}
\caption{The indicator matrix with $K=5$ and $J=50$. The black boxes denote the entry of 1, and the white boxes denote the entry of 0.}
\label{fig:QK5}
 \end{center}
\end{figure}

\begin{figure}[htbp]
\color{black}
 \begin{center}
 %\begin{tabular}{c}
 \centering
 \includegraphics[width=\linewidth]{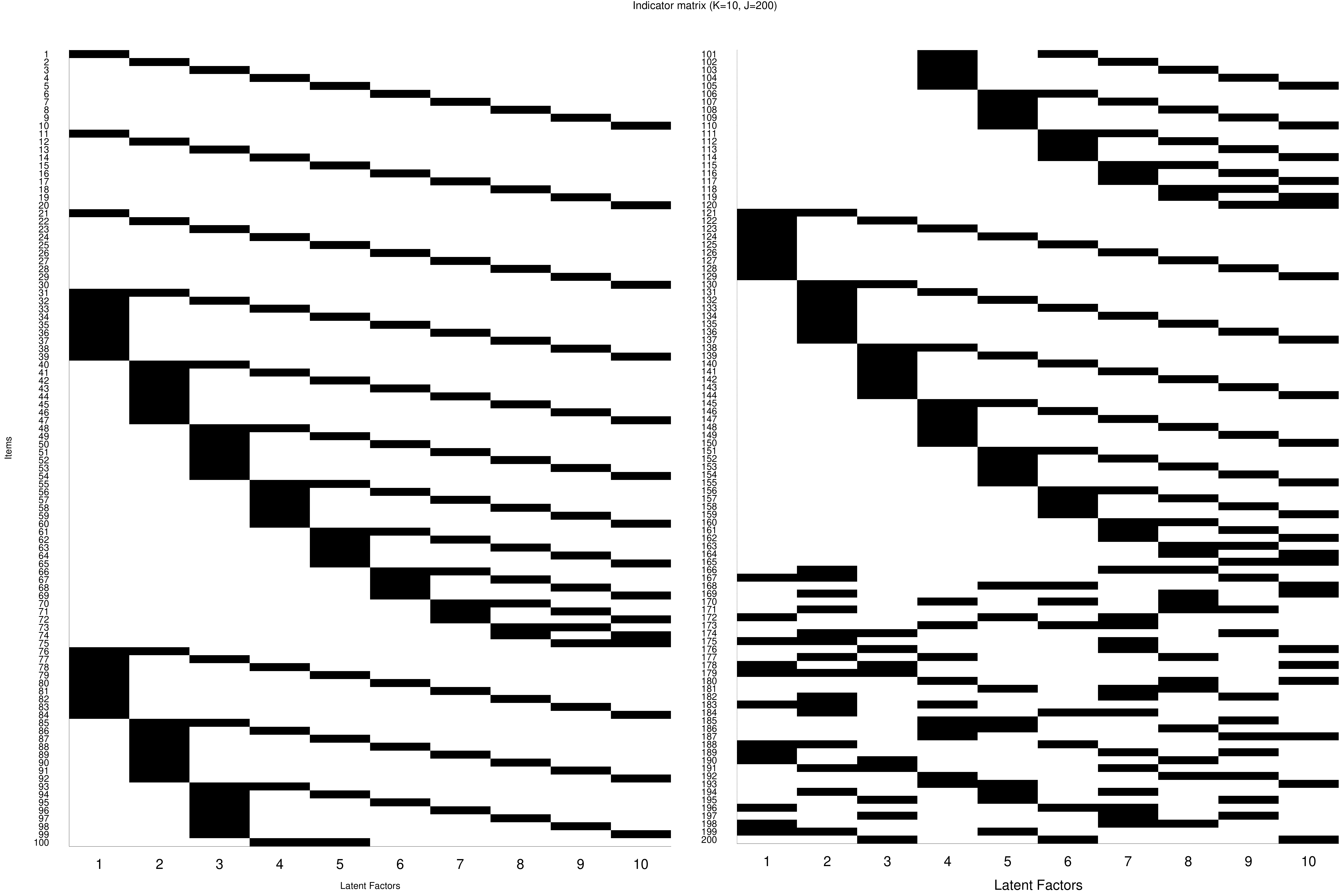}
\caption{The indicator matrix with $K=10$ and $J=200$. The black boxes denote the entry of 1, and the white boxes denote the entry of 0.}
\label{fig:QK10}
 \end{center}
\end{figure}
\clearpage

\addcontentsline{toc}{subsection}{A.2 The Trajectory of the MAE for Different Model Parameters in the Simulation Study}
\subsection*{A.2 The Trajectory of the MAE for Different Model Parameters in the Simulation Study}\label{supm:A3}
Here, we show the trajectories of the averaged MAEs to compare the algorithms that rely on different MCMC samplers while keeping the other specifications fixed (i.e., minibatch size and the use of a quasi-Newton update).

\begin{figure}[htbp]
  \centering
    \includegraphics[width=\linewidth]{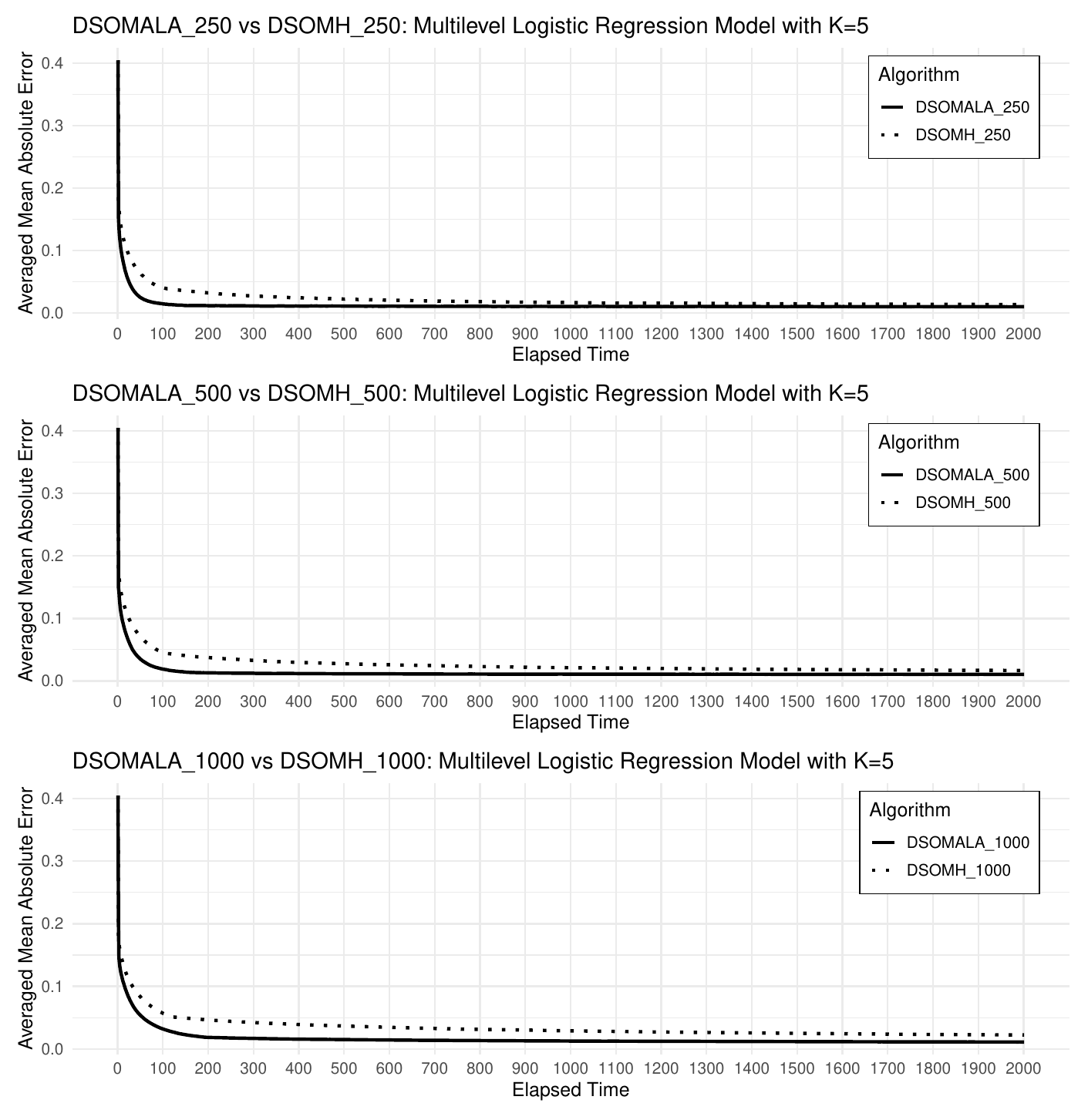}
    \caption{The trajectories of averaged MAEs for the multilevel logistic regression model. The legends indicate the type of algorithm. The number after the underbar ‘‘\_'' represents the minibatch size.}
\end{figure}

\begin{figure}[htbp]
  \centering
    \includegraphics[width=\linewidth]{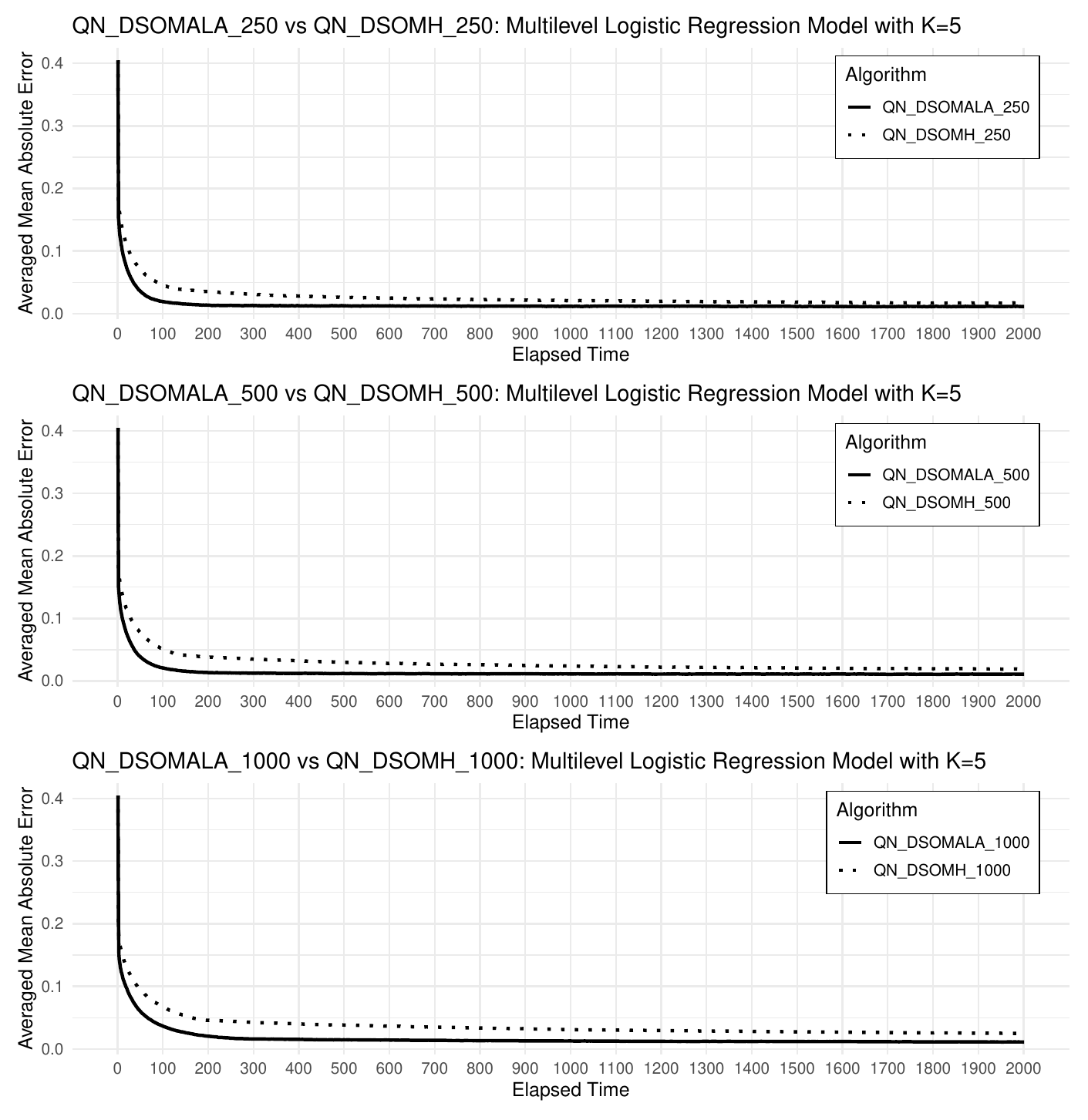}
    \caption{The trajectories of averaged MAEs for the multilevel logistic regression model. The legends indicate the type of algorithm. The number after the underbar ‘‘\_'' represents the minibatch size.}
\end{figure}

\begin{figure}[htbp]
  \centering
    \includegraphics[width=\linewidth]{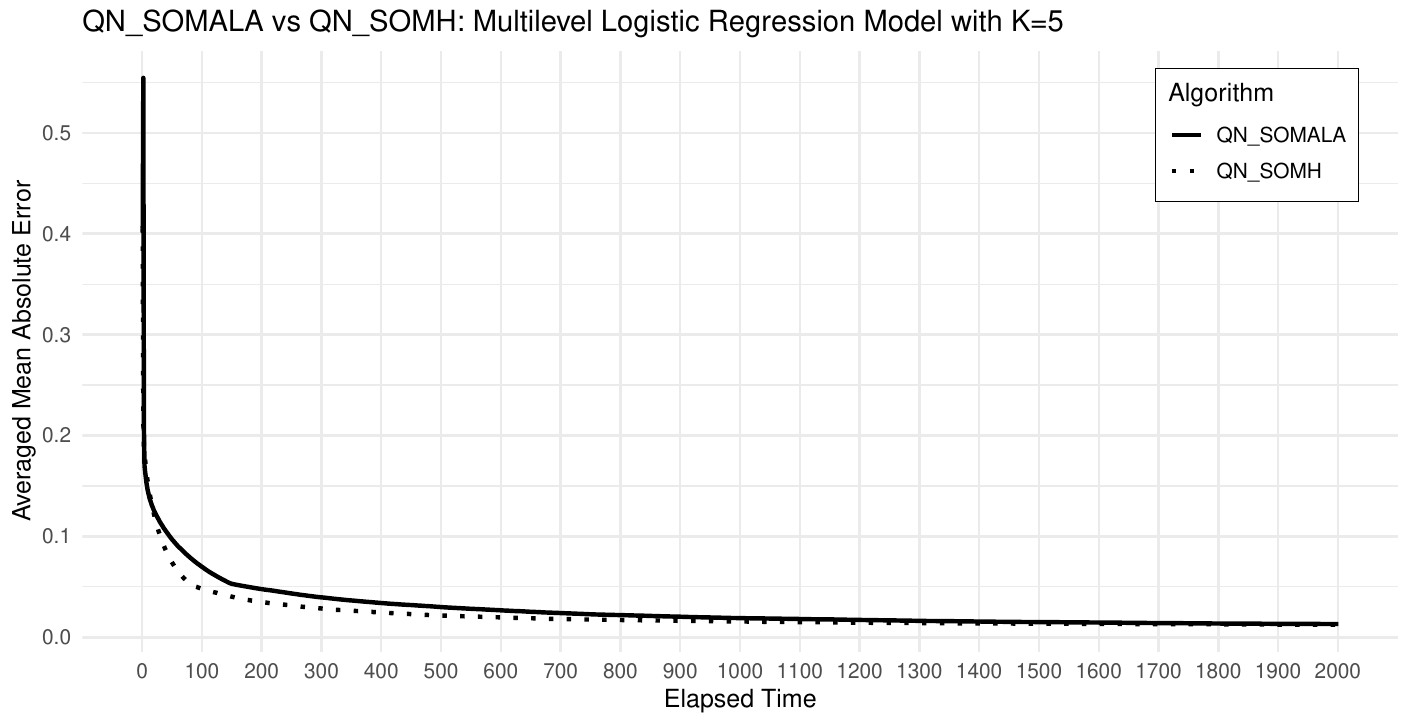}
    \caption{The trajectories of averaged MAEs for the multilevel logistic regression model.}
\end{figure}

\begin{figure}[htbp]
  \centering
    \includegraphics[width=\linewidth]{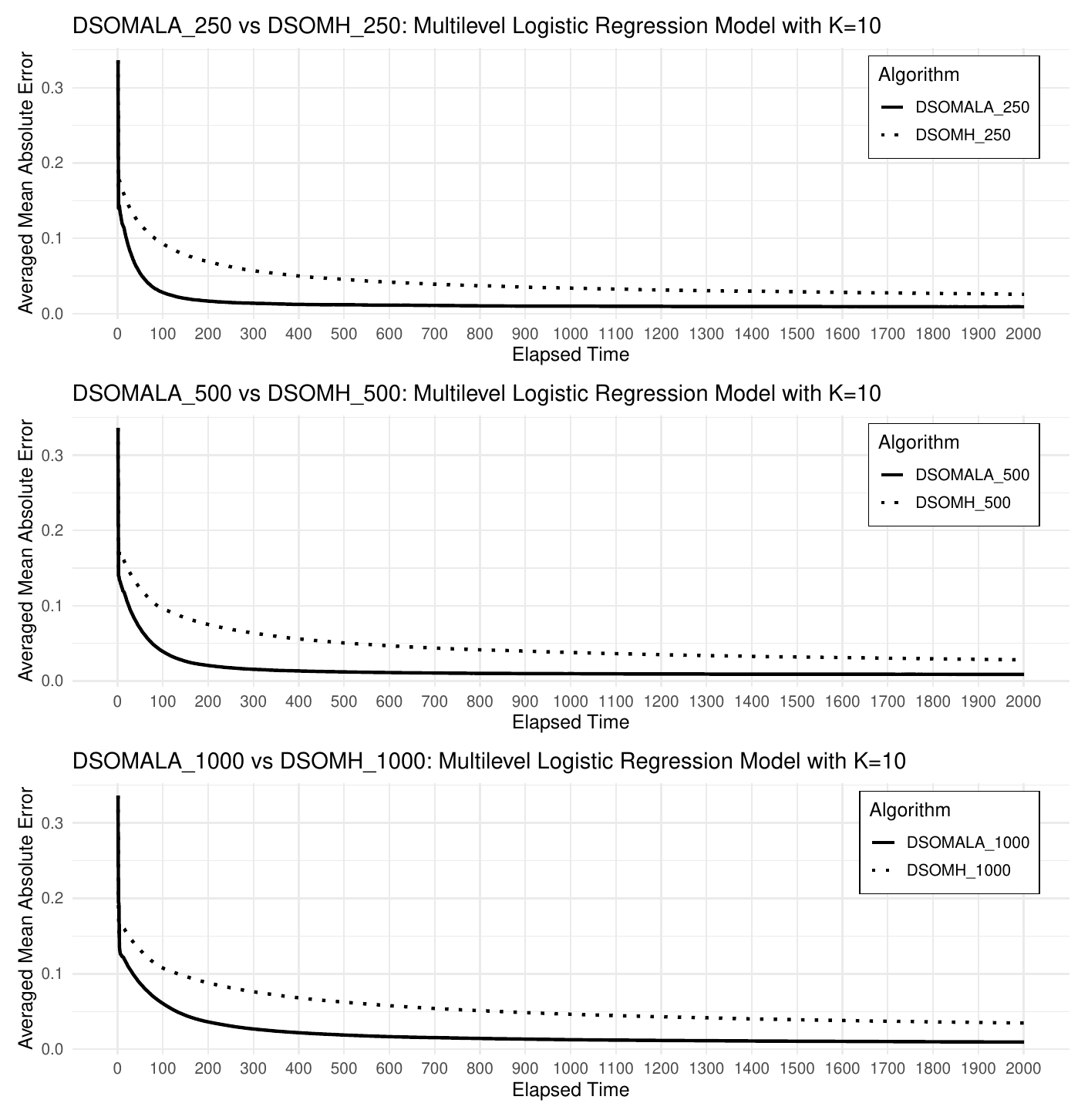}
    \caption{The trajectories of averaged MAEs for the multilevel logistic regression model. The legends indicate the type of algorithm. The number after the underbar ‘‘\_'' represents the minibatch size.}
\end{figure}

\begin{figure}[htbp]
  \centering
    \includegraphics[width=\linewidth]{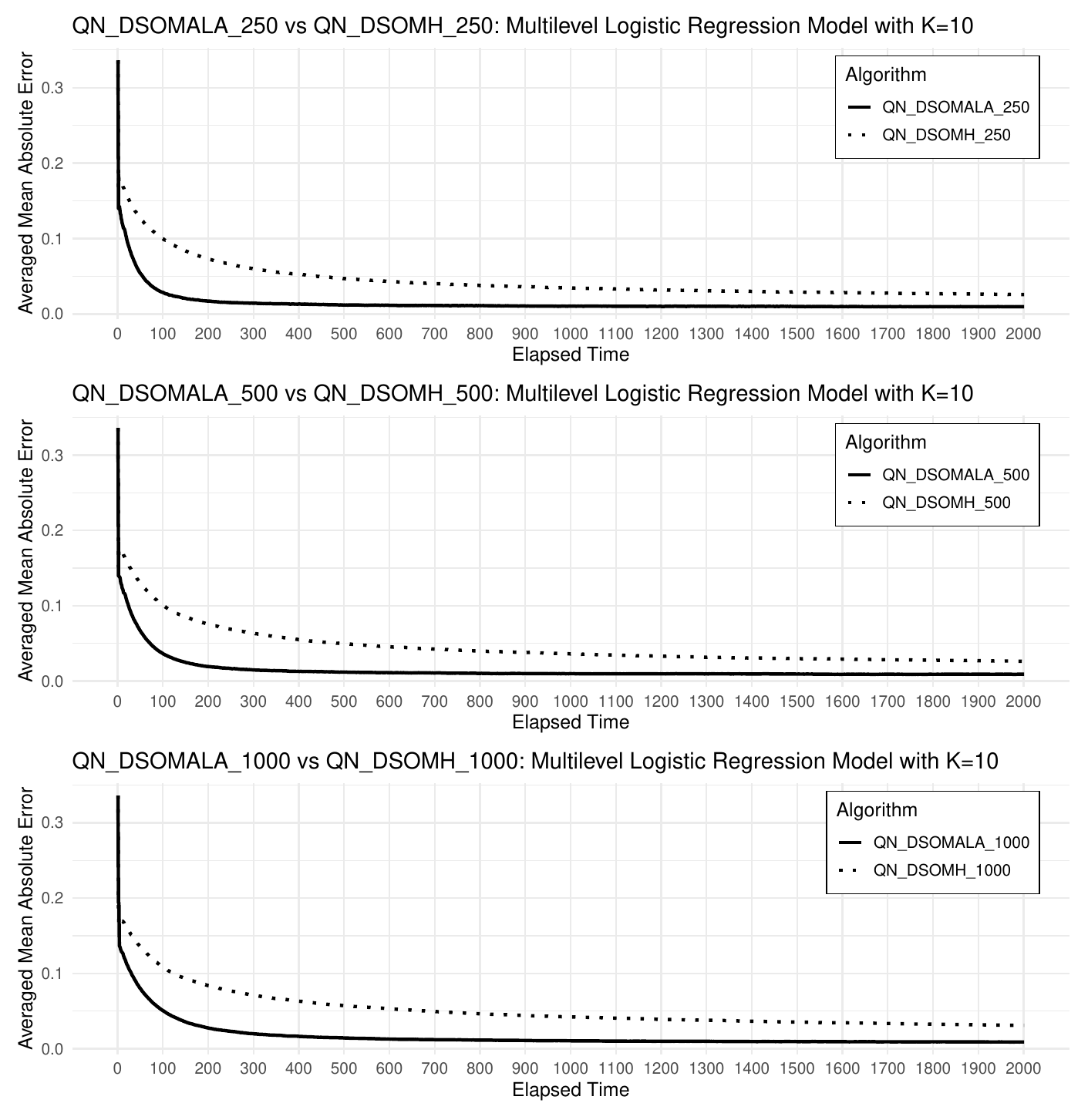}
    \caption{The trajectories of averaged MAEs for the multilevel logistic regression model. The legends indicate the type of algorithm. The number after the underbar ‘‘\_'' represents the minibatch size.}
\end{figure}

\begin{figure}[htbp]
  \centering
    \includegraphics[width=\linewidth]{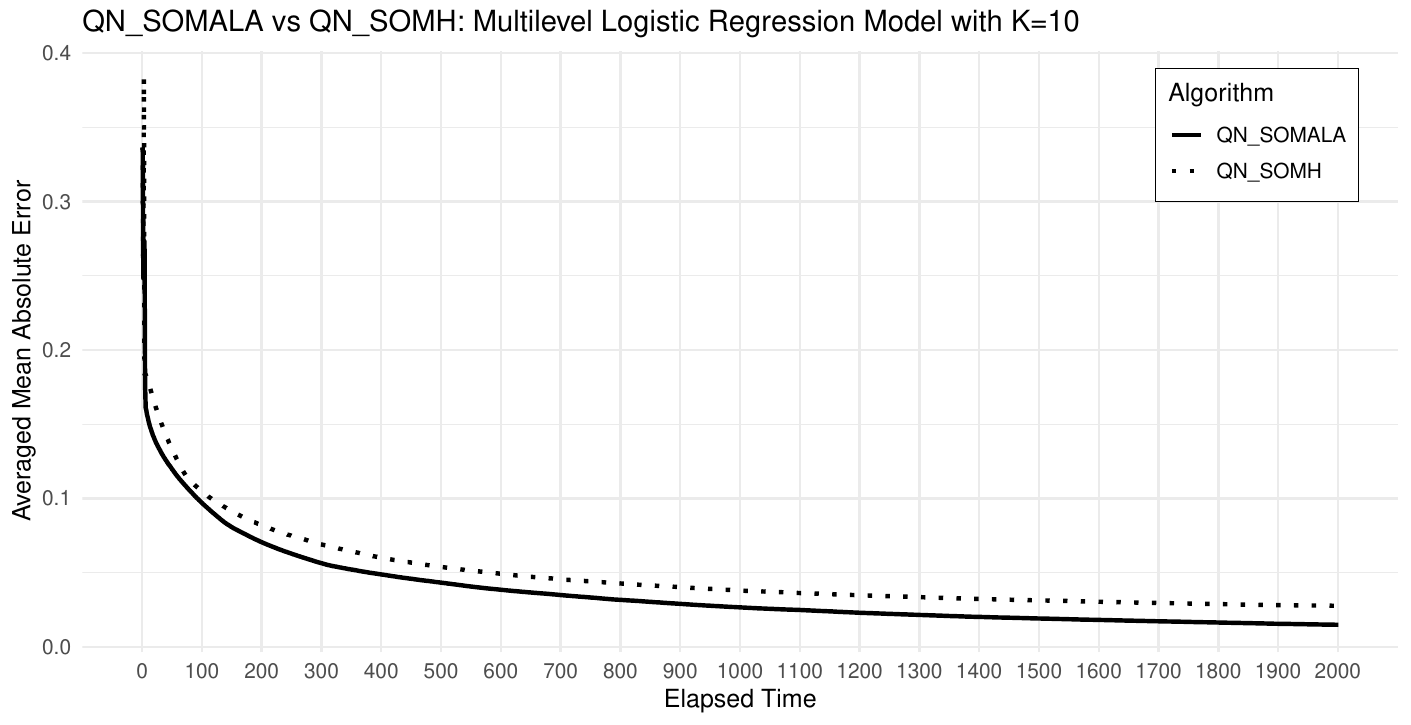}
    \caption{The trajectories of averaged MAEs for the multilevel logistic regression model.}
\end{figure}

\begin{figure}[htbp]
  \centering
    \includegraphics[width=\linewidth]{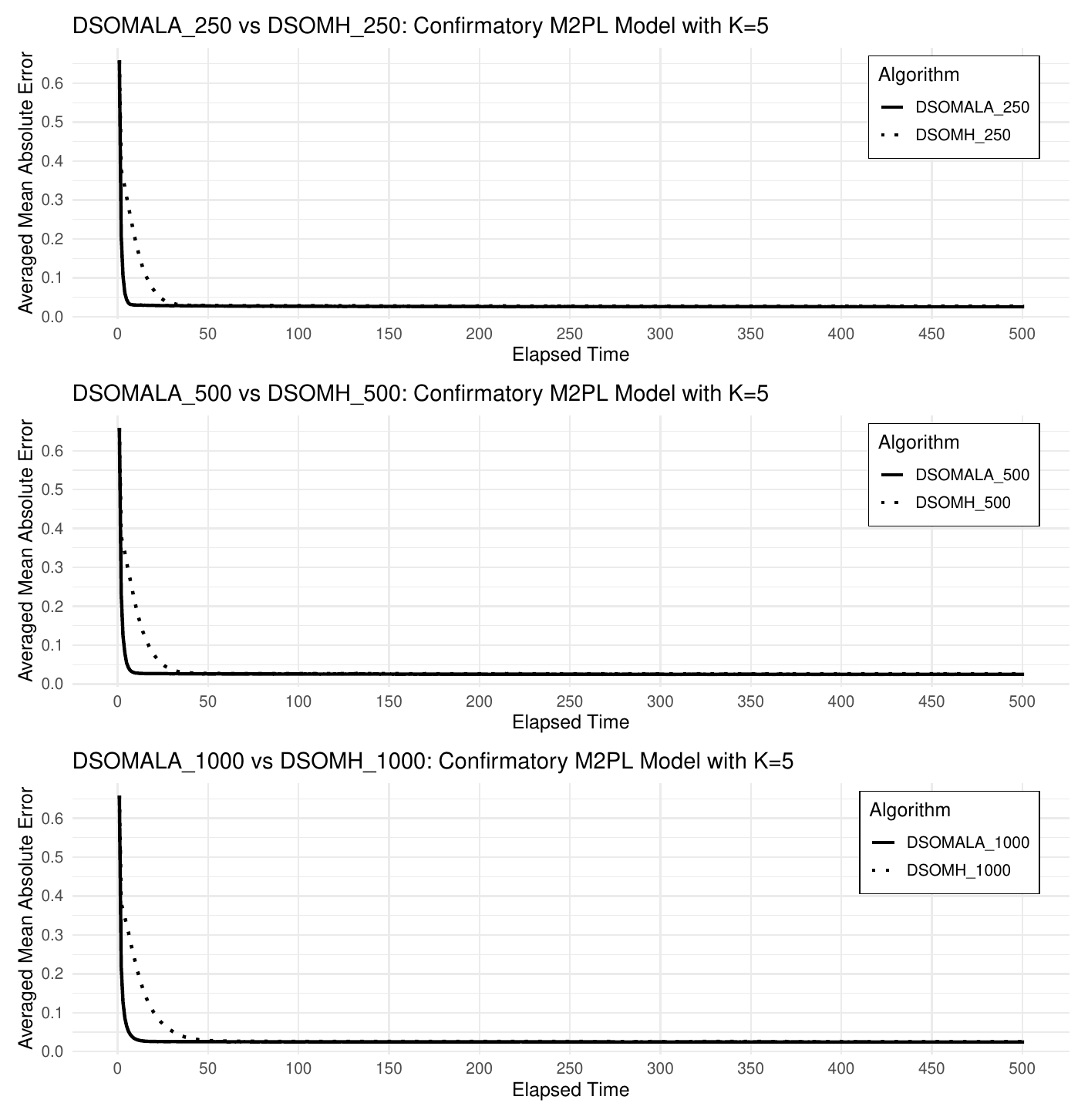}
    \caption{The trajectories of averaged MAEs for the confirmatory M2PL model. The legends indicate the type of algorithm. The number after the underbar ‘‘\_'' represents the minibatch size.}
\end{figure}

\begin{figure}[htbp]
  \centering
    \includegraphics[width=\linewidth]{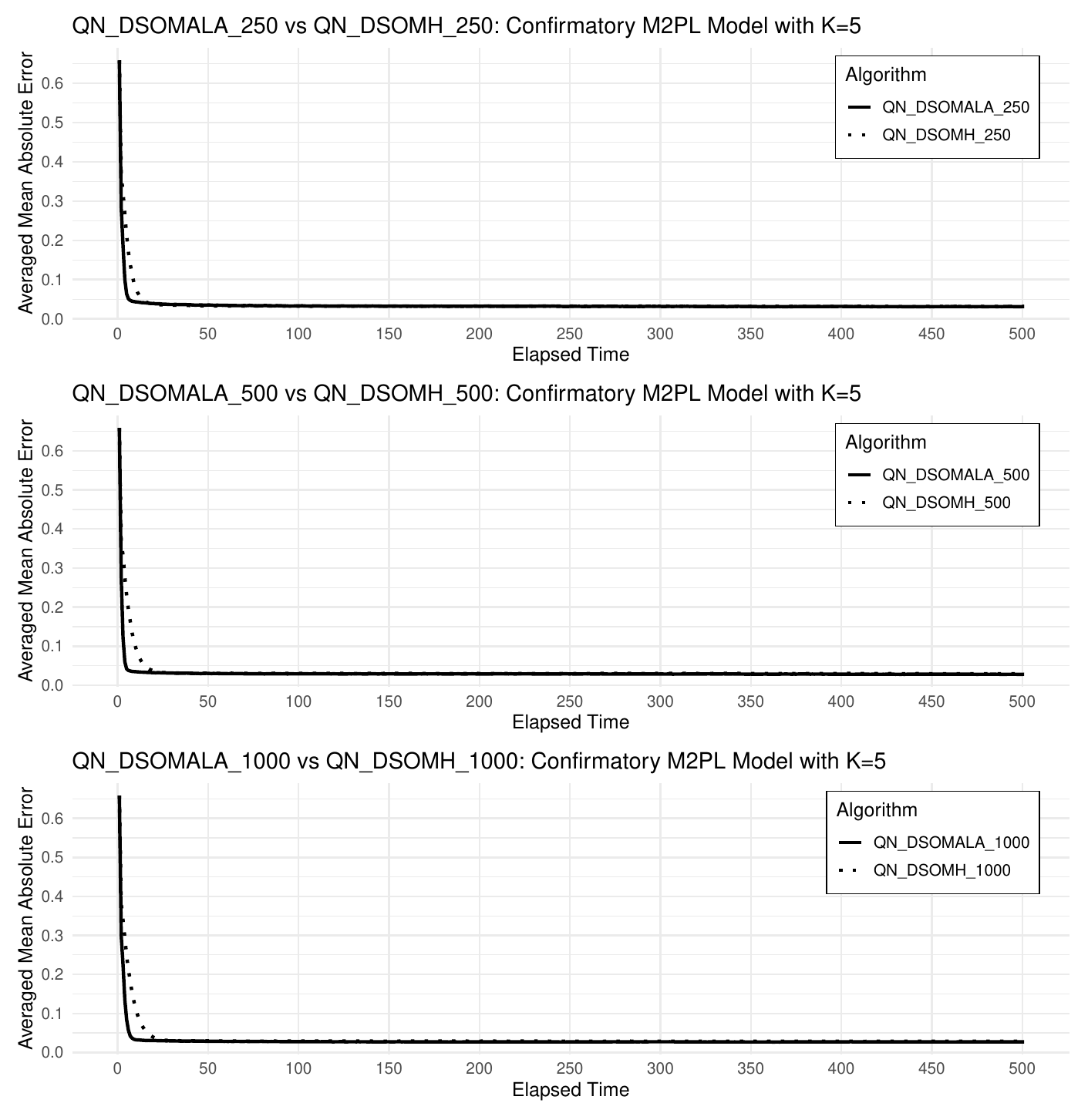}
    \caption{The trajectories of averaged MAEs for the confirmatory M2PL model. The legends indicate the type of algorithm. The number after the underbar ‘‘\_'' represents the minibatch size.}
\end{figure}

\begin{figure}[htbp]
  \centering
    \includegraphics[width=\linewidth]{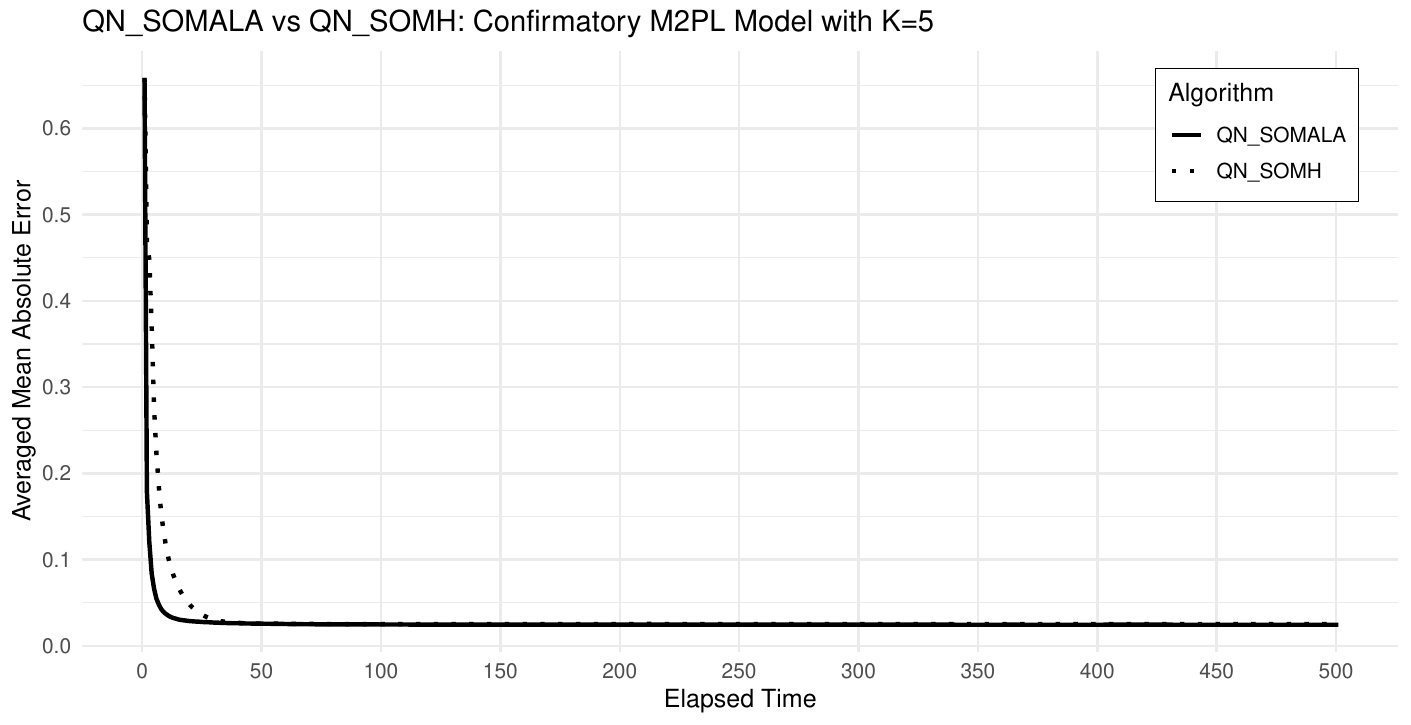}
    \caption{The trajectories of averaged MAEs for the confirmatory M2PL model.}
\end{figure}

\begin{figure}[htbp]
  \centering
    \includegraphics[width=\linewidth]{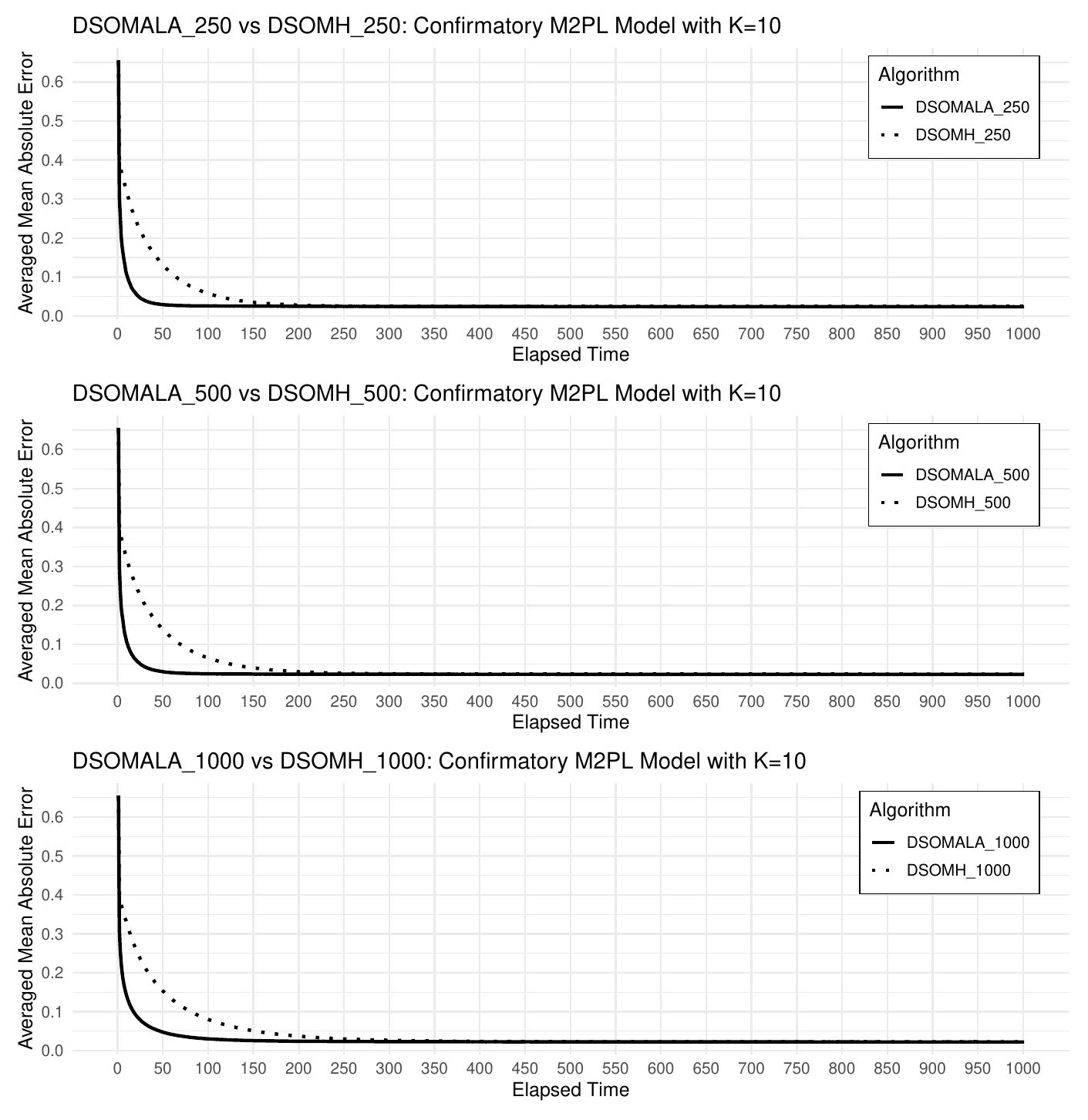}
    \caption{The trajectories of averaged MAEs for the confirmatory M2PL model. The legends indicate the type of algorithm. The number after the underbar ‘‘\_'' represents the minibatch size.}
\end{figure}

\begin{figure}[htbp]
  \centering
    \includegraphics[width=\linewidth]{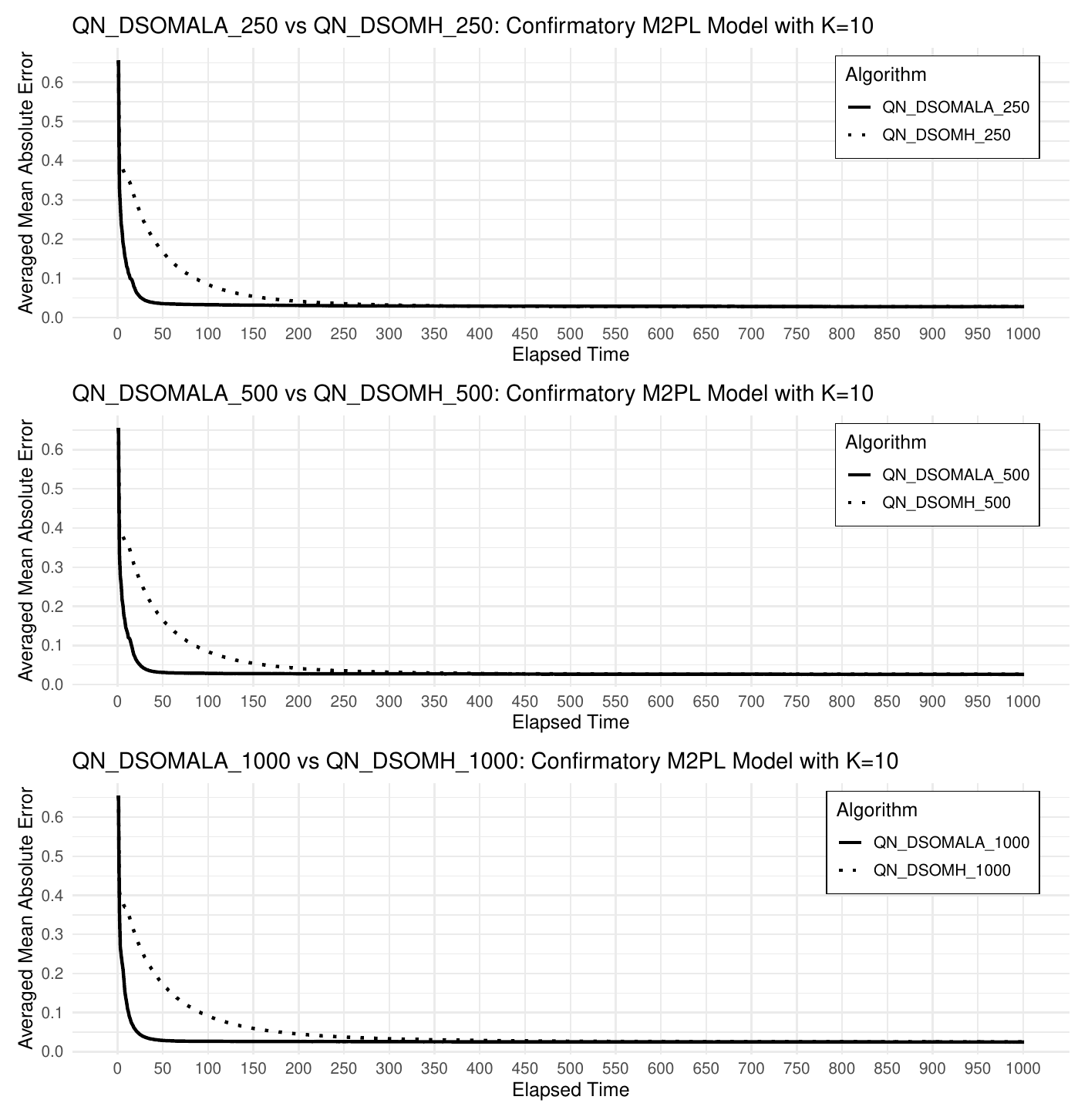}
    \caption{The trajectories of averaged MAEs for the confirmatory M2PL model. The legends indicate the type of algorithm. The number after the underbar ‘‘\_'' represents the minibatch size.}
\end{figure}

\begin{figure}[htbp]
  \centering
    \includegraphics[width=\linewidth]{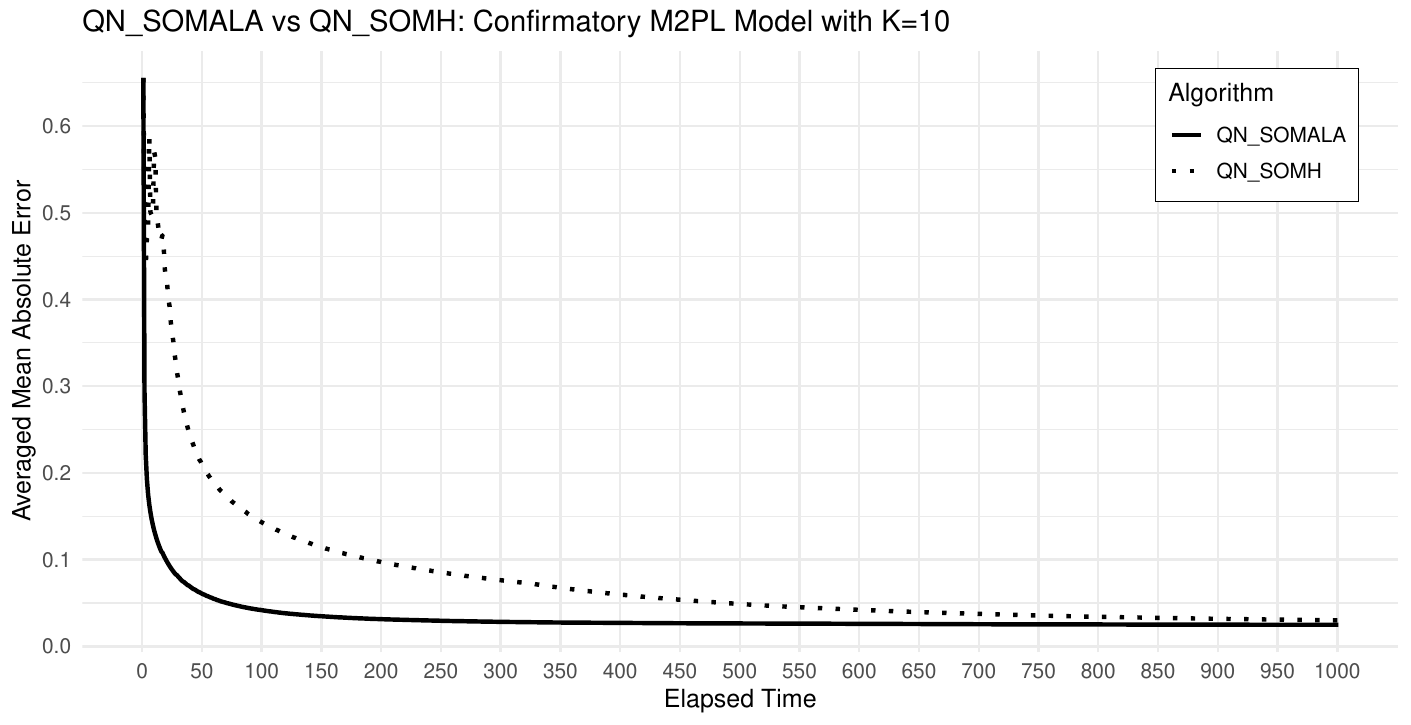}
    \caption{The trajectories of averaged MAEs for the confirmatory M2PL model.}
\end{figure}

\clearpage
Next, we present the trajectories of the MAEs of each of the model parameters for the multilevel logistic regression and confirmatory M2PL models. 

\begin{landscape}

% Table generated by Excel2LaTeX from sheet 'mlogreg_K5'
\begin{table}[htbp]
   \centering
  \caption{The trajectory of the MAEs of the model parameters for the multilevel logistic regression model with $K=5, J=10, N=10,000$}
  \resizebox{\linewidth}{!}{%
    \begin{tabular}{cccccccccccccccccc}
    \toprule
    \multirow{2}[4]{*}{Parameters} & \multicolumn{2}{c}{\multirow{2}[4]{*}{Methods}} & \multicolumn{15}{c}{Elapsed Time (seconds)} \\
\cmidrule{4-18}          & \multicolumn{2}{c}{} & 0     & 40    & 80    & 120   & 160   & 200   & 400   & 600   & 800   & 1000  & 1200  & 1400  & 1600  & 1800  & 2000 \\
    \midrule
    \multirow{14}[2]{*}{\begin{tabular}{c}  Mean\\Structure \end{tabular}} & \multicolumn{1}{l}{D-SOMALA} & \multicolumn{1}{l}{$n=250$} & .4766 & .0325 & .0166 & .0131 & .0119 & .0115 & .0109 & .0107 & .0106 & .0104 & .0101 & .0102 & .0103 & .0102 & .0101 \\
          & \multicolumn{1}{l}{D-SOMALA} & \multicolumn{1}{l}{$n=500$} & .4766 & .0452 & .0233 & .0162 & .0134 & .0127 & .0117 & .0112 & .0107 & .0107 & .0106 & .0103 & .0101 & .0102 & .0102 \\
          & \multicolumn{1}{l}{D-SOMALA} & \multicolumn{1}{l}{$n=1,000$} & .4766 & .0686 & .0415 & .0292 & .0224 & .0191 & .0159 & .0147 & .0137 & .0129 & .0125 & .0122 & .0118 & .0116 & .0113 \\
          \cdashline{2-18} \addlinespace[2pt] 
          & \multicolumn{1}{l}{D-SOMH} & \multicolumn{1}{l}{$n=250$} & .4766 & .0773 & .0484 & .0393 & .0361 & .0330 & .0246 & .0204 & .0182 & .0166 & .0155 & .0151 & .0140 & .0138 & .0132 \\
          & \multicolumn{1}{l}{D-SOMH} & \multicolumn{1}{l}{$n=500$} & .4766 & .0860 & .0555 & .0451 & .0412 & .0388 & .0303 & .0261 & .0232 & .0209 & .0200 & .0185 & .0179 & .0170 & .0163 \\
          & \multicolumn{1}{l}{D-SOMH} & \multicolumn{1}{l}{$n=1,000$} & .4766 & .1008 & .0708 & .0551 & .0523 & .0496 & .0415 & .0364 & .0327 & .0301 & .0279 & .0264 & .0248 & .0234 & .0225 \\
          \cdashline{2-18} \addlinespace[2pt] 
          & \multicolumn{1}{l}{QN-D-SOMALA} & \multicolumn{1}{l}{$n=250$} & .4766 & .0458 & .0238 & .0173 & .0153 & .0139 & .0127 & .0129 & .0117 & .0121 & .0119 & .0123 & .0119 & .0115 & .0113 \\
          & \multicolumn{1}{l}{QN-D-SOMALA} & \multicolumn{1}{l}{$n=500$} & .4766 & .0491 & .0254 & .0186 & .0145 & .0134 & .0124 & .0122 & .0116 & .0115 & .0113 & .0115 & .0114 & .0106 & .0108 \\
          & \multicolumn{1}{l}{QN-D-SOMALA} & \multicolumn{1}{l}{$n=1,000$} & .4766 & .0734 & .0450 & .0322 & .0252 & .0203 & .0154 & .0144 & .0132 & .0130 & .0124 & .0122 & .0123 & .0113 & .0115 \\
          \cdashline{2-18} \addlinespace[2pt] 
          & \multicolumn{1}{l}{QN-D-SOMH} & \multicolumn{1}{l}{$n=250$} & .4766 & .0842 & .0550 & .0415 & .0371 & .0361 & .0282 & .0249 & .0225 & .0209 & .0197 & .0188 & .0179 & .0166 & .0169 \\
          & \multicolumn{1}{l}{QN-D-SOMH} & \multicolumn{1}{l}{$n=500$} & .4766 & .0909 & .0608 & .0464 & .0418 & .0392 & .0328 & .0286 & .0264 & .0242 & .0221 & .0213 & .0203 & .0194 & .0189 \\
          & \multicolumn{1}{l}{QN-D-SOMH} & \multicolumn{1}{l}{$n=1,000$} & .4766 & .1099 & .0792 & .0635 & .0521 & .0475 & .0415 & .0374 & .0345 & .0313 & .0298 & .0285 & .0272 & .0259 & .0251 \\
          \cdashline{2-18} \addlinespace[2pt] 
          & \multicolumn{2}{c}{QN-SOMALA} & .4766 & .1146 & .0867 & .0679 & .0562 & .0516 & .0361 & .0277 & .0224 & .0192 & .0174 & .0155 & .0145 & .0135 & .0131 \\
          & \multicolumn{2}{c}{QN-SOMH} & .4766 & .0915 & .0559 & .0476 & .0412 & .0369 & .0253 & .0198 & .0171 & .0156 & .0143 & .0137 & .0131 & .0127 & .0122 \\
    \midrule
    \multirow{14}[2]{*}{\begin{tabular}{c}  Covariance\\Matrix \end{tabular}} & \multicolumn{1}{l}{D-SOMALA} & \multicolumn{1}{l}{$n=250$} & .3333 & .0296 & .0173 & .0140 & .0127 & .0124 & .0119 & .0116 & .0113 & .0111 & .0110 & .0109 & .0108 & .0106 & .0107 \\
          & \multicolumn{1}{l}{D-SOMALA} & \multicolumn{1}{l}{$n=500$} & .3333 & .0386 & .0217 & .0161 & .0136 & .0131 & .0117 & .0116 & .0111 & .0109 & .0110 & .0109 & .0108 & .0109 & .0107 \\
          & \multicolumn{1}{l}{D-SOMALA} & \multicolumn{1}{l}{$n=1,000$} & .3333 & .0542 & .0344 & .0253 & .0204 & .0177 & .0153 & .0139 & .0130 & .0124 & .0121 & .0116 & .0112 & .0110 & .0108 \\
          \cdashline{2-18} \addlinespace[2pt] 
          & \multicolumn{1}{l}{D-SOMH} & \multicolumn{1}{l}{$n=250$} & .3333 & .0671 & .0438 & .0373 & .0343 & .0319 & .0245 & .0210 & .0187 & .0172 & .0163 & .0156 & .0149 & .0146 & .0142 \\
          & \multicolumn{1}{l}{D-SOMH} & \multicolumn{1}{l}{$n=500$} & .3333 & .0731 & .0493 & .0408 & .0379 & .0357 & .0290 & .0254 & .0227 & .0211 & .0198 & .0187 & .0180 & .0174 & .0168 \\
          & \multicolumn{1}{l}{D-SOMH} & \multicolumn{1}{l}{$n=1,000$} & .3333 & .0823 & .0592 & .0474 & .0450 & .0429 & .0365 & .0325 & .0299 & .0277 & .0261 & .0248 & .0238 & .0228 & .0221 \\
          \cdashline{2-18} \addlinespace[2pt] 
          & \multicolumn{1}{l}{QN-D-SOMALA} & \multicolumn{1}{l}{$n=250$} & .3333 & .0408 & .0216 & .0165 & .0141 & .0129 & .0121 & .0120 & .0114 & .0112 & .0115 & .0111 & .0108 & .0108 & .0109 \\
          & \multicolumn{1}{l}{QN-D-SOMALA} & \multicolumn{1}{l}{$n=500$} & .3333 & .0430 & .0241 & .0179 & .0148 & .0133 & .0118 & .0115 & .0113 & .0111 & .0108 & .0110 & .0110 & .0109 & .0108 \\
          & \multicolumn{1}{l}{QN-D-SOMALA} & \multicolumn{1}{l}{$n=1,000$} & .3333 & .0628 & .0404 & .0299 & .0239 & .0201 & .0151 & .0139 & .0130 & .0125 & .0120 & .0117 & .0114 & .0110 & .0108 \\
          \cdashline{2-18} \addlinespace[2pt] 
          & \multicolumn{1}{l}{QN-D-SOMH} & \multicolumn{1}{l}{$n=250$} & .3333 & .0744 & .0500 & .0390 & .0362 & .0342 & .0276 & .0242 & .0223 & .0205 & .0195 & .0186 & .0176 & .0169 & .0166 \\
          & \multicolumn{1}{l}{QN-D-SOMH} & \multicolumn{1}{l}{$n=500$} & .3333 & .0803 & .0558 & .0433 & .0392 & .0373 & .0316 & .0278 & .0257 & .0237 & .0221 & .0212 & .0204 & .0198 & .0190 \\
          & \multicolumn{1}{l}{QN-D-SOMH} & \multicolumn{1}{l}{$n=1,000$} & .3333 & .0956 & .0704 & .0569 & .0481 & .0440 & .0386 & .0351 & .0326 & .0304 & .0289 & .0277 & .0264 & .0254 & .0246 \\
          \cdashline{2-18} \addlinespace[2pt] 
          & \multicolumn{2}{c}{QN-SOMALA} & .3333 & .0939 & .0706 & .0559 & .0473 & .0438 & .0319 & .0257 & .0216 & .0189 & .0171 & .0158 & .0148 & .0139 & .0133 \\
          & \multicolumn{2}{c}{QN-SOMH} & .3333 & .0773 & .0485 & .0416 & .0365 & .0330 & .0237 & .0197 & .0172 & .0159 & .0148 & .0140 & .0136 & .0132 & .0127 \\
    \bottomrule
    \end{tabular}%
    }
  \label{tab:addlabel}%
\end{table}%

\end{landscape}

\begin{landscape}

% Table generated by Excel2LaTeX from sheet 'mlogreg_K10'
\begin{table}[htbp]
  \centering
  \caption{The trajectory of the MAEs of the model parameters for the multilevel logistic regression model with $K=10, J=20, N=10,000$}
    \resizebox{\linewidth}{!}{%
    \begin{tabular}{cccccccccccccccccc}
    \toprule
    \multirow{2}[4]{*}{Parameters} & \multicolumn{2}{c}{\multirow{2}[4]{*}{Methods}} & \multicolumn{15}{c}{Elapsed Time (seconds)} \\
\cmidrule{4-18}          & \multicolumn{2}{c}{} & 0     & 40    & 80    & 120   & 160   & 200   & 400   & 600   & 800   & 1000  & 1200  & 1400  & 1600  & 1800  & 2000 \\
    \midrule
    \multirow{14}[2]{*}{\begin{tabular}{c}  Mean\\Structure \end{tabular}} & \multicolumn{1}{l}{D-SOMALA} & \multicolumn{1}{l}{$n=250$} & .4681 & .0881 & .0470 & .0325 & .0252 & .0215 & .0148 & .0132 & .0118 & .0112 & .0108 & .0104 & .0101 & .0100 & .0098 \\
          & \multicolumn{1}{l}{D-SOMALA} & \multicolumn{1}{l}{$n=500$} & .4681 & .1128 & .0683 & .0460 & .0340 & .0280 & .0169 & .0136 & .0118 & .0110 & .0104 & .0099 & .0099 & .0098 & .0095 \\
          & \multicolumn{1}{l}{D-SOMALA} & \multicolumn{1}{l}{$n=1,000$} & .4681 & .1368 & .1015 & .0781 & .0625 & .0528 & .0309 & .0227 & .0190 & .0164 & .0148 & .0136 & .0128 & .0119 & .0115 \\
          & \multicolumn{1}{l}{D-SOMH} & \multicolumn{1}{l}{$n=250$} & .4681 & .1762 & .1420 & .1211 & .1072 & .0977 & .0704 & .0586 & .0518 & .0471 & .0433 & .0410 & .0385 & .0365 & .0347 \\
          & \multicolumn{1}{l}{D-SOMH} & \multicolumn{1}{l}{$n=500$} & .4681 & .1834 & .1467 & .1290 & .1165 & .1071 & .0795 & .0662 & .0585 & .0533 & .0489 & .0457 & .0433 & .0408 & .0390 \\
          & \multicolumn{1}{l}{D-SOMH} & \multicolumn{1}{l}{$n=1,000$} & .4681 & .1937 & .1637 & .1475 & .1362 & .1266 & .0981 & .0832 & .0736 & .0666 & .0615 & .0573 & .0542 & .0515 & .0492 \\
          & \multicolumn{1}{l}{QN-D-SOMALA} & \multicolumn{1}{l}{$n=250$} & .4681 & .0885 & .0473 & .0325 & .0256 & .0221 & .0164 & .0137 & .0135 & .0121 & .0116 & .0117 & .0114 & .0108 & .0108 \\
          & \multicolumn{1}{l}{QN-D-SOMALA} & \multicolumn{1}{l}{$n=500$} & .4681 & .1066 & .0627 & .0427 & .0314 & .0256 & .0160 & .0132 & .0119 & .0113 & .0109 & .0107 & .0100 & .0099 & .0100 \\
          & \multicolumn{1}{l}{QN-D-SOMALA} & \multicolumn{1}{l}{$n=1,000$} & .4681 & .1250 & .0844 & .0613 & .0473 & .0383 & .0221 & .0165 & .0144 & .0129 & .0120 & .0115 & .0113 & .0110 & .0105 \\
          & \multicolumn{1}{l}{QN-D-SOMH} & \multicolumn{1}{l}{$n=250$} & .4681 & .1869 & .1519 & .1284 & .1148 & .1021 & .0744 & .0602 & .0529 & .0478 & .0440 & .0412 & .0385 & .0366 & .0351 \\
          & \multicolumn{1}{l}{QN-D-SOMH} & \multicolumn{1}{l}{$n=500$} & .4681 & .1914 & .1542 & .1303 & .1172 & .1066 & .0775 & .0641 & .0558 & .0502 & .0456 & .0422 & .0402 & .0377 & .0356 \\
          & \multicolumn{1}{l}{QN-D-SOMH} & \multicolumn{1}{l}{$n=1,000$} & .4681 & .1968 & .1645 & .1419 & .1286 & .1187 & .0894 & .0753 & .0656 & .0596 & .0548 & .0513 & .0484 & .0456 & .0435 \\
          & \multicolumn{2}{c}{QN-SOMALA} & .4681 & .1717 & .1465 & .1265 & .1112 & .1002 & .0696 & .0545 & .0446 & .0369 & .0315 & .0273 & .0242 & .0215 & .0191 \\
          & \multicolumn{2}{c}{QN-SOMH} & .4681 & .1989 & .1587 & .1402 & .1271 & .1169 & .0859 & .0703 & .0608 & .0537 & .0490 & .0452 & .0423 & .0399 & .0382 \\
    \midrule
    \multirow{14}[2]{*}{\begin{tabular}{c}  Covariance\\Matrix \end{tabular}} & \multicolumn{1}{l}{D-SOMALA} & \multicolumn{1}{l}{$n=250$} & .2045 & .0369 & .0207 & .0156 & .0131 & .0118 & .0097 & .0092 & .0088 & .0087 & .0085 & .0086 & .0085 & .0083 & .0084 \\
          & \multicolumn{1}{l}{D-SOMALA} & \multicolumn{1}{l}{$n=500$} & .2045 & .0449 & .0267 & .0189 & .0149 & .0131 & .0098 & .0089 & .0085 & .0085 & .0083 & .0082 & .0082 & .0082 & .0081 \\
          & \multicolumn{1}{l}{D-SOMALA} & \multicolumn{1}{l}{$n=1,000$} & .2045 & .0515 & .0362 & .0279 & .0225 & .0194 & .0129 & .0107 & .0096 & .0089 & .0085 & .0083 & .0080 & .0079 & .0077 \\
          & \multicolumn{1}{l}{D-SOMH} & \multicolumn{1}{l}{$n=250$} & .2045 & .0754 & .0592 & .0501 & .0442 & .0402 & .0294 & .0250 & .0224 & .0207 & .0195 & .0187 & .0179 & .0172 & .0167 \\
          & \multicolumn{1}{l}{D-SOMH} & \multicolumn{1}{l}{$n=500$} & .2045 & .0773 & .0598 & .0521 & .0469 & .0430 & .0321 & .0272 & .0244 & .0225 & .0210 & .0198 & .0189 & .0181 & .0175 \\
          & \multicolumn{1}{l}{D-SOMH} & \multicolumn{1}{l}{$n=1,000$} & .2045 & .0811 & .0652 & .0578 & .0530 & .0490 & .0380 & .0323 & .0289 & .0262 & .0245 & .0231 & .0220 & .0211 & .0202 \\
          & \multicolumn{1}{l}{QN-D-SOMALA} & \multicolumn{1}{l}{$n=250$} & .2045 & .0393 & .0218 & .0159 & .0134 & .0121 & .0099 & .0091 & .0089 & .0087 & .0085 & .0085 & .0086 & .0085 & .0085 \\
          & \multicolumn{1}{l}{QN-D-SOMALA} & \multicolumn{1}{l}{$n=500$} & .2045 & .0465 & .0269 & .0188 & .0149 & .0128 & .0097 & .0089 & .0085 & .0083 & .0082 & .0081 & .0081 & .0081 & .0081 \\
          & \multicolumn{1}{l}{QN-D-SOMALA} & \multicolumn{1}{l}{$n=1,000$} & .2045 & .0523 & .0345 & .0252 & .0197 & .0166 & .0110 & .0093 & .0087 & .0082 & .0079 & .0077 & .0077 & .0076 & .0076 \\
          & \multicolumn{1}{l}{QN-D-SOMH} & \multicolumn{1}{l}{$n=250$} & .2045 & .0825 & .0651 & .0547 & .0481 & .0432 & .0311 & .0259 & .0231 & .0212 & .0199 & .0188 & .0181 & .0173 & .0167 \\
          & \multicolumn{1}{l}{QN-D-SOMH} & \multicolumn{1}{l}{$n=500$} & .2045 & .0845 & .0662 & .0551 & .0490 & .0443 & .0325 & .0271 & .0240 & .0220 & .0204 & .0192 & .0182 & .0175 & .0169 \\
          & \multicolumn{1}{l}{QN-D-SOMH} & \multicolumn{1}{l}{$n=1,000$} & .2045 & .0875 & .0705 & .0592 & .0533 & .0490 & .0369 & .0309 & .0274 & .0249 & .0233 & .0218 & .0207 & .0197 & .0189 \\
          & \multicolumn{2}{c}{QN-SOMALA} & .2045 & .0783 & .0629 & .0525 & .0452 & .0405 & .0278 & .0223 & .0187 & .0162 & .0144 & .0130 & .0120 & .0112 & .0106 \\
          & \multicolumn{2}{c}{QN-SOMH} & .2045 & .0843 & .0647 & .0567 & .0509 & .0466 & .0341 & .0280 & .0245 & .0221 & .0205 & .0193 & .0184 & .0177 & .0171 \\
    \bottomrule
    \end{tabular}%
    }
  \label{tab:addlabel}%
\end{table}%
\end{landscape}

\clearpage

\begin{landscape}
% Table generated by Excel2LaTeX from sheet 'm2pl_K5'
\begin{table}[htbp]
  \centering
  \caption{The trajectory of the MAEs of the factor loading and intercept parameters for the confirmatory M2PL model with $K=5, J=50, N=10,000$}
    \resizebox{\linewidth}{!}{%
    \begin{tabular}{cccccccccccccccccc}
    \toprule
    \multirow{2}[4]{*}{Parameters} & \multicolumn{2}{c}{\multirow{2}[4]{*}{Methods}} & \multicolumn{15}{c}{Elapsed Time (seconds)} \\
\cmidrule{4-18}          & \multicolumn{2}{c}{} & 0     & 10    & 20    & 30    & 40    & 50    & 60    & 70    & 80    & 90    & 100   & 200   & 300   & 400   & 500 \\
    \midrule
    \multirow{14}[2]{*}{\begin{tabular}{c}          Factor \\Loadings     \end{tabular}} & \multicolumn{1}{l}{D-SOMALA} & \multicolumn{1}{l}{$n=250$} & .5419 & .0417 & .0410 & .0405 & .0395 & .0394 & .0395 & .0391 & .0389 & .0388 & .0385 & .0382 & .0380 & .0375 & .0376 \\
          & \multicolumn{1}{l}{D-SOMALA} & \multicolumn{1}{l}{$n=500$} & .5419 & .0404 & .0388 & .0387 & .0385 & .0382 & .0379 & .0378 & .0377 & .0379 & .0379 & .0374 & .0373 & .0372 & .0371 \\
          & \multicolumn{1}{l}{D-SOMALA} & \multicolumn{1}{l}{$n=1,000$} & .5419 & .0468 & .0376 & .0377 & .0374 & .0375 & .0372 & .0373 & .0372 & .0372 & .0371 & .0370 & .0369 & .0369 & .0369 \\
          \cdashline{2-18} \addlinespace[2pt] 
          & \multicolumn{1}{l}{D-SOMH} & \multicolumn{1}{l}{$n=250$} & .5419 & .3307 & .1071 & .0498 & .0427 & .0422 & .0420 & .0419 & .0417 & .0417 & .0412 & .0405 & .0396 & .0398 & .0398 \\
          & \multicolumn{1}{l}{D-SOMH} & \multicolumn{1}{l}{$n=500$} & .5419 & .3488 & .1292 & .0558 & .0429 & .0409 & .0403 & .0400 & .0403 & .0402 & .0398 & .0393 & .0386 & .0387 & .0387 \\
          & \multicolumn{1}{l}{D-SOMH} & \multicolumn{1}{l}{$n=1,000$} & .5419 & .3967 & .1877 & .0926 & .0552 & .0436 & .0403 & .0392 & .0387 & .0386 & .0386 & .0381 & .0379 & .0377 & .0378 \\
          \cdashline{2-18} \addlinespace[2pt] 
          & \multicolumn{1}{l}{QN-D-SOMALA} & \multicolumn{1}{l}{$n=250$} & .5419 & .0638 & .0573 & .0546 & .0533 & .0524 & .0510 & .0502 & .0505 & .0494 & .0488 & .0475 & .0472 & .0469 & .0461 \\
          & \multicolumn{1}{l}{QN-D-SOMALA} & \multicolumn{1}{l}{$n=500$} & .5419 & .0512 & .0480 & .0473 & .0457 & .0455 & .0446 & .0445 & .0440 & .0443 & .0437 & .0432 & .0433 & .0428 & .0424 \\
          & \multicolumn{1}{l}{QN-D-SOMALA} & \multicolumn{1}{l}{$n=1,000$} & .5419 & .0479 & .0457 & .0441 & .0446 & .0431 & .0424 & .0425 & .0419 & .0416 & .0418 & .0409 & .0411 & .0406 & .0405 \\
          \cdashline{2-18} \addlinespace[2pt] 
          & \multicolumn{1}{l}{QN-D-SOMH} & \multicolumn{1}{l}{$n=250$} & .5419 & .1015 & .0545 & .0517 & .0513 & .0495 & .0495 & .0489 & .0491 & .0493 & .0488 & .0489 & .0482 & .0487 & .0473 \\
          & \multicolumn{1}{l}{QN-D-SOMH} & \multicolumn{1}{l}{$n=500$} & .5419 & .1338 & .0512 & .0476 & .0469 & .0466 & .0464 & .0461 & .0465 & .0464 & .0460 & .0457 & .0456 & .0457 & .0450 \\
          & \multicolumn{1}{l}{QN-D-SOMH} & \multicolumn{1}{l}{$n=1,000$} & .5419 & .1698 & .0526 & .0451 & .0450 & .0447 & .0438 & .0436 & .0437 & .0438 & .0447 & .0435 & .0431 & .0430 & .0429 \\
          \cdashline{2-18} \addlinespace[2pt] 
          & \multicolumn{2}{c}{QN-SOMALA} & .5419 & .0435 & .0375 & .0373 & .0372 & .0372 & .0371 & .0370 & .0370 & .0370 & .0369 & .0370 & .0370 & .0370 & .0369 \\
          & \multicolumn{2}{c}{QN-SOMH} & .5419 & .2124 & .0810 & .0468 & .0397 & .0386 & .0383 & .0382 & .0379 & .0378 & .0377 & .0381 & .0377 & .0378 & .0376 \\
    \midrule
    \multirow{14}[2]{*}{Intercepts} & \multicolumn{1}{l}{D-SOMALA} & \multicolumn{1}{l}{$n=250$} & .9339 & .0288 & .0275 & .0281 & .0276 & .0271 & .0271 & .0268 & .0267 & .0268 & .0265 & .0259 & .0256 & .0254 & .0255 \\
          & \multicolumn{1}{l}{D-SOMALA} & \multicolumn{1}{l}{$n=500$} & .9339 & .0265 & .0262 & .0258 & .0256 & .0257 & .0258 & .0254 & .0255 & .0253 & .0253 & .0255 & .0249 & .0250 & .0254 \\
          & \multicolumn{1}{l}{D-SOMALA} & \multicolumn{1}{l}{$n=1,000$} & .9339 & .0263 & .0254 & .0254 & .0253 & .0253 & .0249 & .0250 & .0248 & .0248 & .0248 & .0253 & .0250 & .0249 & .0248 \\
          \cdashline{2-18} \addlinespace[2pt] 
          & \multicolumn{1}{l}{D-SOMH} & \multicolumn{1}{l}{$n=250$} & .9339 & .0315 & .0269 & .0275 & .0280 & .0283 & .0283 & .0278 & .0286 & .0283 & .0277 & .0274 & .0264 & .0266 & .0266 \\
          & \multicolumn{1}{l}{D-SOMH} & \multicolumn{1}{l}{$n=500$} & .9339 & .0332 & .0262 & .0261 & .0264 & .0262 & .0265 & .0262 & .0260 & .0260 & .0264 & .0263 & .0257 & .0264 & .0264 \\
          & \multicolumn{1}{l}{D-SOMH} & \multicolumn{1}{l}{$n=1,000$} & .9339 & .0394 & .0271 & .0251 & .0248 & .0254 & .0257 & .0255 & .0257 & .0257 & .0256 & .0262 & .0259 & .0256 & .0256 \\
          \cdashline{2-18} \addlinespace[2pt] 
          & \multicolumn{1}{l}{QN-D-SOMALA} & \multicolumn{1}{l}{$n=250$} & .9339 & .0444 & .0390 & .0377 & .0353 & .0358 & .0344 & .0342 & .0342 & .0336 & .0336 & .0326 & .0317 & .0316 & .0317 \\
          & \multicolumn{1}{l}{QN-D-SOMALA} & \multicolumn{1}{l}{$n=500$} & .9339 & .0352 & .0325 & .0310 & .0307 & .0302 & .0302 & .0296 & .0290 & .0286 & .0291 & .0294 & .0292 & .0288 & .0281 \\
          & \multicolumn{1}{l}{QN-D-SOMALA} & \multicolumn{1}{l}{$n=1,000$} & .9339 & .0322 & .0301 & .0300 & .0292 & .0282 & .0279 & .0285 & .0283 & .0278 & .0276 & .0269 & .0268 & .0271 & .0275 \\
          \cdashline{2-18} \addlinespace[2pt] 
          & \multicolumn{1}{l}{QN-D-SOMH} & \multicolumn{1}{l}{$n=250$} & .9339 & .0391 & .0358 & .0353 & .0348 & .0338 & .0334 & .0331 & .0334 & .0332 & .0332 & .0332 & .0327 & .0308 & .0308 \\
          & \multicolumn{1}{l}{QN-D-SOMH} & \multicolumn{1}{l}{$n=500$} & .9339 & .0336 & .0320 & .0312 & .0311 & .0309 & .0312 & .0308 & .0307 & .0307 & .0306 & .0301 & .0305 & .0295 & .0290 \\
          & \multicolumn{1}{l}{QN-D-SOMH} & \multicolumn{1}{l}{$n=1,000$} & .9339 & .0294 & .0296 & .0289 & .0286 & .0285 & .0282 & .0285 & .0281 & .0285 & .0287 & .0288 & .0286 & .0290 & .0283 \\
          \cdashline{2-18} \addlinespace[2pt] 
          & \multicolumn{2}{c}{QN-SOMALA} & .9339 & .0438 & .0354 & .0316 & .0294 & .0281 & .0271 & .0265 & .0260 & .0258 & .0258 & .0250 & .0250 & .0250 & .0249 \\
          & \multicolumn{2}{c}{QN-SOMH} & .9339 & .0330 & .0264 & .0256 & .0256 & .0253 & .0254 & .0254 & .0251 & .0251 & .0252 & .0255 & .0256 & .0255 & .0252 \\
    \bottomrule
    \end{tabular}%
    }
  \label{tab:addlabel}%
\end{table}%

\end{landscape}

\begin{landscape}
\begin{table}[htbp]
  \centering
  \caption{The trajectory of the MAEs of the correlation matrix for the confirmatory M2PL model with $K=5, J=50, N=10,000$}
    \resizebox{\linewidth}{!}{%
    \begin{tabular}{cccccccccccccccccc}
    \toprule
    \multirow{2}[4]{*}{Parameters} & \multicolumn{2}{c}{\multirow{2}[4]{*}{Methods}} & \multicolumn{15}{c}{Elapsed Time (seconds)} \\
\cmidrule{4-18}          & \multicolumn{2}{c}{} & 0     & 10    & 20    & 30    & 40    & 50    & 60    & 70    & 80    & 90    & 100   & 200   & 300   & 400   & 500 \\
    \midrule
    \multirow{14}[2]{*}{\begin{tabular}{c}          Correlation \\Matrix     \end{tabular}} & \multicolumn{1}{l}{D-SOMALA} & \multicolumn{1}{l}{$n=250$} & .5000 & .0189 & .0183 & .0167 & .0166 & .0166 & .0155 & .0157 & .0154 & .0156 & .0151 & .0142 & .0148 & .0140 & .0136 \\
          & \multicolumn{1}{l}{D-SOMALA} & \multicolumn{1}{l}{$n=500$} & .5000 & .0175 & .0156 & .0150 & .0154 & .0144 & .0145 & .0144 & .0142 & .0141 & .0147 & .0141 & .0138 & .0132 & .0130 \\
          & \multicolumn{1}{l}{D-SOMALA} & \multicolumn{1}{l}{$n=1,000$} & .5000 & .0172 & .0144 & .0141 & .0137 & .0141 & .0132 & .0133 & .0134 & .0137 & .0132 & .0128 & .0122 & .0122 & .0127 \\
          \cdashline{2-18} \addlinespace[2pt] 
          & \multicolumn{1}{l}{D-SOMH} & \multicolumn{1}{l}{$n=250$} & .5000 & .1573 & .0527 & .0234 & .0175 & .0168 & .0161 & .0157 & .0155 & .0156 & .0155 & .0150 & .0149 & .0141 & .0141 \\
          & \multicolumn{1}{l}{D-SOMH} & \multicolumn{1}{l}{$n=500$} & .5000 & .1614 & .0539 & .0246 & .0162 & .0146 & .0148 & .0148 & .0148 & .0149 & .0150 & .0143 & .0137 & .0143 & .0143 \\
          & \multicolumn{1}{l}{D-SOMH} & \multicolumn{1}{l}{$n=1,000$} & .5000 & .1843 & .0649 & .0329 & .0205 & .0153 & .0147 & .0140 & .0144 & .0143 & .0139 & .0138 & .0135 & .0136 & .0135 \\
          \cdashline{2-18} \addlinespace[2pt] 
          & \multicolumn{1}{l}{QN-D-SOMALA} & \multicolumn{1}{l}{$n=250$} & .5000 & .0202 & .0182 & .0175 & .0179 & .0172 & .0172 & .0162 & .0166 & .0161 & .0168 & .0161 & .0153 & .0162 & .0158 \\
          & \multicolumn{1}{l}{QN-D-SOMALA} & \multicolumn{1}{l}{$n=500$} & .5000 & .0169 & .0155 & .0155 & .0147 & .0151 & .0144 & .0145 & .0143 & .0147 & .0149 & .0144 & .0144 & .0141 & .0145 \\
          & \multicolumn{1}{l}{QN-D-SOMALA} & \multicolumn{1}{l}{$n=1,000$} & .5000 & .0159 & .0157 & .0150 & .0147 & .0146 & .0143 & .0143 & .0138 & .0139 & .0139 & .0134 & .0143 & .0132 & .0134 \\
          \cdashline{2-18} \addlinespace[2pt] 
          & \multicolumn{1}{l}{QN-D-SOMH} & \multicolumn{1}{l}{$n=250$} & .5000 & .0618 & .0192 & .0167 & .0165 & .0160 & .0163 & .0153 & .0160 & .0154 & .0154 & .0153 & .0156 & .0152 & .0157 \\
          & \multicolumn{1}{l}{QN-D-SOMH} & \multicolumn{1}{l}{$n=500$} & .5000 & .0786 & .0230 & .0160 & .0153 & .0149 & .0147 & .0152 & .0157 & .0150 & .0149 & .0150 & .0153 & .0151 & .0144 \\
          & \multicolumn{1}{l}{QN-D-SOMH} & \multicolumn{1}{l}{$n=1,000$} & .5000 & .0908 & .0277 & .0161 & .0146 & .0149 & .0149 & .0149 & .0146 & .0149 & .0146 & .0145 & .0146 & .0138 & .0141 \\
          \cdashline{2-18} \addlinespace[2pt] 
          & \multicolumn{2}{c}{QN-SOMALA} & .5000 & .0173 & .0123 & .0116 & .0115 & .0114 & .0113 & .0113 & .0113 & .0113 & .0113 & .0113 & .0114 & .0112 & .0113 \\
          & \multicolumn{2}{c}{QN-SOMH} & .5000 & .0600 & .0265 & .0165 & .0140 & .0132 & .0127 & .0126 & .0125 & .0123 & .0122 & .0122 & .0124 & .0121 & .0122 \\
    \bottomrule
    \end{tabular}%
    }
  \label{tab:addlabel}%
\end{table}%
\end{landscape}

\begin{landscape}
% Table generated by Excel2LaTeX from sheet 'm2pl_K10'
\begin{table}[htbp]
  \centering
  \caption{The trajectory of the MAEs of the factor loading and intercept parameters for the confirmatory M2PL model with $K=10, J=200, N=10,000$}
    \resizebox{\linewidth}{!}{%
    \begin{tabular}{cccccccccccccccccc}
    \toprule
    \multirow{2}[4]{*}{Parameters} & \multicolumn{2}{c}{\multirow{2}[4]{*}{Methods}} & \multicolumn{15}{c}{Elapsed Time (seconds)} \\
\cmidrule{4-18}          & \multicolumn{2}{c}{} & 0     & 20    & 40    & 60    & 80    & 100   & 200   & 300   & 400   & 500   & 600   & 700   & 800   & 900   & 1000 \\
    \midrule
    \multirow{14}[2]{*}{\begin{tabular}{c}          Factor \\Loadings     \end{tabular}} & \multicolumn{1}{l}{D-SOMALA} & \multicolumn{1}{l}{$n=250$} & .5423 & .0738 & .0416 & .0364 & .0353 & .0351 & .0346 & .0342 & .0339 & .0338 & .0338 & .0337 & .0335 & .0335 & .0335 \\
          & \multicolumn{1}{l}{D-SOMALA} & \multicolumn{1}{l}{$n=500$} & .5423 & .0832 & .0446 & .0367 & .0346 & .0340 & .0334 & .0333 & .0332 & .0330 & .0330 & .0330 & .0329 & .0329 & .0329 \\
          & \multicolumn{1}{l}{D-SOMALA} & \multicolumn{1}{l}{$n=1,000$} & .5423 & .1408 & .0795 & .0563 & .0457 & .0401 & .0334 & .0327 & .0327 & .0327 & .0327 & .0326 & .0326 & .0326 & .0325 \\
          \cdashline{2-18} \addlinespace[2pt] 
          & \multicolumn{1}{l}{D-SOMH} & \multicolumn{1}{l}{$n=250$} & .5423 & .4645 & .2786 & .1688 & .1116 & .0792 & .0378 & .0354 & .0352 & .0351 & .0351 & .0349 & .0350 & .0349 & .0349 \\
          & \multicolumn{1}{l}{D-SOMH} & \multicolumn{1}{l}{$n=500$} & .5423 & .4772 & .2967 & .1874 & .1287 & .0937 & .0402 & .0348 & .0342 & .0341 & .0341 & .0341 & .0341 & .0339 & .0339 \\
          & \multicolumn{1}{l}{D-SOMH} & \multicolumn{1}{l}{$n=1,000$} & .5423 & .4972 & .3320 & .2255 & .1634 & .1245 & .0516 & .0373 & .0343 & .0335 & .0334 & .0333 & .0333 & .0332 & .0332 \\
          \cdashline{2-18} \addlinespace[2pt] 
          & \multicolumn{1}{l}{QN-D-SOMALA} & \multicolumn{1}{l}{$n=250$} & .5423 & .0810 & .0539 & .0506 & .0491 & .0485 & .0454 & .0440 & .0431 & .0423 & .0421 & .0417 & .0414 & .0411 & .0410 \\
          & \multicolumn{1}{l}{QN-D-SOMALA} & \multicolumn{1}{l}{$n=500$} & .5423 & .0806 & .0458 & .0432 & .0424 & .0419 & .0399 & .0394 & .0389 & .0389 & .0384 & .0384 & .0382 & .0380 & .0378 \\
          & \multicolumn{1}{l}{QN-D-SOMALA} & \multicolumn{1}{l}{$n=1,000$} & .5423 & .0675 & .0417 & .0390 & .0384 & .0381 & .0373 & .0370 & .0365 & .0361 & .0362 & .0360 & .0359 & .0359 & .0359 \\
          \cdashline{2-18} \addlinespace[2pt] 
          & \multicolumn{1}{l}{QN-D-SOMH} & \multicolumn{1}{l}{$n=250$} & .5423 & .5419 & .3605 & .2315 & .1616 & .1189 & .0550 & .0450 & .0424 & .0421 & .0418 & .0415 & .0417 & .0413 & .0415 \\
          & \multicolumn{1}{l}{QN-D-SOMH} & \multicolumn{1}{l}{$n=500$} & .5423 & .5405 & .3606 & .2328 & .1618 & .1207 & .0537 & .0430 & .0400 & .0392 & .0389 & .0388 & .0386 & .0388 & .0389 \\
          & \multicolumn{1}{l}{QN-D-SOMH} & \multicolumn{1}{l}{$n=1,000$} & .5423 & .5452 & .3763 & .2536 & .1809 & .1360 & .0592 & .0436 & .0393 & .0378 & .0371 & .0371 & .0369 & .0370 & .0370 \\
          \cdashline{2-18} \addlinespace[2pt] 
          & \multicolumn{2}{c}{QN-SOMALA} & .5423 & .1398 & .0897 & .0676 & .0554 & .0483 & .0364 & .0340 & .0332 & .0328 & .0327 & .0326 & .0325 & .0325 & .0325 \\
          & \multicolumn{2}{c}{QN-SOMH} & .5423 & .7666 & .4346 & .3480 & .2904 & .2502 & .1546 & .1150 & .0852 & .0669 & .0559 & .0489 & .0443 & .0413 & .0391 \\
    \midrule
    \multirow{14}[2]{*}{Intercepts} & \multicolumn{1}{l}{D-SOMALA} & \multicolumn{1}{l}{$n=250$} & .9245 & .0301 & .0288 & .0287 & .0282 & .0282 & .0277 & .0275 & .0274 & .0271 & .0271 & .0273 & .0274 & .0272 & .0267 \\
          & \multicolumn{1}{l}{D-SOMALA} & \multicolumn{1}{l}{$n=500$} & .9245 & .0281 & .0274 & .0271 & .0268 & .0262 & .0262 & .0261 & .0259 & .0260 & .0259 & .0260 & .0261 & .0260 & .0260 \\
          & \multicolumn{1}{l}{D-SOMALA} & \multicolumn{1}{l}{$n=1,000$} & .9245 & .0340 & .0318 & .0308 & .0299 & .0294 & .0279 & .0273 & .0267 & .0264 & .0261 & .0260 & .0259 & .0257 & .0258 \\
          \cdashline{2-18} \addlinespace[2pt] 
          & \multicolumn{1}{l}{D-SOMH} & \multicolumn{1}{l}{$n=250$} & .9245 & .0318 & .0271 & .0273 & .0271 & .0273 & .0280 & .0283 & .0285 & .0282 & .0284 & .0284 & .0284 & .0285 & .0285 \\
          & \multicolumn{1}{l}{D-SOMH} & \multicolumn{1}{l}{$n=500$} & .9245 & .0327 & .0260 & .0257 & .0261 & .0260 & .0264 & .0268 & .0270 & .0273 & .0270 & .0267 & .0269 & .0269 & .0269 \\
          & \multicolumn{1}{l}{D-SOMH} & \multicolumn{1}{l}{$n=1,000$} & .9245 & .0360 & .0267 & .0252 & .0251 & .0250 & .0253 & .0255 & .0253 & .0254 & .0252 & .0253 & .0256 & .0255 & .0257 \\
          \cdashline{2-18} \addlinespace[2pt] 
          & \multicolumn{1}{l}{QN-D-SOMALA} & \multicolumn{1}{l}{$n=250$} & .9245 & .0482 & .0415 & .0397 & .0382 & .0373 & .0353 & .0340 & .0336 & .0332 & .0333 & .0327 & .0322 & .0322 & .0331 \\
          & \multicolumn{1}{l}{QN-D-SOMALA} & \multicolumn{1}{l}{$n=500$} & .9245 & .0391 & .0351 & .0340 & .0333 & .0323 & .0313 & .0321 & .0312 & .0306 & .0308 & .0309 & .0301 & .0301 & .0304 \\
          & \multicolumn{1}{l}{QN-D-SOMALA} & \multicolumn{1}{l}{$n=1,000$} & .9245 & .0337 & .0325 & .0315 & .0304 & .0305 & .0297 & .0290 & .0285 & .0290 & .0291 & .0292 & .0288 & .0290 & .0284 \\
          \cdashline{2-18} \addlinespace[2pt] 
          & \multicolumn{1}{l}{QN-D-SOMH} & \multicolumn{1}{l}{$n=250$} & .9245 & .0521 & .0378 & .0354 & .0349 & .0343 & .0328 & .0324 & .0321 & .0319 & .0321 & .0321 & .0319 & .0316 & .0316 \\
          & \multicolumn{1}{l}{QN-D-SOMH} & \multicolumn{1}{l}{$n=500$} & .9245 & .0454 & .0317 & .0306 & .0308 & .0302 & .0305 & .0309 & .0305 & .0305 & .0305 & .0304 & .0307 & .0305 & .0310 \\
          & \multicolumn{1}{l}{QN-D-SOMH} & \multicolumn{1}{l}{$n=1,000$} & .9245 & .0438 & .0291 & .0278 & .0276 & .0278 & .0292 & .0294 & .0290 & .0288 & .0288 & .0291 & .0289 & .0285 & .0284 \\
          \cdashline{2-18} \addlinespace[2pt] 
          & \multicolumn{2}{c}{QN-SOMALA} & .9245 & .0531 & .0476 & .0457 & .0443 & .0433 & .0404 & .0386 & .0376 & .0370 & .0363 & .0355 & .0348 & .0344 & .0340 \\
          & \multicolumn{2}{c}{QN-SOMH} & .9245 & .1383 & .0610 & .0444 & .0396 & .0374 & .0342 & .0329 & .0318 & .0313 & .0310 & .0307 & .0304 & .0302 & .0301 \\
    \bottomrule
    \end{tabular}%
    }
  \label{tab:addlabel}%
\end{table}%

% Table generated by Excel2LaTeX from sheet 'm2pl_K10'
\begin{table}[htbp]
  \centering
  \caption{The trajectory of the MAEs of the correlation matrix for the confirmatory M2PL model with $K=10, J=200, N=10,000$}
    \resizebox{\linewidth}{!}{%
    \begin{tabular}{cccccccccccccccccc}
    \toprule
    \multirow{2}[4]{*}{Parameters} & \multicolumn{2}{c}{\multirow{2}[4]{*}{Methods}} & \multicolumn{15}{c}{Elapsed Time (seconds)} \\
\cmidrule{4-18}          & \multicolumn{2}{c}{} & 0     & 20    & 40    & 60    & 80    & 100   & 200   & 300   & 400   & 500   & 600   & 700   & 800   & 900   & 1000 \\
    \midrule
    \multirow{14}[2]{*}{\begin{tabular}{c}          Correlation \\Matrix     \end{tabular}} & \multicolumn{1}{l}{D-SOMALA} & \multicolumn{1}{l}{$n=250$} & .5000 & .0600 & .0253 & .0175 & .0145 & .0136 & .0122 & .0113 & .0111 & .0112 & .0111 & .0107 & .0106 & .0105 & .0105 \\
          & \multicolumn{1}{l}{D-SOMALA} & \multicolumn{1}{l}{$n=500$} & .5000 & .0620 & .0281 & .0175 & .0137 & .0123 & .0102 & .0102 & .0101 & .0099 & .0098 & .0099 & .0095 & .0095 & .0095 \\
          & \multicolumn{1}{l}{D-SOMALA} & \multicolumn{1}{l}{$n=1,000$} & .5000 & .0924 & .0548 & .0373 & .0266 & .0208 & .0116 & .0098 & .0094 & .0092 & .0094 & .0093 & .0093 & .0091 & .0090 \\
          \cdashline{2-18} \addlinespace[2pt] 
          & \multicolumn{1}{l}{D-SOMH} & \multicolumn{1}{l}{$n=250$} & .5000 & .2263 & .1655 & .1237 & .0882 & .0642 & .0176 & .0117 & .0110 & .0108 & .0109 & .0106 & .0106 & .0106 & .0106 \\
          & \multicolumn{1}{l}{D-SOMH} & \multicolumn{1}{l}{$n=500$} & .5000 & .2305 & .1672 & .1288 & .0960 & .0732 & .0215 & .0116 & .0103 & .0099 & .0100 & .0101 & .0099 & .0100 & .0100 \\
          & \multicolumn{1}{l}{D-SOMH} & \multicolumn{1}{l}{$n=1,000$} & .5000 & .2389 & .1715 & .1396 & .1115 & .0889 & .0348 & .0172 & .0122 & .0102 & .0098 & .0096 & .0094 & .0096 & .0095 \\
          \cdashline{2-18} \addlinespace[2pt] 
          & \multicolumn{1}{l}{QN-D-SOMALA} & \multicolumn{1}{l}{$n=250$} & .5000 & .0622 & .0183 & .0144 & .0141 & .0139 & .0131 & .0124 & .0122 & .0121 & .0121 & .0116 & .0114 & .0118 & .0115 \\
          & \multicolumn{1}{l}{QN-D-SOMALA} & \multicolumn{1}{l}{$n=500$} & .5000 & .0657 & .0175 & .0127 & .0118 & .0116 & .0113 & .0111 & .0107 & .0108 & .0105 & .0110 & .0106 & .0106 & .0103 \\
          & \multicolumn{1}{l}{QN-D-SOMALA} & \multicolumn{1}{l}{$n=1,000$} & .5000 & .0545 & .0185 & .0118 & .0107 & .0105 & .0103 & .0103 & .0100 & .0099 & .0098 & .0099 & .0099 & .0098 & .0097 \\
          \cdashline{2-18} \addlinespace[2pt] 
          & \multicolumn{1}{l}{QN-D-SOMH} & \multicolumn{1}{l}{$n=250$} & .5000 & .2804 & .1892 & .1544 & .1236 & .0973 & .0368 & .0200 & .0140 & .0121 & .0114 & .0114 & .0114 & .0113 & .0113 \\
          & \multicolumn{1}{l}{QN-D-SOMH} & \multicolumn{1}{l}{$n=500$} & .5000 & .2768 & .1879 & .1526 & .1229 & .0981 & .0392 & .0209 & .0143 & .0120 & .0112 & .0106 & .0104 & .0105 & .0105 \\
          & \multicolumn{1}{l}{QN-D-SOMH} & \multicolumn{1}{l}{$n=1,000$} & .5000 & .2793 & .1908 & .1586 & .1314 & .1069 & .0460 & .0260 & .0175 & .0134 & .0115 & .0106 & .0105 & .0102 & .0102 \\
          \cdashline{2-18} \addlinespace[2pt] 
          & \multicolumn{2}{c}{QN-SOMALA} & .5000 & .0988 & .0668 & .0502 & .0401 & .0333 & .0173 & .0125 & .0106 & .0096 & .0091 & .0086 & .0086 & .0084 & .0084 \\
          & \multicolumn{2}{c}{QN-SOMH} & .5000 & .3458 & .2245 & .1693 & .1514 & .1397 & .1024 & .0806 & .0618 & .0483 & .0395 & .0328 & .0277 & .0239 & .0209 \\
    \bottomrule
    \end{tabular}%
    }
  \label{tab:addlabel}%
\end{table}%

\end{landscape}

\clearpage

\addcontentsline{toc}{subsection}{A.3 Pseudocode of the SOMALA and D-SOMALA Algorithms}
\subsection*{A.3 Pseudocode of the SOMALA and D-SOMALA Algorithms}\label{supm:A3}

\RestyleAlgo{ruled}

\begin{algorithm} [htbp]
\caption{SOMALA Algorithm}\label{alg:SOMALA}

\textbf{Input:} The initial values of model parameter $\bbeta^{(0)}$ and latent variables $\bxi^{(0)}$, and step sizes $h$ and $\gamma_1, \gamma_2, \ldots$.

\vspace{0.2cm}
\textbf{for} iterations $t = 1, 2..., $ \textbf{do}

\vspace{0.2cm}
\begin{enumerate}
\item For $i = 1, \ldots, N$, sample $\bZ_i^\iter{t}$
    from $K$-variate standard normal distribution and update $\bxi_i^{(t)}$ by 
     $$\bxi_i^{(*)} = \bxi_i^{(t-1)} -h \nabla U_i^\iter{t-1} + \sqrt{2h} \bZ_i^\iter{t}, $$
     where $\nabla U_i^\iter{t-1}$ is computed by \eqref{eq:ulsgd}. 
\item Obtain the posterior sample at the $t$-th iteration by 
$$\bxi_i^\iter{t} = \bxi_i^{(t-1)} + \mathbb{I}_{\bbR^+}\nami{ \alpha_h\maru{\bxi_i^{(*)}, \bxi_i^{(t-1)}}  - W_i^\iter{t}}\maru{ \bxi_i^{(*)} -  \bxi_i^{(t-1)}},$$
where $W_i^\iter{t}$ is sampled from the uniform distribution on $[0,1]$ and the MH ratio $\alpha_h\maru{\bxi_i^{(*)}, \bxi_i^{(t-1)}}$ is computed by \eqref{eq:mhratio}.
\item Compute $\bD^{(t)}$ based on \textcite{zhang_computation_2022} if a quasi-Newton update is applied. Otherwise, $\bD^{(t)} = \btI_{p}$.
\item Update the model parameter by the following SG ascent update
$$\bbeta^{(t)} = \bbeta^{(t-1)} + \gamma_t  {\left(\bD^{(t)}\right)}^{-1}\bG_{\bbeta^\iter{t-1}}(\bxi^\iter{t}, S^\iter{t})$$
when the parameter space is unconstrained, and by the fullbatch quasi-Newton proximal update 
$$\bbeta^{(t)} = \argmin_{\bbeta \in {\mathcal B}} \Vert \bbeta - \bbeta^{(t-1)} - \gamma_t  {(\bD^{(t)})}^{-1}\bG_{\bbeta^\iter{t-1}}(\bxi^\iter{t}, S^\iter{t}) \Vert_{\bD^{(t)}}^2. $$
\end{enumerate}
\textbf{until} stopping decision is met. Let $T$ be the final iteration number.

\vspace{0.2cm}
\textbf{Output}:  $\hat \bbeta = \bbeta^{(T)}$. If the Polyak-Ruppert trajectory average is adopted, compute it from $\nami{\bbeta^\iter{T'}, \ldots, \bbeta^\iter{T}}$, where $T'$ denotes the iteration number from which the sequence of updated model parameters is averaged, i.e., the average is computed over iterations $T', T'+1, \ldots, T$.
\end{algorithm}

\RestyleAlgo{ruled}

\begin{algorithm}[htbp]
\caption{D-SOMALA Algorithm}\label{alg:D-SOMALA}

\scalebox{0.95}{

\begin{minipage}{\linewidth}
\textbf{Input:} Minibatch size $n$, the initial values of model parameter $\bbeta^{(0)}$ and latent variables $\bxi^{(0)}$, step sizes $h$ and $\gamma_1, \gamma_2, \ldots$, and $\bD^{(1)} = \btI_p$.

\vspace{0.2cm}
\textbf{for} iterations $t = 1, 2..., $ \textbf{do}

\vspace{0.2cm}
Let $\bbeta^\iter{s-1} = \bbeta^\iter{t-1}$, $\bD^\iter{s-1} = \bD^\iter{t-1}$, and $\bxi_i^\iter{s-1} = \bxi_i^\iter{t-1}$ for all $i = 1, \ldots, N$.

\vspace{0.2cm}
\hspace{\parindent}\textbf{for} $s = 1, 2..., \tilde{t}  (\tilde{t}=\floor{\frac{N}{n}})$ \textbf{do}

\vspace{0.2cm}
\begin{enumerate}
    \item Sample minibatch $S^{(s)} \subset\{1, ..., N\}$. 
    \item For all $i \in S^{(s)}$, sample $\bZ_i^\iter{s}$
    from $K$-variate standard normal distribution and generate $\bxi_i^{(*)}$ by 
     $$\bxi_i^{(*)} = \bxi_i^{(s-1)} -h \nabla U_i^\iter{s-1} + \sqrt{2h} \bZ_i^\iter{s}, $$
     where $\nabla U_i^\iter{s-1}$ is computed by \eqref{eq:ulsgd}. 
\item Obtain the posterior sample at the $t$-th iteration by 
$$\bxi_i^\iter{s} = \bxi_i^{(s-1)} + \mathbb{I}_{\bbR^+}\nami{ \alpha_h\maru{\bxi_i^{(*)}, \bxi_i^{(s-1)}}  - W_i^\iter{s}}\maru{ \bxi_i^{(*)} -  \bxi_i^{(s-1)}},$$
where $W_i^\iter{s}$ is sampled from the uniform distribution on $[0,1]$ and the MH ratio $\alpha_h\maru{\bxi_i^{(*)}, \bxi_i^{(s-1)}}$ is computed by \eqref{eq:mhratio}. For $i \notin S^{(s)}$, $\bxi_i^\iter{s} = \bxi_i^\iter{s-1}$.
\item Construct the minibatch SG 
$$\bG_{\bbeta^\iter{s-1}}^{\text{mini}}(\bxi^\iter{s}, S^\iter{s}) = \frac{N}{n} \summa{i}{1}{N} \mathbbm{1}_{\nami{i \in S^\iter{s}}} \left.\left( \frac{\partial \log\nami{f_i(\bY_i, \bxi_i^\iter{s}\vert \bbeta)}}{\partial \bbeta}\right)\right\vert_{\bbeta = \bbeta^\iter{s-1}}.$$
\item Using $\bxi_i^\iter{s}$ whose $i$ is in $S^\iter{s}$, compute $\bD^{(s)}$ based on \textcite{zhang_computation_2022} if a quasi-Newton update is applied. Otherwise, $\bD^{(s)} = \btI_{p}$.
\item  Update the model parameter by the minibatch SG update
$$\bbeta^{(s)} = \bbeta^{(s-1)} + \gamma_t  {\left(\bD^{(s)}\right)}^{-1}\bG_{\bbeta^\iter{s-1}}^{\text{mini}}(\bxi^\iter{s}, S^\iter{s})$$
when the parameter space is unconstrained, and by the minibatch 
quasi-Newton proximal update 
$$\bbeta^{(s)} = \argmin_{\bbeta \in {\mathcal B}} \Vert \bbeta - \bbeta^{(s-1)} - \gamma_t  {(\bD^{(s)})}^{-1}\bG_{\bbeta^\iter{s-1}}^{\text{mini}}(\bxi^\iter{s}, S^\iter{s}) \Vert_{\bD^{(t)}}^2. $$
\end{enumerate}

\textbf{end}

Let $\bbeta^\iter{t} = \bbeta^\iter{\tilde{t}}$, $\bD^\iter{t} = \bD^\iter{\tilde{t}}$, and $\bxi_i^\iter{t} = \bxi_i^\iter{\tilde{t}}$ for all $i = 1, \ldots, N$.

\textbf{until} stopping decision is met. Let $T$ be the final iteration number.

\vspace{0.2cm}
\textbf{Output}:  $\hat \bbeta = \bbeta^{(T)}$. If the Polyak-Ruppert trajectory average is adopted, compute it from $\nami{\bbeta^\iter{T'}, \ldots, \bbeta^\iter{T}}$, where $T'$ denotes the iteration number from which the sequence of updated model parameters is averaged, i.e., the average is computed over iterations $T', T'+1, \ldots, T$.

\end{minipage}
}
\end{algorithm}

\clearpage
\addcontentsline{toc}{subsection}{A.4 Implementation Details}
\subsection*{A.4 Implementation Details}\label{supm:A4}

We provide the implementation details of the stochastic optimisation algorithms. 

First, there are some approaches to a stopping criterion of estimation algorithms with stochastic optimisation, such as monitoring a window of consecutive differences of $\bbeta^\iter{t}$ given a certain window size \parencite[e.g., 3;][]{zhang_computation_2022}. In this study, we introduce the following stopping criterion. Let the window size be denoted as $\tilde{w}$. Denote $\bar{\bbeta}^\iter{\tilde{w}}$ as the average of model parameters updated within the last $\tilde{w}$ epochs. Note that the number of epochs means the cumulative number of observations used to construct the minibatch SG, divided by $n$. For instance, one epoch implies running the algorithm until the number of observations used to construct the minibatch SG during the iterations reaches $N$.  

Every after $\tilde{w}$ epochs, we sequentially obtain the average of updated model parameters given window size $\tilde{w}$, i.e., $\nami{\bar{\bbeta}^\iter{\tilde{w}}, \bar{\bbeta}^\iter{2\tilde{w}}, \ldots}$. For instance, when $t = 2\tilde{w}$, we have $\bar{\bbeta}^\iter{2\tilde{w}} = \tilde{w}^{-1}\sum_{w = \tilde{w} + 1}^{2\tilde{w}} \bbeta^\iter{w}$.
Then, we compute the absolute difference between the averaged value from epoch $t-\tilde{w}+1$ to epoch $t$ and the one from epoch $t-2\tilde{w}+1$ to epoch $t-\tilde{w}$ every after $\tilde{w}$ epochs and save the maximum of the absolute difference of model parameters in such a manner that, for example, when the algorithm reaches $\tilde{w}$ and $2\tilde{w}$ epochs, the maximum value is computed as
\begin{align}
    &\text{DIFF}_{\text{\scriptsize MAX}}(\tilde{w}) = \max_{q \in \nami{1, \ldots, p}} \abs{\bar{\beta}_{q}^\iter{\tilde{w}} - \beta_q^\iter{0}},\\
    &\text{DIFF}_{\text{\scriptsize MAX}}(2\tilde{w}) = \max_{q \in \nami{1, \ldots, p}} \abs{\bar{\beta}_{q}^\iter{2\tilde{w}} - \bar{\beta}_{q}^\iter{\tilde{w}}},
\end{align}
respectively. Subsequently, we monitor the trajectory of the maximum absolute difference and terminate the algorithm once the value of $\text{DIFF}_{\text{\scriptsize MAX}}(\cdot)$ remains below a pre-specified threshold consecutively. This convergence criterion is based on the idea that, after a stochastic optimisation algorithm has converged, the iterate-averaged estimates computed over two consecutive windows of size $\tilde{w}$ should be nearly identical. The convergence criterion was applied to real data analysis. 

Second, the observed information matrix can be computed via the Fisher identity and stochastic approximation procedure. More precisely, by the Fisher identity, the $i$-th contribution of the score function of the marginal log-likelihood function given the model parameter updated at the $t$-th iteration can be written as
\begin{align}
 \nabla  \ell_i \maru{\bbeta^\iter{t}} &= \mathbb{E}_{p_{\bbeta^\iter{t}, i}}\kagi{ \nabla \log f_i \maru{\bY_i, \bxi_i \vert \bbeta^\iter{t}} },
\end{align}
where $\ell_i\maru{\bbeta}$ is the $i$-th contribution of the marginal log-likelihood function:
\begin{align}
    \ell_i\maru{\bbeta} = \log \maru{\int f_i \maru{\bY_i, \bxi_i \vert \bbeta } d\bxi_i}.
\end{align}
The observed information given the model parameter updated at the $t$-th iteration can be computed as
\begin{align}
     I\maru{\bbeta^\iter{t}} &= \frac{1}{N} \sum_{i=1}^N \mathbb{E}_{p_{\bbeta^\iter{t}, i}}\kagi{ \nabla \log f_i \maru{\bY_i, \bxi_i \vert \bbeta^\iter{t}} }\maru{\mathbb{E}_{p_{\bbeta^\iter{t}, i}}\kagi{ \nabla \log f_i \maru{\bY_i, \bxi_i \vert \bbeta^\iter{t}} }}^\top.
\end{align}
However, since the posterior sample of latent variables given $\bbeta^\iter{t}$, i.e., $\bxi_i^\iter{t+1}$, is available from the sampling step, by approximating $\mathbb{E}_{p_{\bbeta^\iter{t}, i}}\kagi{ \nabla \log f_i \maru{\bY_i, \bxi_i \vert \bbeta^\iter{t}} }$ with its Monte Carlo approximation, i.e., $\mathbb{E}_{p_{\bbeta^\iter{t}, i}}\kagi{ \nabla \log f_i \maru{\bY_i, \bxi_i \vert \bbeta^\iter{t}} } \approx \nabla \log f_i \maru{\bY_i, \bxi_i^\iter{t+1} \vert \bbeta^\iter{t}}$, the observed information matrix given the model parameter updated at the $t$-th iteration can be approximated as
\begin{align}
    \tilde{I}\maru{\bbeta^\iter{t}, \bxi_1^\iter{t+1}, \ldots, \bxi_N^\iter{t+1}} &= \frac{1}{N} \sum_{i=1}^N \nabla \log f_i \maru{\bY_i, \bxi_i^\iter{t+1} \vert \bbeta^\iter{t}} \maru{\nabla \log f_i \maru{\bY_i, \bxi_i^\iter{t+1} \vert \bbeta^\iter{t}}}^\top.
\end{align}
Subsequently, the recursive approximation of the observed information matrix can be constructed as
\begin{align}
    I^\iter{t} &= I^\iter{t-1} + \gamma_{t}\maru{\tilde{I}\maru{\bbeta^\iter{t}, \bxi_1^\iter{t+1}, \ldots, \bxi_N^\iter{t+1}} - I^\iter{t-1}}.
\end{align}
Then, its iterate-averaged value is the estimate of the observed information matrix: $\hat{I} =  \frac{1}{T - t'}\sum_{t=t'+1}^T I^\iter{t}$, where $t'$ is the iteration number at which the burn-in period ends.  

Lastly, estimating the marginal log-likelihood value under the stochastic optimisation algorithms can be performed by the importance sampling estimator \parencite[Chapter 11; ][]{bishop_pattern_2006}. The details of the proposed estimation procedure is provided in Supplementary Materials A.4. 
  
\clearpage

\addcontentsline{toc}{subsection}{A.5 Estimating the Log Marginal Likelihood Value under the Stochastic Optimisation Algorithms}
\subsection*{A.5 Estimating the Log Marginal Likelihood Value under the Stochastic Optimisation Algorithms}\label{supm:A4}

\addcontentsline{toc}{subsubsection}{A.5.1 Importance Sampling Estimator for the Log Marginal Likelihood Value}
\subsubsection*{A.5.1 Importance Sampling Estimator for the Log Marginal Likelihood Value}

Although there are a wide variety of methods to approximate the intractable log marginal likelihood \parencite[e.g., ][]{gronau_tutorial_2017,newton_approximate_1994}, our focus is placed on the importance sampling estimator for the ease of implementation. The importance sampling estimator for the $i$-th contribution of the log marginal likelihood is given as
\begin{align}
 \bxi_i^{*\iter{1}}, \ldots, \bxi_i^{*\iter{T}} &\sim \pi^*(\bxi_i^*)  \\
 w_i^\iter{t} &= \frac{\pi\maru{\bxi_i^{*\iter{t}}}}{ \pi^*(\bxi_i^{*\iter{t}}) } \\
 \log \int f_i\maru{\bY_i, \bxi_i \vert \bbeta}d\bxi_i &\approx \log  \maru{ \frac{1}{\sum_{t=1}^T w_i^\iter{t} }\nami{ \sum_{t=1}^T w_i^\iter{t} f_i\maru{\bY_i \vert \bxi_i^{*\iter{t}}, \bbeta } } }\label{eq:estimatinglogmarginallik},
\end{align}
where $T$ is the number of samples used to approximate the log marginal likelihood; $\pi(\bxi_i)$ and $\pi^*(\bxi_i)$ are the prior distribution of latent variables and the corresponding importance density from which $\bxi_i^{*\iter{1}}, \ldots, \bxi_i^{*\iter{T}}$ are sampled; and $w_i^\iter{t}$ is the importance ratio. 

The key question in the importance sampling estimator is the choice of the importance density $\pi^*(\bxi_i)$. According to \textcite{gronau_tutorial_2017,neal_annealed_2001,busemeyer_model_2015}, a proper importance density in our context should be easy to evaluate; its domain is the same as the posterior distribution of latent variables; the importance density of latent variables has a shape relatively similar to their posterior distribution; and the importance density has heavier tails than the posterior distribution. 
Based on these points, the importance density of latent variables is specified by finding a parametric form of the empirical distribution based on the sampled latent variables during the optimisation of the stochastic optimisation algorithm and making this distribution have heavier tails. For example, assuming the posterior normality of latent variables, the multivariate normal distribution can be assigned as the importance density of latent variables, and its parameters can be estimated by the sampled latent variables during the optimisation. Then, one can rescale the diagonal elements of the estimated covariance matrix make the importance density have heavier tails.

\end{document}